Istituto Nazionale di Fisica Nucleare

# STAR HE-Linac
# Complete Detailed Design Report

A. Bacci[1], L. Faillace[2], L. Pellegrino[2], D. Alesini[2], S. Bini[2], F. Cardelli[2], G. Catuscelli[2], F. Chiarelli[2], I. Drebot[1], A. Esposito[2], A. Gallo[2], A. Ghigo[2], D. Giannotti[1], V. Petrillo[1,3], L. Piersanti[2], E. Puppin[1,4], M. Rossetti Conti[1], L. Serafini[1], A. Stella[2], A. Vannozzi[2], S. Vescovi[2]

1 - INFN, Sezione di Milano e LASA
2 - INFN, Laboratori Nazionali di Frascati
3 - Università degli Studi di Milano
4 - Politecnico di Milano

*Approved by:* Luca Serafini, Andrea Ghigo

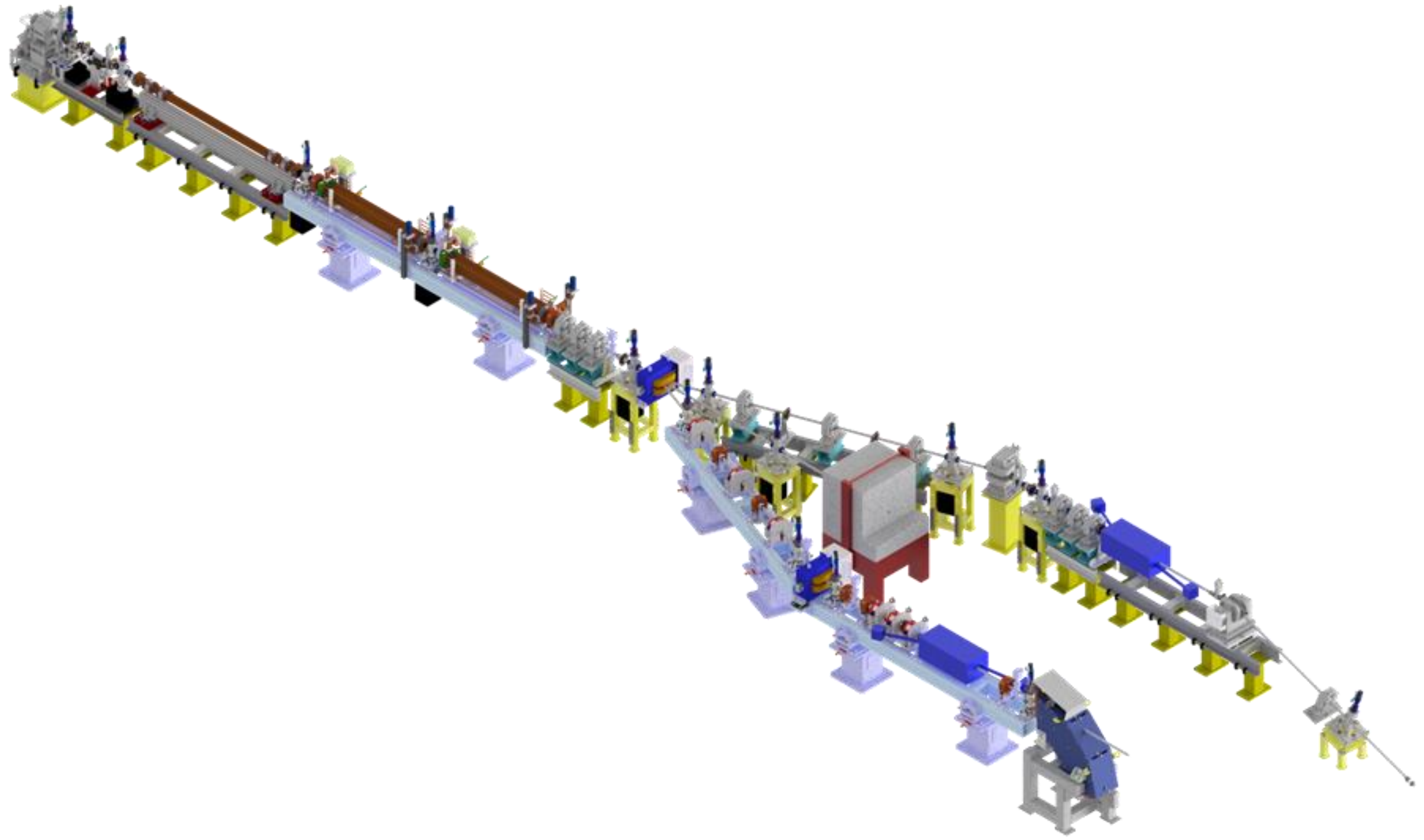

| DATE | STATUS | DISTRIBUTION LIST | |
|---|---|---|---|
| 19/7/2021 | FINAL-R | UNICAL | |

| REVISION | DATE | COMMENT | APPROVED |
|---|---|---|---|
| post-review | 04/08/21 | revised version after UniCal review/comments | A. Ghigo<br>L. Serafini |
| | | | |
| | | | |
| | | | |
| | | | |
| | | | |
| | | | |
| | | | |

## INTRODUCTION

This Document contains a complete technical description of the system devoted to the upgrade of the STAR Linear Accelerator (STAR Linac). According to the Contract signed between Università della Calabria (UniCal) and Istituto Nazionale di Fisica Nucleare (INFN) on May 7th, 2021, INFN is committed to install, test and commission the upgrade of the STAR Linac denominated STAR-HE-Linac (STAR High Energy Linac), hereafter STAR-HEL. The technical components as well as the installation/test procedures and the ancillary equipment involved in such an upgrade are the object of this Document, named Complete Detailed Design Report.
A technical offer was submitted by INFN in the frame of its participation to the tender issued by UniCal, describing a possible energy upgrade of the STAR Linac, with an electron beam energy boosted from 65 MeV up to 150 MeV by means of Radiofrequency (RF) accelerating sections and power stations based on S-band technology (i.e., 2856 MHz RF frequency).
Following the Contract signature, INFN conceived and conceptually designed a technology change that offers several advantages both on performances and on operational reliability of STAR-HEL, based on adopting C-band technology (i.e., 5712 MHz RF frequency) for accelerating sections and RF power stations. Such a technology change was illustrated in a dedicated document named "opzione migliorativa per STAR-2 HE-Linac", addressed to UniCal STAR Management Board and tender R.U.P. for approval on June 9th, 2021. INFN received a formal letter of approval on June 16th, 2021.
Just to briefly remind the main advantages of the change of RF technology from S-band to C-band we list: i) C-band accelerating sections can be operated at a higher accelerating gradient, making the Linac upgrade more compact, ii) they have a dedicated and effective damping of high order RF modes, allowing a possible future upgrade of the STAR Linac to multi-bunch operation, iii) they will be driven by 2 independent RF power stations of lower peak power (nominally 42 MW), making the whole STAR-HEL more stable and reliable for long-run operations typical of user facilities, and iv) their acquisition time-scale is shorter.

Therefore, this document will describe the STAR-HEL technical design based on C-band RF technology as conceptually delineated in the approved document named "opzione migliorativa per STAR-2 HE-Linac".

Accordingly, STAR-HEL will schematically comprise two C-band 1.8 m long accelerating sections, followed by a quadrupole triplet, a H-shape dipole magnet to alternatively deflect the electron beam either to the low energy beam line (located on the North side of the STAR bunker) or to the high energy beam line (located on the South side). The low energy beam line is based on same dogleg lay-out and instrumentation of present STAR-1 machine, just flipped over and relocated to the North side, according to a specific request from UniCal. The high energy beam line is also based on a dogleg lay-out, with quadrupoles and dipoles arranging as described in the following sections. As extensively discussed in the literature, such an arrangement based on doglegs guarantees no contamination of the Compton X-ray photon beam lines from the bremsstrahlung radiation generated by the Linac dark current, that can be efficiently shielded without reaching the two Compton photon beam line of view. The two doglegs properly transport, match and focus the electron beam at the two interaction points, inside dedicated chambers hosting the diagnostics for the collision of the electron and the laser beams. A dedicated 90-degree dipole magnet takes the electron beam after collision down vertically to a beam dump buried inside the concrete floor of the bunker, allowing for maximum shielding of the generated radiation (in particular e.m. cascade and neutrons). Outside the bunker, two RF Power Stations, delivering 42 MW peak power RF pulses to drive the C-band accelerating sections, will be located aside of the present S-band RF Power Station. An adequate number of racks for electronics and controls will also be installed, as extensively illustrated below.

This Document is organized in Sections as follows: Sections 1 to 9 describe the devices and instrumentation that will be delivered, installed and tested by INFN for STAR-HEL at STAR site, while Sections 10 and 11 illustrate the time schedule, WBS and the procedures applied for site acceptance tests of STAR-HEL.

*Acknowledgments:* we are deeply grateful to Angela Campanale, Pier Paolo Deminicis, Giorgio Fornasier, Dino Franciotti, Attilio Sequi and Marta Solinas for their great support and assistance with contract/financial/administration tasks on the project

## Contents

# 1. Lay-out, start-to-end simulations

STAR high-energy linac lay-out upgrade is shown in Figure 1; it is based on two beam lines injected by an improved linac, which is capable to accelerate electron bunches up to 150 MeV. The two beam lines are the low energy line (LE-line, STAR-1 existing one) and the new high energy line (HE-line), that is illustrated in this CDDR. For STAR-1 we refer to the present STAR layout, i.e., with only one beam line able to work at the maximum energy of 65 MeV.

The new HE-line will drive an ICS source at maximum photon energy of 350 keV therefore the linac exit beam energy have to be at least ~140 MeV. The STAR-1 linac acceleration capability will be upgraded from 65 MeV up to 150 MeV, as maximum available beam energy, by two additional C-band acceleration cavities, which are presented in the following chapter. The value of 150 MeV takes into account some safety margins due to beam injection phases into the RF cavities; this to better compensate the beam energy spread.

A short solenoid (8 cm long) is added just before the first S-band cavity, with capability to reach a maximum $B_z$ field peak of ~ 0.2 T. This solenoid is under engineering and presented in chapter 4. This new device permits a better control of the beam divergence and emittance compensation in a larger range of energy operation.

Because the Linac has been improved with two additional acceleration cavities, the STAR-1 working points (WPs) have been modified so that now it is possible to suppress the beam energy spread with a technique that in STAR-1 with only one RF cavity was not feasible. This methodology is valid for both the LE-line and the HE-line, considering the different working energy ranges LE-line: 23-65 MeV and HE-line: 40-150 MeV.

In this chapter, for the LE-line WP we show the beam dynamics (BD) simulation of the new LINAC configuration set to operate at 60 MeV; all the other machine concepts and configurations for this LE-line remain unchanged. The WP simulation for the HE-line shows the BD from the cathode up to the dump, highlighting beam performances at the interaction point IP of an electron beam accelerated up to 140 MeV, i.e. corresponding to 350 keV ICS source HE-line WP.

A list of main beam characteristics for the full dynamic range is reported at the end of this section.

All the BD simulations shown in this CDDR have been done using the tracking code Astra [http://www.desy.de/~mpyflo/], which can simulate the photocathode electron beam extraction, consider full 3D space charge effects and performs off-axis beam tracking. Furthermore, in order to perform optimization on the BD and to find beam optics best parameters, the genetic algorithms-based code GIOTTO [IPAC2016, Busan (Korea): WEPOY039] was used, that works running Astra code simulations on parallel CPUs.

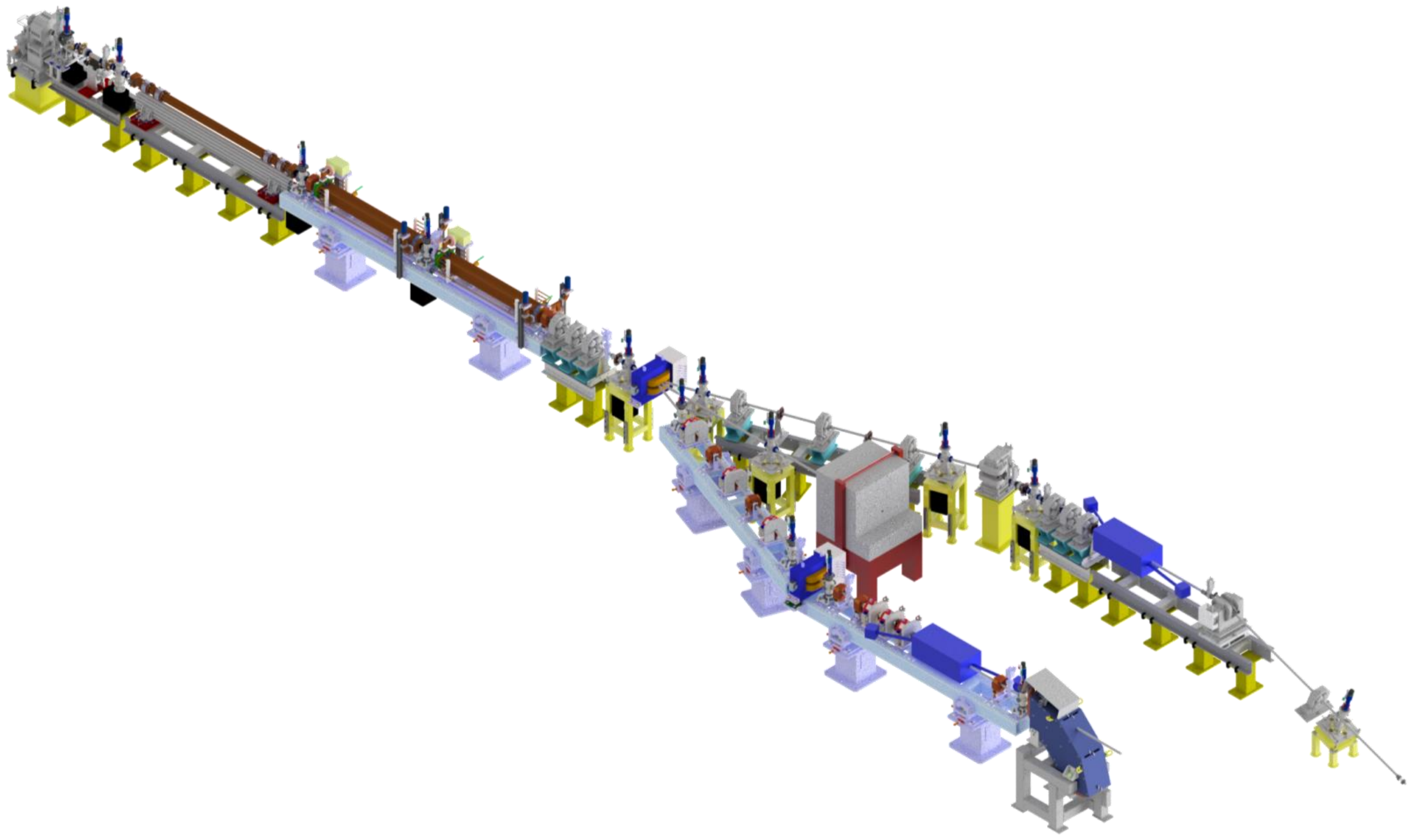

***Figure 1: STAR-HEL upgraded layout.***

## 1.1. Cathode laser pulse shaping

The pulse shaping of the extracting laser hitting the photocathode is strictly connected to the BD performances. The main shaping parameters are reported in Table 1. These values have been kept equal for both the LE-line and the HE-line WPs. The extracted bunch charge is related to the photocathode quantum efficiency (QE) and to the laser pulse energy (we referred to STAR-1 values).

***Table 1: Laser pulse shape data at the gun photocathode***

| | |
|---|---|
| Gaussian laser pulse, $\sigma_t$ [ps] | 1.5 |
| $\sigma_x = \sigma_y$, transverse uniform [mm] | 0.365 |
| Extracted bunch charge [pC] | 500 |

### 1.2. LE-line and BD into the linac

The new linac configuration, now designed with three accelerating cavities, guarantees a better performance respect to the previous one, i.e., the previous linac based on one cavity was incompatible to perform an optimal energy spread compensations.
In Figure 2 the upper plot shows the rms beam envelope (in blue) and the beam normalized emittance (in yellow); the lower plot shows the electron bunch length (in yellow), the energy spread (in black) and the energy gain (in red). The optimal electron beam energy spread compensation is visible in Figure 2 (lower plot), where using ad hoc injection phases, off crest, in all the three acceleration cavities, (one S-band, two C-band) it was possible to recover the energy spread at the same value that the beam had at the linac entrance. Main beam parameters at the linac exit are reported in Table 2.

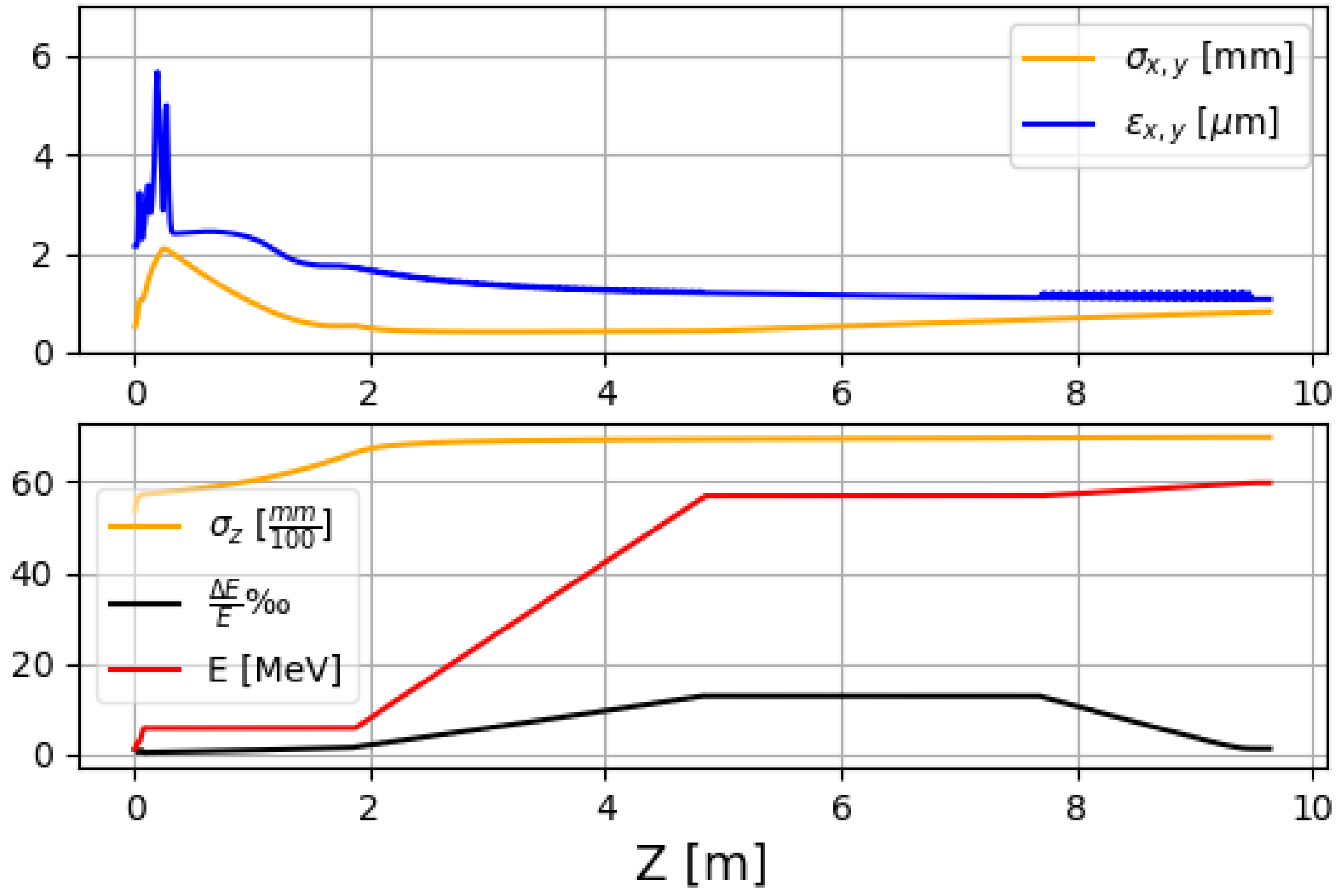

***Figure 2: BD linac simulation for the 60 MeV LE-line WP. The upper plot shows the normalized emittance (in blue) and the beam envelope (in yellow), because here the beam is in cylindrical symmetry it is shown only one plane. The lower plot shows the bunch length (in yellow), the relative energy spread (in black) and the energy gain (in red).***

***Table 2: Main BD data of the bunch exiting the linac for the LE-line WP***

| | |
|---|---|
| rms x,y envelopes [mm] | 0.830 |
| Normalized x,y emittances [mm-mrad] | 1.0 |
| rms bunch length [mm] | 0.7 |
| relative energy spread [%] | 0.1 |

### 1.3. HE-line BD

The STAR high energy line represents the main scope of this CDDR. Changes and upgrades made to the original STAR layout (STAR-1) have the final goal to deliver an electron beam with maximum energy of 150 MeV at the ICS interaction point, preserving or improving STAR-1 performances, which are related to the LE-line, i.e., 60 MeV WP.
A BD simulation is reported in Figure 3 for the upgraded STAR-HEL linac (two additional C-band TW cavities) where the bunch is accelerated at 140 MeV, electron energy compatible for an ICS source producing 350 keV photons.

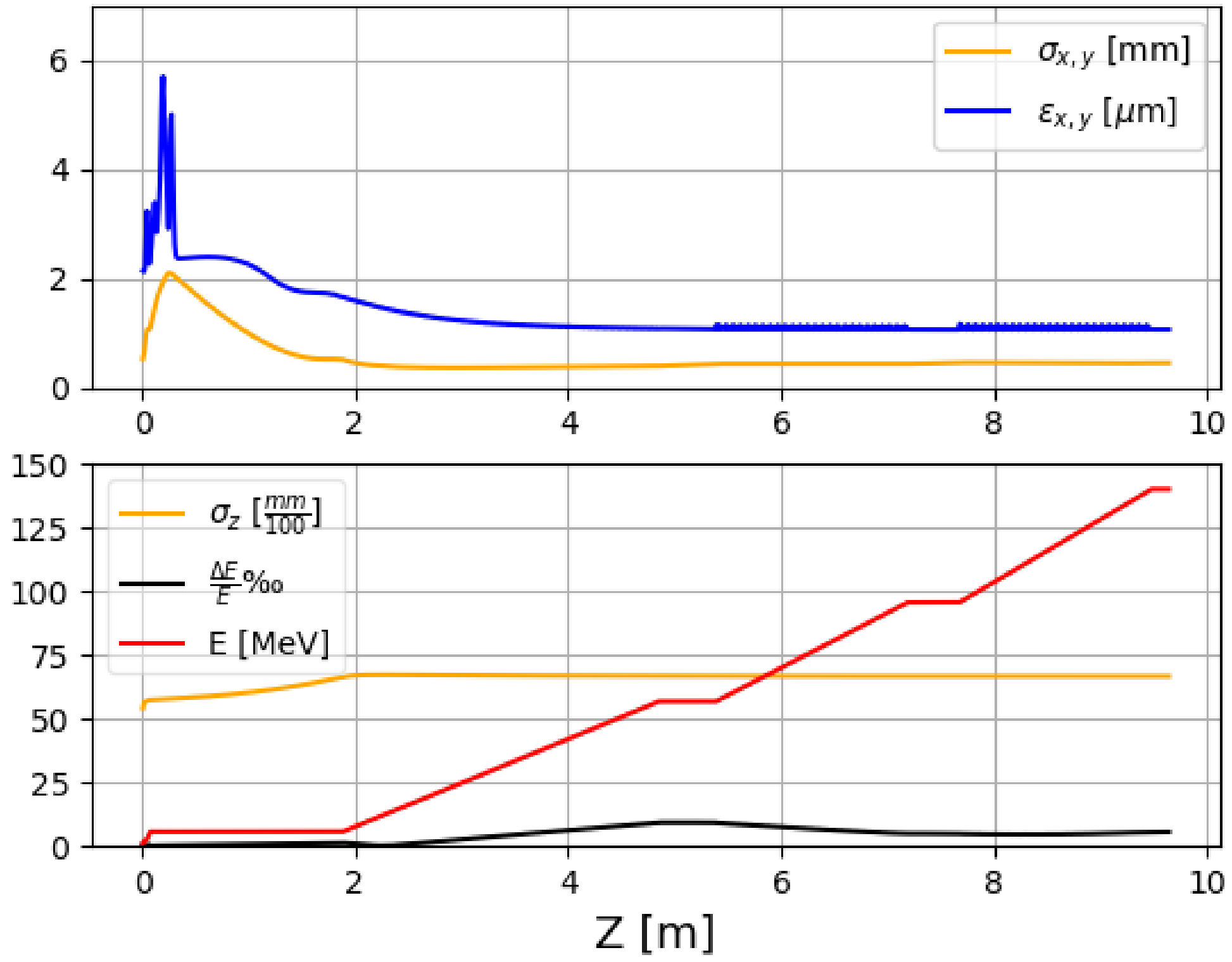

***Figure 3: BD linac simulation for a 140 MeV HE-line WP. The upper plot shows the normalized emittance (in blue) and the beam envelope (in yellow), because here the beam is in cylindrical symmetry it is shown only one plane. The lower plot shows the bunch length (in yellow), the relative energy spread (in black) and the energy gain (in red).***

As for the LE-line the acceleration cavities are injected out of RF crest to compensate the beam energy spread; this compensation is clearly shown in Figure 3 by the black curve of the lower plot. Main beam parameters at the linac exit are reported in Table 3, while linac main setting is reported in Table 4.

***Table 3: Main BD data of the bunch exiting the linac for the LE-line WP***

| | |
|---|---|
| rms x,y envelopes [mm] | 0.450 |
| Normalized x,y emittances [mm-mrad] | 1.0 |
| rms bunch length [mm] | 0.670 |
| relative energy spread [%] | 0.05 |

*Table 4: Relevant linac setting to operate at 140 MeV with energy spread compensation*

| Elements | Parameter | Values |
|---|---|---|
| RF-Gun | Ez peak Field (MV/m), phase (deg) | 120, -6.1 |
| Gun solenoid | Bz peak filed (T) | 0.3120 |
| S-band cavity Solenoid | Bz peak filed (T) | 0.1580 |
| S-band cavity | Ez Peak Field (MV/m), phase (deg) | 24.0, -19.0 |
| C-band cavity 1 | Ez Peak Field (MV/m), phase (deg) | 29.0, 3.7 |
| C-band cavity 2 | Ez Peak Field (MV/m), phase (deg) | 33.0, 1.7 |

Downstream the linac, the beam enters the dogleg dispersive beamline to reach the IP region. Figure 4 shows in the upper plot the x, y normalized emittances, that are well preserved along the dispersive path up to the exit. The lower plot shows η and η' for both the transverse planes, which are typical parameters for dispersive beam-lines, that must be taken down to zero a matched into the non -dispersive beam-line regions. These is a critical issue of the beam dynamics study.

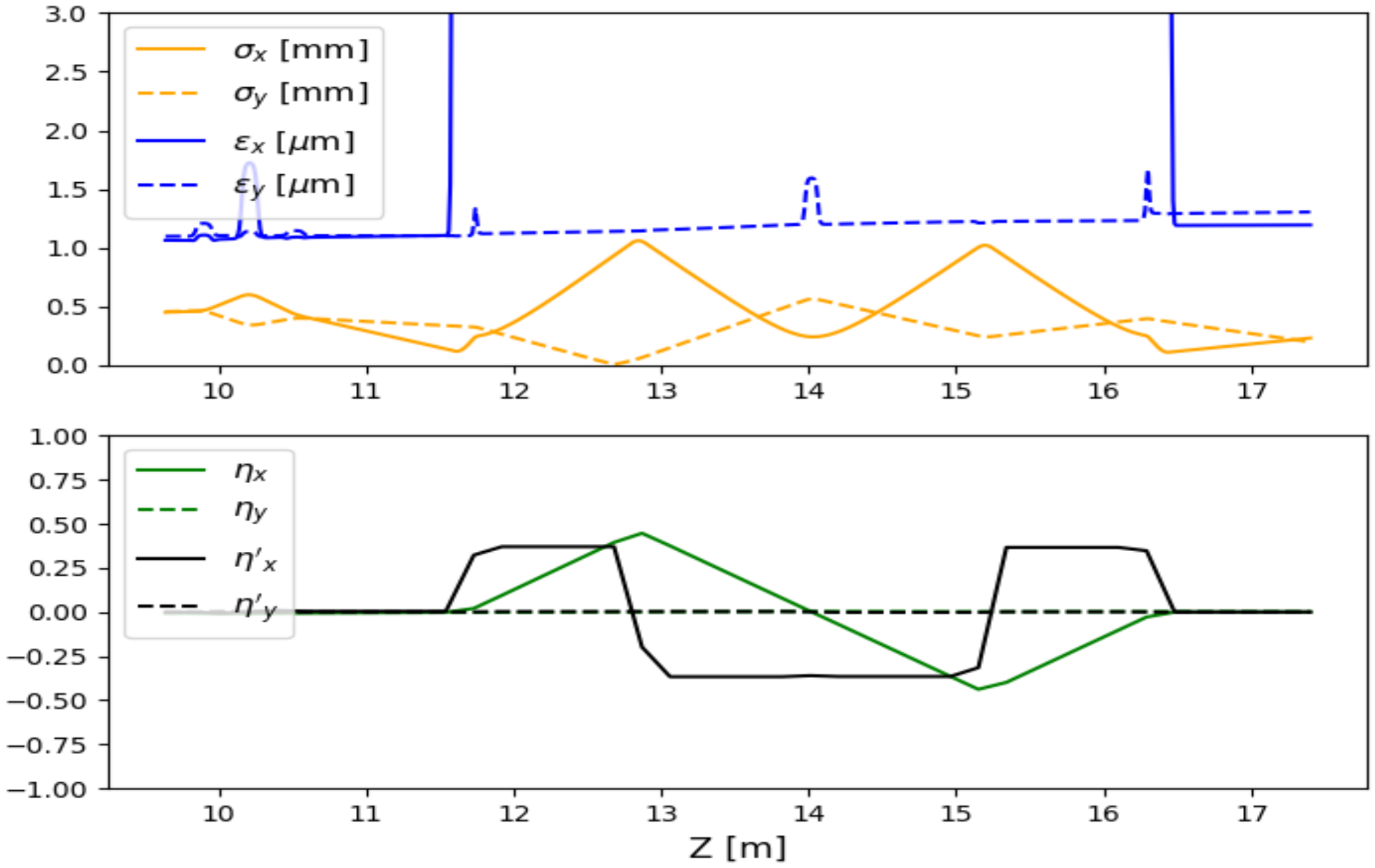

***Figure 4: In the upper plot are shown the normalize emittance (blue solid and dashed lines) and the beam envelopes (yellow solid and dashed lines) for the x, y planes. In the lower plot are shown the η (green solid and dash lines) and η' (black solid and dashed lines) again for the x, y planes.***

Downstream the dogleg the beam enters the focusing channel (a quadrupole triplet) reaching the IP (at 18.9 m from the cathode) with the design spot size (~ 30 μm). Beam envelopes and emittances curves produced by the Astra simulation are reported in Figure 5, the quadrupoles gradients of the focusing triplet are reported in Table 5. For this WP only two of the three quadrupoles are needed to focus the electron bunch. The main bunch characteristics are reported in Table 6.

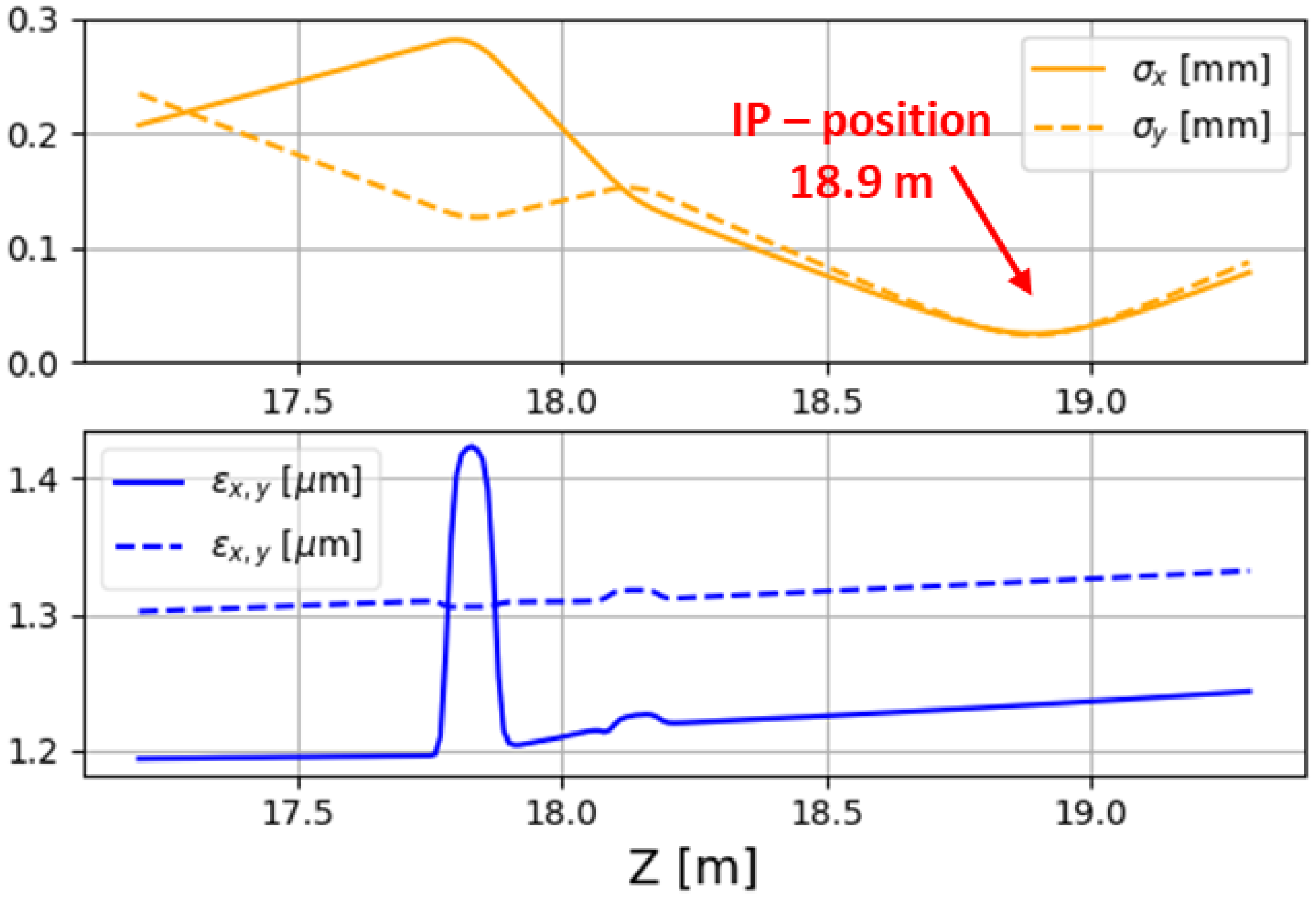

***Figure 5: In the upper plot are shown the x,y rms beam envelops, in the lower plot the x,y rms normalized emittance at the IP.***

***Table 5: Quadrupole gradients of the HE-line focusing triplet***

| **Quadrupole** | **Gradient (T/m)** |
|---|---|
| QUAHEL04 | 0.00 |
| QUAHEL05 | 10.30 |
| QUAHEL06 | -9.70 |

***Table 6: Electron beam characteristics at the IP***

| | |
|---|---|
| Beam energy [MeV] | 140.05 |
| Beam charge [pC] | 500 |
| σx, σy [μm] | 25.0, 23.7 |
| σz [μm] | 667 |
| εx, εy [μm] | 1.2, 1.3 |
| relative energy spread [%] | 0.2 |

As already presented in the technical offer, the STAR-HEL will be able to work in different energy range, assuring the performances reported in Table 7.

*Table 7: STAR-HEL electron beam quality parameters at IP. By "(bf)" we report the best effort for some of the parameters. This represents the best possible reachable value considering theoretical limitations for such beam line; from a contractual point of view these bf values are not guaranteed.*

| | HE-linac | LE-linac |
|---|---|---|
| Energy range [MeV] | 40-150 | 23-65 |
| Rap. rate [Hz] | 100 | |
| Bunch charge range [pC] | 100 – 500 (bf:2000) | |
| Normalized emittance (x,y) [μm] | 2.0 (bf: 1.0) | |
| Bunch energy spread [%] | 0.5 (bf: 0.2) | |
| Bunch length – rms [ps] | ≤ 5 | |
| Bunch spot dimension at IP [μm] | 40 (bf: 20) | |

### 1.4. HE-line dump

Downstream the IP the beam enters directly into the dumping dipole which has been designed to dump the beam perpendicularly to the floor, with a bending angle of 90 degrees. The beam starts the curved orbit at about 1.8 m after the IP where it is still almost collimated. There is no need of additional optics to keep its envelopes under control. Main beam parameters (orbit and transverse envelopes) are shown into Figure 6.

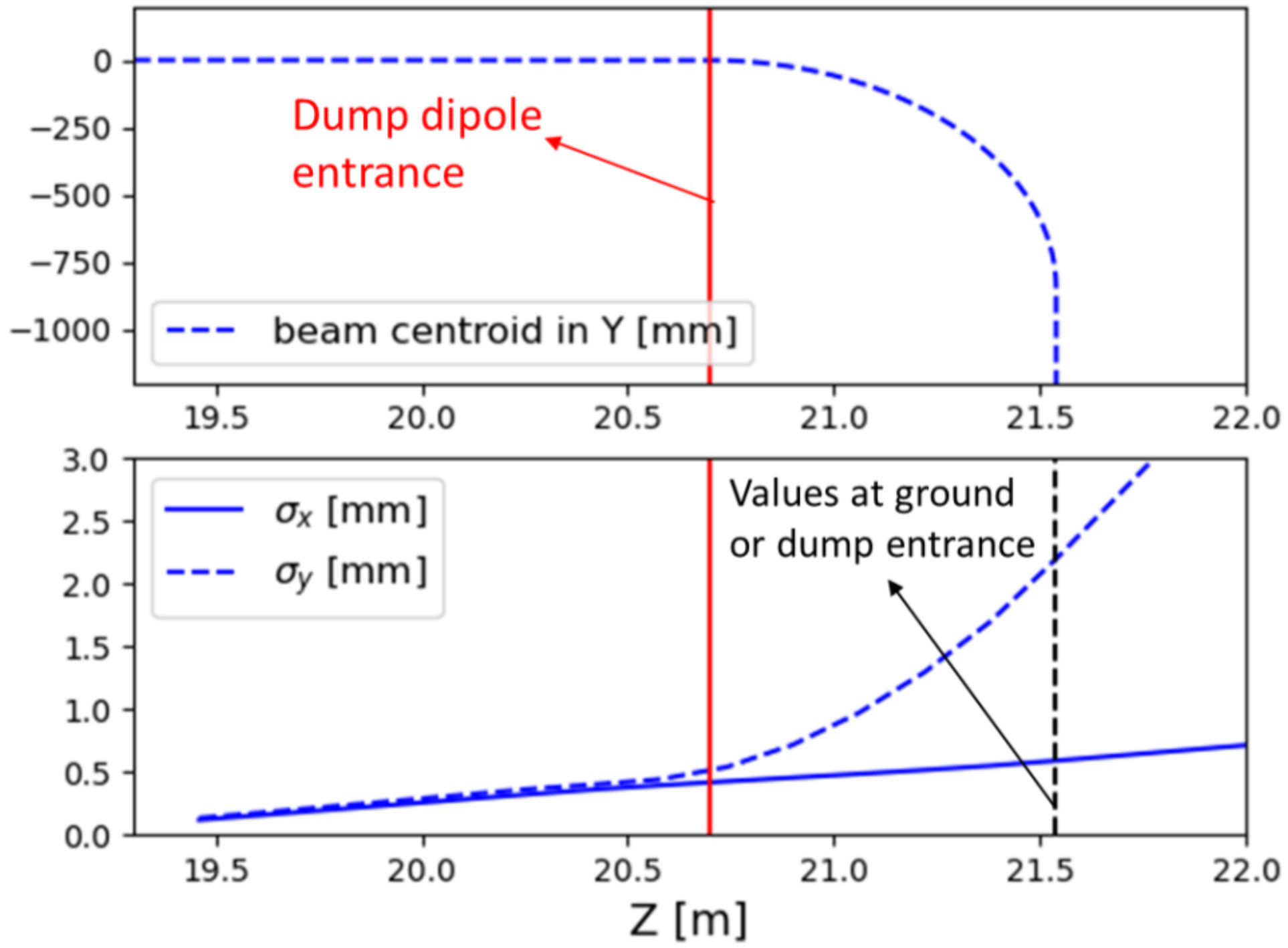

***Figure 6: In the upper plot is shown the beam centroid vertical position, the beam line is set at vertical coordinate of 0, the ground at -1200 mm; it is highlighted the 90° dump dipole entrance where the beam starts to bend. In the lower plot are shown the x, y rms beam envelopes when the beam enters into the dump dipole and goes perpendicular to the ground, to the dump.***

## 1.5. Stability, jitters, risk

About the main jitters and fluctuations that can affect the STAR-HE-Linac, impacting the quality and performances of the delivered electron beam at 150 MeV, we have considered the following, in agreement with the expected performances of the LLRF, discussed in Chapter 3.2:

1) Regarding the C-band accelerating sections, the maximum accelerating gradient relative jitters is +/- 0.2%, as typical of these RF systems.
2) Regarding the phase jitter of the C-band accelerating sections the target value is +/- 0.2 RF degrees

Running the beam dynamics simulations by considering a statistical variation of the above listed jitters, we obtain the following results concerning the standard deviation of the listed beam parameters:

***Table 8: Statistical standard deviations of main beam parameters vs. jitters and fluctuations.***

| PARAMETER | std. dev. |
|---|---|
| energy | 0.06 MeV |
| energy spread | 6.65 keV |
| sig_x | 0.2 um |
| emitn-x | 0.013 um |
| sig_y | 0.2 um |
| emitn-y | 0.010 um |

Showing that the sensitivity of the beam dynamics to jitters is very low, so the beam quality at the interaction point is very well preserved even in presence of jitters. We point out that these jitters combine in an uncorrelated way with the jitters effective on STAR-1 (S-band Gun and acc. sections, magnets, lasers, etc.). About the quadrupole current ripple into the focusing channel, we consider a max ripple of $5 \cdot 10^{-4}$ relative to the maximum gradient (19 T/m), i.e. +/- 0.001 T/m. A statistical random variation of the gradient in the final set of quadrupoles before the IP results in a negligible variation of the beam spot size at IP. In order to see an effect on the spot size the ripple should be a factor 10 higher (+/- 0.01 T/m), that results in 0.2um spot size maximum change in the x plane and 0.04 um in the y plane.
Therefore, we confirm the expected performances for the electron beam quality delivered by STAR-HE-Linac as listed in the technical offer, clearly subordinate to the delivery of the specified beam at the exit of the S-band accelerating section, as extensively discussed in the technical offer. Concerning a risk analysis for STAR-HE-Linac we point out that the components of this project are state of the art: the C-band accelerating sections have been already tested as prototypes, and the C-band power stations have been delivered by industrial companies to other projects world-wide. The magnets listed in this document are state of the art, proven technology, available on the market. The interaction chamber replicates many others in use at ICS, with very limited risk thanks to state-of-the-art components in the laser optics and stabilization hardware.

## 2. IP chamber Lay-out

The STAR-HEL upgrade foresees the production of two optimized interaction chambers, one for the LE-line and the other for the HE-line. Because the lay-out of the two interaction zones are designed with an identical electron beam optical scheme, the two chambers have been conceived equal. The preliminary scheme of the working principle of the interaction chambers has already been presented in the technical offer. In that circumstance, the two essential requirements that the interaction chambers must guarantee were highlighted: 1) the possibility to tune the angle of interaction between electrons and photons in the 175–178-degree range, by means of a motorized gimbal mounted mirror; 2) the possibility to focus the laser spot by changing the focal length using an adjustable telescope. By keeping in mind these constraints, the conceptual scheme of the technical offer has been implemented in a truly feasible device as shown in fig. 7, where the overall dimensions, and beam angles, are in scale (collision angle 5° as correctly defined in ICS established nomenclature, where 0° is meant as head-on collision).
An important issue arises from the upgrade of the STAR interaction laser, with an expected increase of the laser pulse energy from 100 mJ up to 1J: optics damage concerns must be carefully taken into consideration, because STAR must be operated in the context of a user facility - operation reliability and stability is a major concern. There is no way to dump the laser inside the interaction chamber, therefore it will be present an optical line for bringing the laser pulses out of the chamber where they will be safely dumped.
The 1J laser pulse needs a focusing system based on 2-inch lenses to avoid damage to the optics and for the same reason the high grazing mirror guarantees the laser extraction with a damage threshold (laser energy density) about a factor two lower than the mirror robustness. The laser pointing stability is designed to be lower than 5 μm.

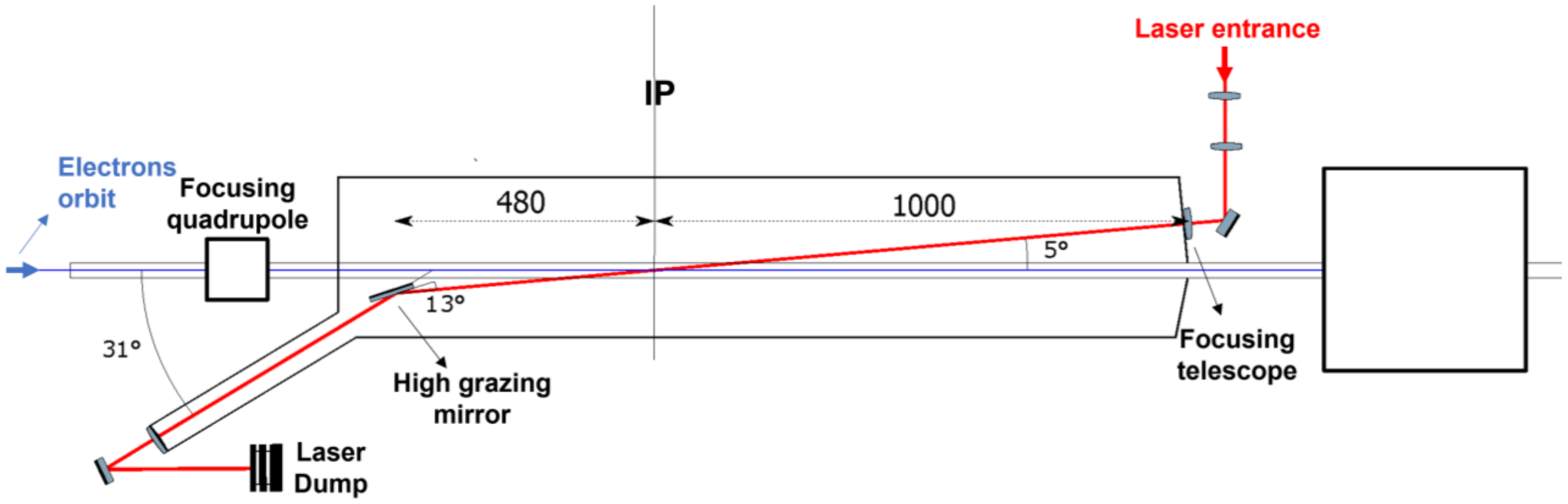

***Figure 7: Laser beam line scheme with main overall dimensions. The figure quotes are in mm.***

# 3. The RF System: Accelerating Sections, Power and Network.

## 3.1. Linac

### 3.1.1. C-band accelerating structures

The STAR-HEL upgrade foresees the installation of two traveling-wave (TW) disk-loaded linear accelerating structures (linacs) working in C-Band at 5.712 GHz. Each linac is 1.8 m long and works in a quasi-constant gradient regime with a field phase advance of 120 degrees. The main RF parameters of the C-Band TW structures are listed in Table 9.

***Table 9: Main parameters of the C-Band TW accelerating structures.***

| **Structure type** | **Constant impedance, TW** |
|---|---|
| Working frequency (fRF) | 5.712 GHz |
| Number of cells | 102 |
| Structure length | 1.8 m |
| Working mode | $TM_{01}$-like |
| Phase advance between cells | 2π/3 |
| Nominal RF input power ($P_{IN}$) | 40 MW |
| Average accelerating ($E_{acc}$) | 33 MV/m |
| Quality factor ($Q_0$) | 8800 |
| Shunt Impedance per unit length | 67-73 MΩ/m |
| Phase velocity | *c* |
| Normalized group velocity ($v_g/c$) | 0.025-0.014 |
| Filling time (τ) | 310 ns |
| Structure attenuation constant | 0.58 neper |
| Operating vacuum pressure (typical) | $10^{-8}$-$10^{-9}$ mbar |
| Max RF input pulse length, flat top ($\tau_P$) | 1 µs |
| Pulse duration for beam acceleration ($\tau_{BEAM}$) | <500 ns |
| Iris half aperture (a) | 6.8-5.8 mm |
| Rep. Rate ($f_{rep}$) | 100 Hz |
| Average dissipated power | 2.3 kW |

The number of accelerating cells per structure is 100 plus 2 coupler cells for a total of 102 cells. The input coupler is a matched RF power splitter for dual-feed setup which results in the elimination of the field dipole component. A fat-lip geometry is used for the coupling slot hole to reduce the surface RF pulsed heating due to the high magnetic surface field. Each cell geometry is mainly optimized to lower the peak surface electric field on the cell-to-cell irises and to obtain an average accelerating gradient up to 33 MV/m using 40 MW RF power at the linac input coupler. Four tuning pins are placed around each cell and eight cooling pipes to sustain the 100 Hz operation. The linac is also optimized for multi-bunch operation. As a matter of fact, each accelerating cell is connected to four dummy waveguides where dissipative media, consisting in silicon carbides (SiC) tiles, are inserted. These absorbing media will absorb and dissipate higher-order modes (HOMs) thus minimizing the detrimental beam break-up effects during a multi-bunch operation, that may seriously impact the beam emittance and energy spread along the multi-bunch train.

The simulation of the HFSS 3D surface model of the C-band accelerating structure is shown in Figure 8 together with on-axis accelerating electric field profile distribution.

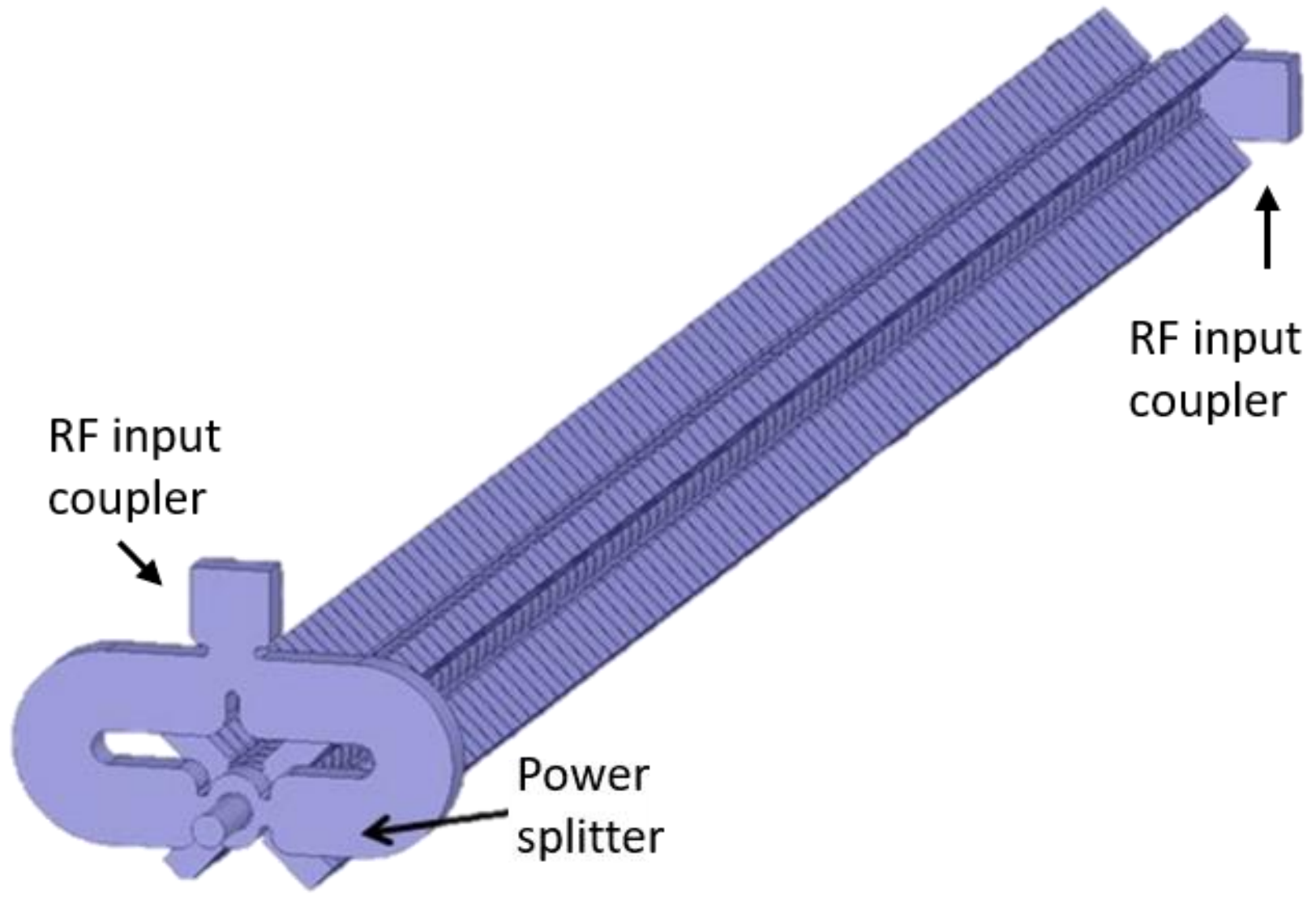

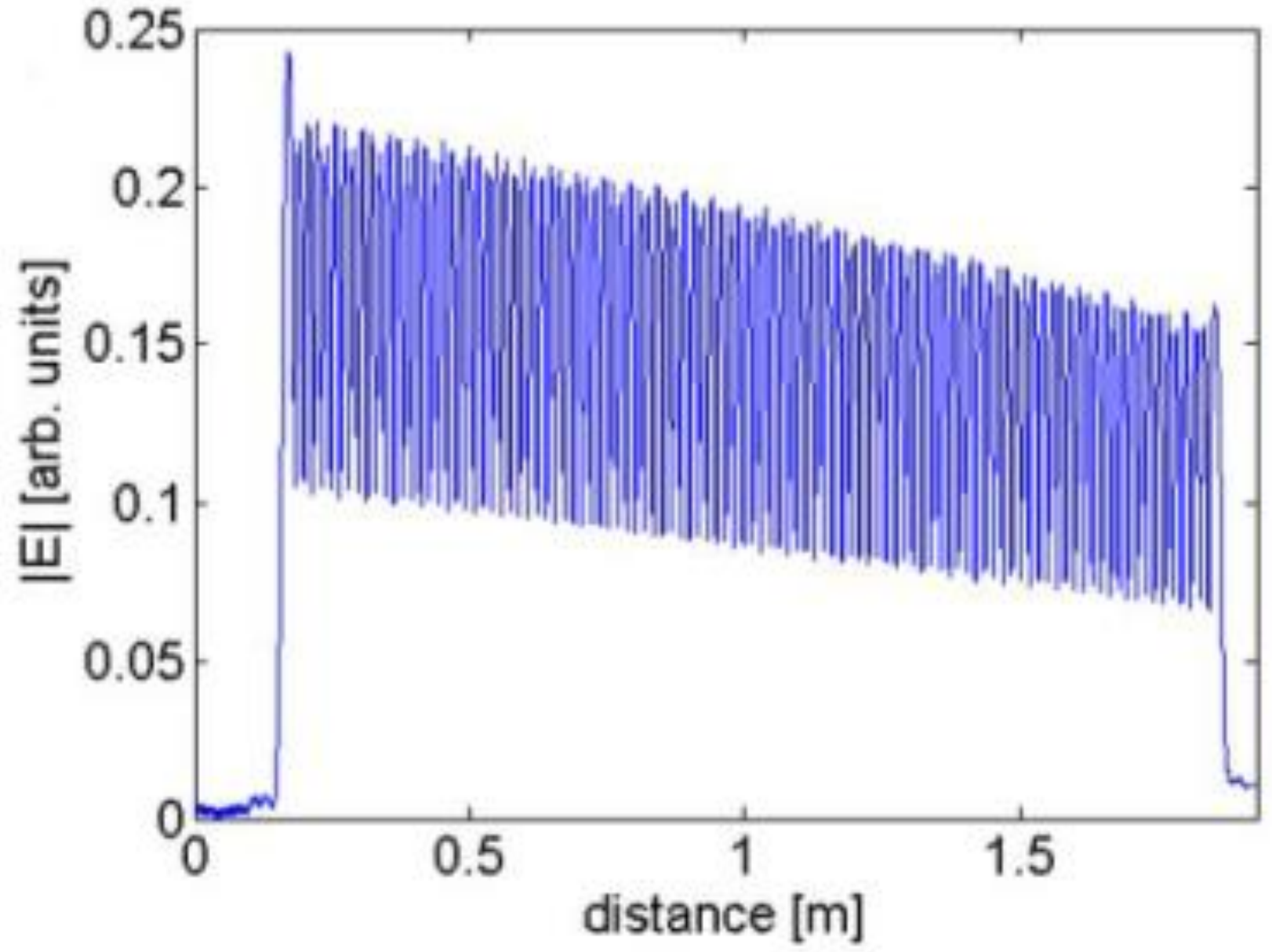

***Figure 8: Top, preliminary HFSS 3D surface model of the C-band accelerating structure; bottom, on-axis accelerating electric field profile.***

### 3.1.2. RF waveguide network and power distribution

The RF power is provided by two C-Band power units working at 5.712 GHz. The RF power will be distributed through a network of copper waveguides connected with LIL type flanges. The waveguides will be standard WR187 working in C-Band. In order to reduce arcing due to the high RF power, the waveguides will be operated in ultra-high vacuum, below $10^{-8}$ mbar. Therefore, ion pumps will be placed along the network every nearly 1.5 - 2 m. We foresee three ion pumps with pumping units on each C-Band waveguide network from the klystron output to the linac input. An RF window will be located at each entrance of the accelerating structures.

The RF microwave sources are high-power klystrons with a nominal value of 42 MW. Each klystron is energized by a solid-state pulsed high-voltage modulator. The RF klystron drivers are solid-state amplifiers (SSA's) with an output power > 400W.

A schematic layout of the C-Band RF system is shown in Figure 9.

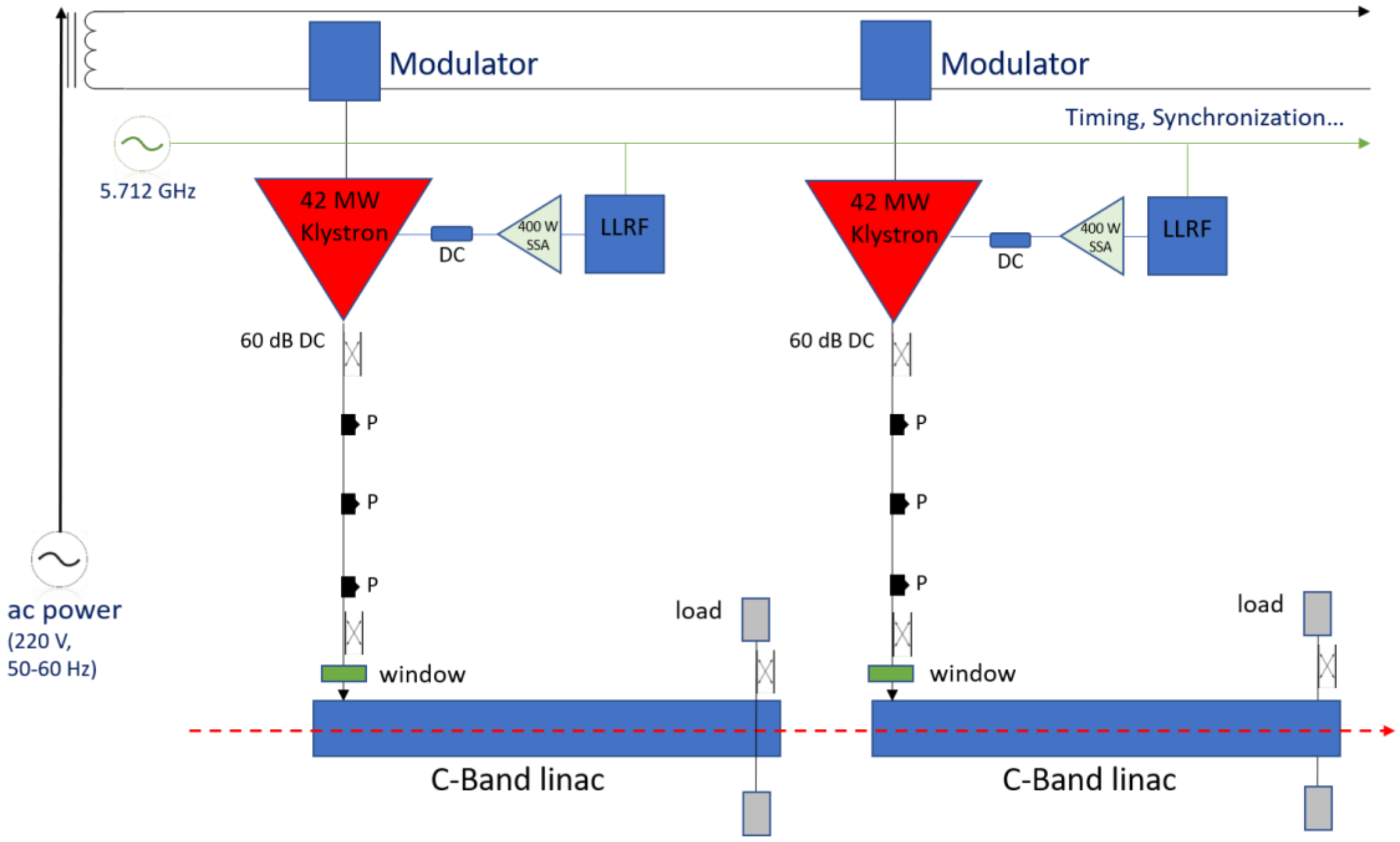

*Figure 9: Schematic layout of the C-Band RF system.*

The klystron specifications are reported in Table 10.

*Table 10: Specifications of the klystron.*

| Parameters | Specifications | Units |
|---|---|---|
| RF Frequency | 5712 | MHz |
| Peak RF power | 42 | MW |
| RF Average power | 4.2 | kW |
| RF gain | > 50 | dB |
| Efficiency | > 40 | % |
| RF pulse length (flat top) | 1 | µs |
| Pulse repetition rate (PRF) range | 0-100 | Hz |

The main parameters of the pulsed high-voltage modulators, based on solid-state technology, are listed in Table 11.

***Table 11: Specifications of the pulsed high-voltage solid-state modulator.***

| Parameters | Specifications | units |
|---|---|---|
| Peak Voltage Range | 0-370 | kV |
| Peak Current Range | 0-344 | A |
| Modulator Peak Power | 127 | MW |
| Modulator Average power | 12.7 | kW |
| Pulse repetition rate (PRF) range | 0-100 | Hz |
| RF pulse length (flat top) | 1 | µs |

The forward and reflected power signals will be monitored by means of RF monitors. These monitors are commercially available 60 dB directional couplers (DCs). Mainly four DCs will be used in the C-band RF system; one will be placed on the output waveguide of each klystron and one on the input waveguide of each linac input.

Since the RF losses in the WR187 waveguides is approximately 0.035 dB/m, which means 4.2% of RF power loss per 5 m length, the RF power routinely available at each C-band linac input is about 40 MW, which confirms the data shown in Table 9 for reaching up to 33 MV/m accelerating gradient, assuming the klystron works at the nominal power of 42 MW.

### 3.2. LLRF System: Control Racks.

In order to achieve the synchronization requests for the STAR-HEL upgrade the LLRF System will work alongside the already existing system for the S band, using the same type of components so that it can be integrated into the control system already in operation, achieving a synchronization level consistent with the specifications listed in Chapter 1.5.
Two 19-inch racks will be used for controlling the signals for the two C-band power units (klystron plus modulator), the waveguide networks including the direction couplers and linacs with high-power loads. In alternative, we will evaluate the possibility to integrate the C-Band LLRF system into the existing racks. The rack composition and distribution of the rack units (RU) are listed in Table 12. Each rack will contain a Libera LLRF system, as shown in Figure 10, able to handle up to 7 signals in order to control the electromagnetic field amplitudes and phases inside the accelerating structures. One coaxial 35 dB directional coupler can be placed between the solid-state driver and the klystron. The signals from this DC and the one located after the klystron can monitor the VSWR value and can be used by the modulator as interlock. The cables from the accelerator will be connected to a patch panel. Two RUs per rack will be used to handle and process the various triggers and interlocks.

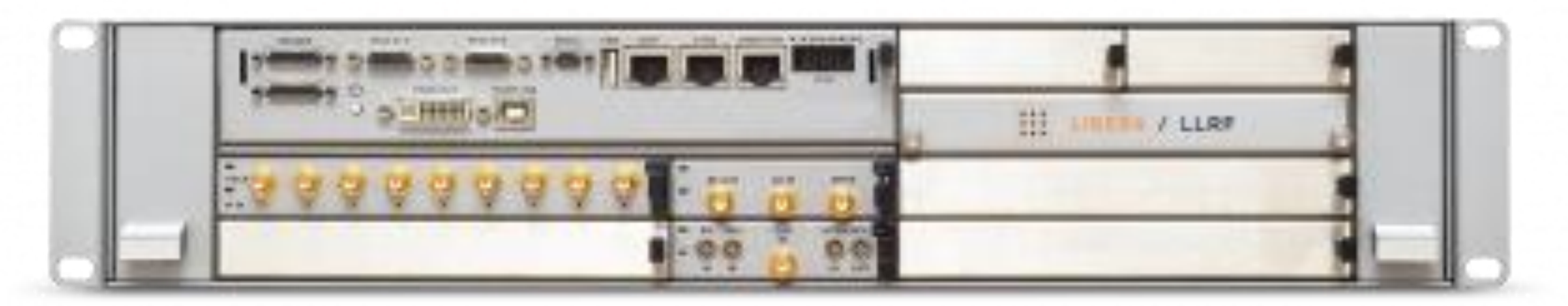

*Table 12: Rack Numbers and Distribution.*

*Figure 10: Libera LLRF System Module (Ref. Instrumentation Technology).*

| LLRF System, Synchronization and Timing | Rack 1 | Rack 2 |
|---|---|---|
| Libera LLRF System | 5 RU | 5 RU |
| Patch Panel | 5 RU | 5 RU |
| Ethernet Switch | 1 RU | 5 RU |
| Triggers and Interlocks | 2 RU | 2 RU |
| Reference Frequency Distribution | 3 RU | - |

The reference frequency (5.712GHz) is sent to the RF clients (driver and klystron) by an ad-hoc distribution board designed and implemented by INFN. The reference frequency distribution sketch is given in Figure 11. The existing reference master oscillator will be split into two lines. The first one will distribute the original 2856 MHz reference signal and the second one will deliver the 5712 MHz reference signal to the LLRF system for the C-Band network.

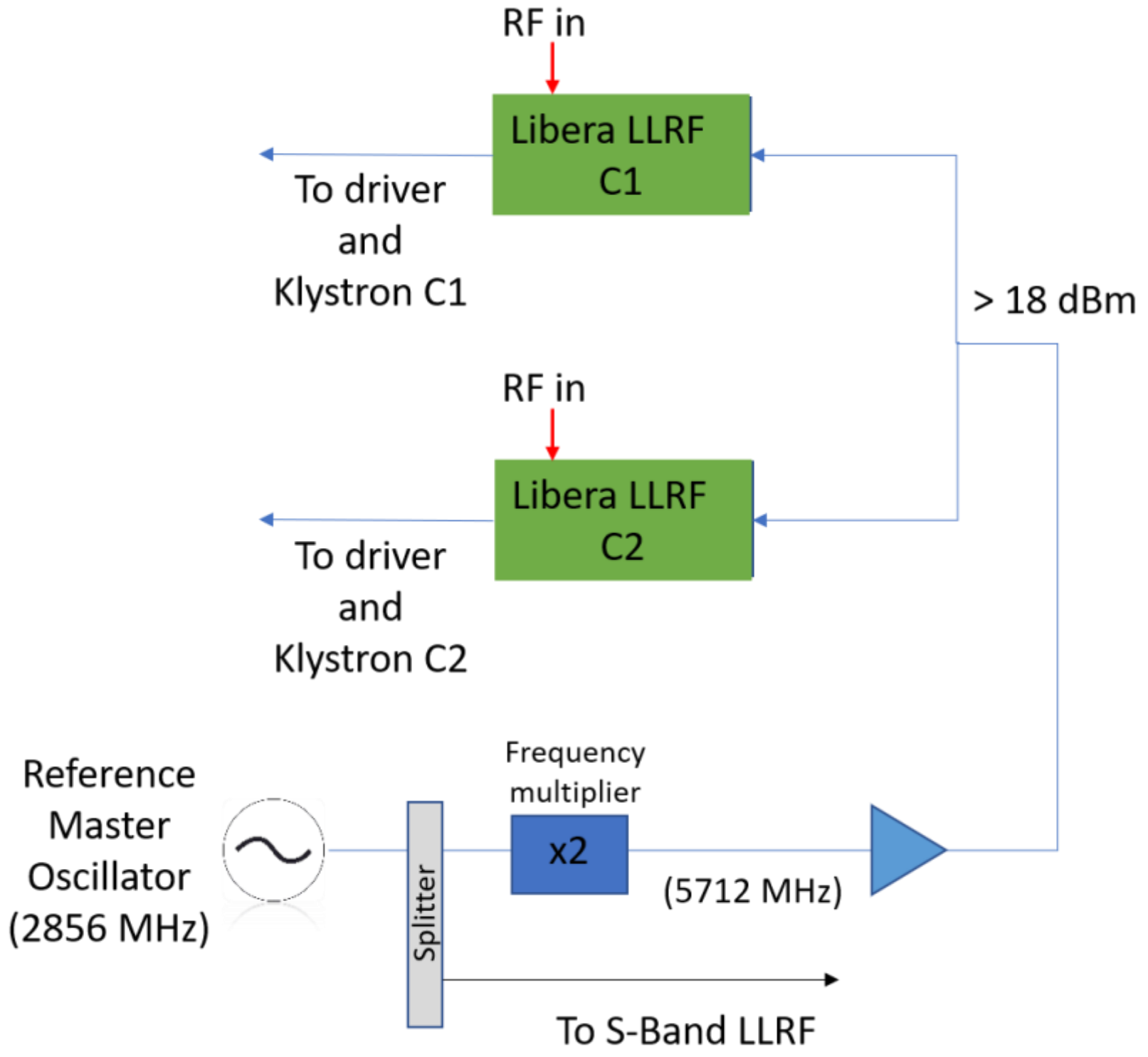

*Figure 11: Reference signal distribution.*

It is important to point out that the LLRF system for the C-Band network will be compatible with the existing one at STAR-1.

# 4. Magnets and Power Supplies

## 4.1. General Remarks

Magnets and Power Supplies (PSs) are core systems of the STAR-HEL upgrade. The realization of a new beamline requires the installation of some electromagnets for an accurate beam transport towards the interaction points (IPs) with the laser beam. For this purpose, it is necessary to add several dipoles, steerers, quadrupoles and one solenoid. From the beam dynamics studies reported in section 1, all the magnets will be normal conducting magnets (maximum field experienced by the beam within a 1.6 T range) and they will be powered by direct current (DC) power supplies.
All the magnets will be made of solid steel yokes with accurately machined pole faces designed to maximize the field quality, according to the beam dynamics requirements, and by oxygen free copper coil. Except for the air-cooled steerers, all the magnet coils will be water-cooled in order to reach higher power density and therefore a higher compactness.
The current state of art for magnets manufacturing and the DC power supplies performances for particle accelerators allow to completely fulfill the beam dynamics requirements in terms of field quality and time field stability.

## 4.2. Magnets

The STAR-HEL upgrade foresees the installation of several new magnets distributed both along the Linac section and the new high energy line. The definition of main parameters and the design phase is completed. During the detailed design of the magnets system emerged, in order to achieve the better possible performances and guarantee the requested stability, the need for an additional solenoid (solenoid AC1). The parameters of this extra device, not foreseen in the technical offer, have been calculated in order to integrate it with the existing layout.
This solenoid will be placed upstream the first accelerating structure (AC1) to permit a better control of the beam divergence and emittance compensation in a larger range of energy operation.
The outcome of this improvement of the magnets system allowed to define detailed mechanical 3D models based on Finite Element Analysis (FEA) 2D and 3D software simulations. This result, that typically would have been requested a significant amount of time, has been achieved within two months from the beginning of the activity. The definition of the main parameters physically characterizing the magnets (gradients, physical length, etc) has been extensively reported in the technical offer, and the design phase is completed.

All the magnets have been designed to fulfill a field quality of $5 \cdot 10^{-4}$ in terms of field uniformity and maximum multipole content allowed.

In detail, in the Linac section, the above-mentioned solenoid AC1 and two steerers will be added, while at the end of the Linac a dipole (type A) will be installed to switch the beam in the Low Energy Line (LE-line) or in the High Energy Line (HE-line) respectively.
In the new HE-line beam line, five steerers, six quadrupoles and one dipole type A will be installed together with the 90-degree beam dump dipole. With this configuration it will be possible to reduce the radiation impact of the beam dump.
In Table 13 the quantities of all the STAR-HEL magnets are listed together with the power dissipated both in air and water.

***Table 13: STAR-HEL Magnets. The power dissipations of these magnets, both in air and in water are also listed.***

| Magnet | QTY | P_Water [kW] | P_Air [kW] |
|---|---|---|---|
| Solenoid AC1 | 1 | 0.53 | 0.05 |
| Dipole type A | 2 | 5.02 | 0.50 |
| Dipole Dump HEL | 1 | 4.29 | 1.20 |
| Quadrupoles | 6 | 0.31 | 0.03 |
| Steerers | 7 | | 0.03 |

Figure 12 and Figure 13 depict a 3D view of both the dipoles from Opera 3D software simulations and a sketch of a mechanical design while Table 14 lists all the main parameters for both dipoles.

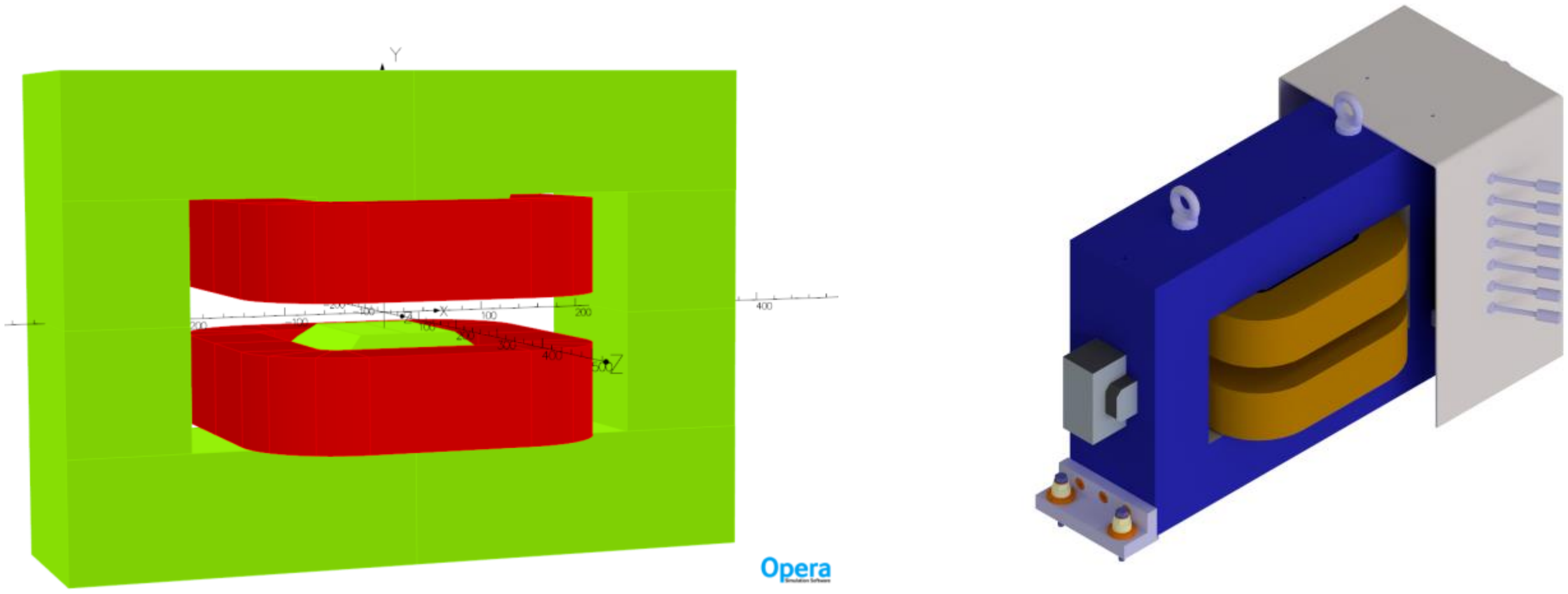

***Figure 12: Dipole type A view from an Opera 3D software simulation (left) and mechanical drawing sketch (right)***

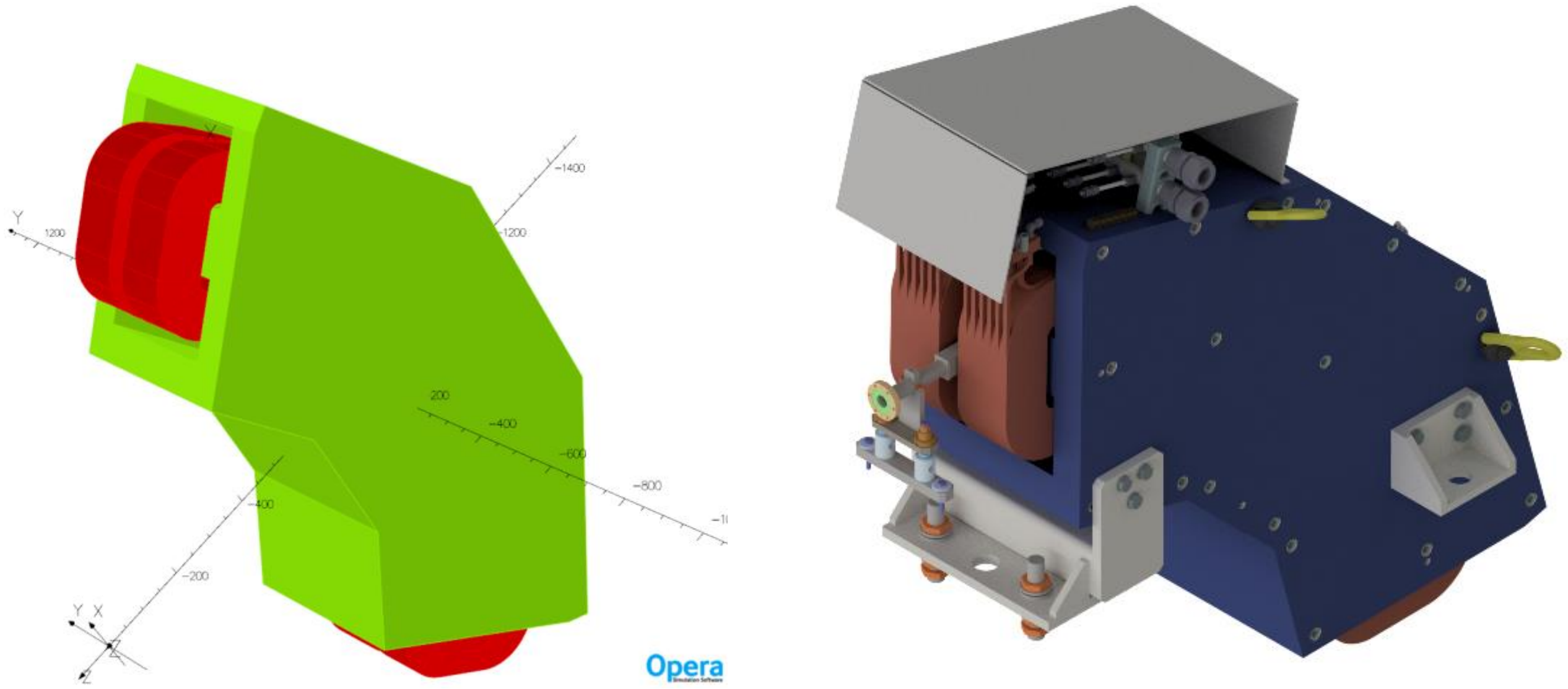

*Figure 13: Dipole Dump HEL view from an Opera 3D software simulation (left) and mechanical drawing sketch (right).*

*Table 14: STAR HEL Dipoles main parameters.*

| Parameters | Dipole type A | Dipole Dump HEL |
|---|---|---|
| Bending Radius | 677 mm | 846 mm |
| Bending Angle | 17.18° | 90° |
| Bmax | 1.6T | 1.3T |
| Overall Dim. WxHxL (including coil ends) | 660x465x340 mm | 1241x414x1241 mm |
| Magnetic Length | 201 mm | 1329 mm |
| Radial Field Homogeneity | 5E-4 over ±10mm | 1E-3 over ±10mm |
| Pole Gap | 30 mm | 30 mm |
| **COIL SPECIFICATIONS** | | |
| Number of Turns per coil | 120 | 120 |
| Conductor dimensions | 6x6 / bore 3 mm | 8.5x8.5 / bore 5.5 mm |
| **ELECTRICAL INTERFACE** | | |
| Nominal Current | 175 A | 140 A |
| Nominal Voltage | 30 V | 37 V |
| Inductance | 74 mH | 300 mH |
| Resistance | 120 mΩ | 266 mΩ |
| **WATER COOLING** | | |
| Water Flow Rate | 9 l/min | 19 l/min |
| Temperature drop | 8 °C | 8 °C |
| Pressure drop | 4 bar | 4 bar |

Figure 14 shows a 3D view of the STAR-HEL quadrupoles from Opera 3D software simulations and a sketch of a mechanical design while Table 15 lists all their main parameters.

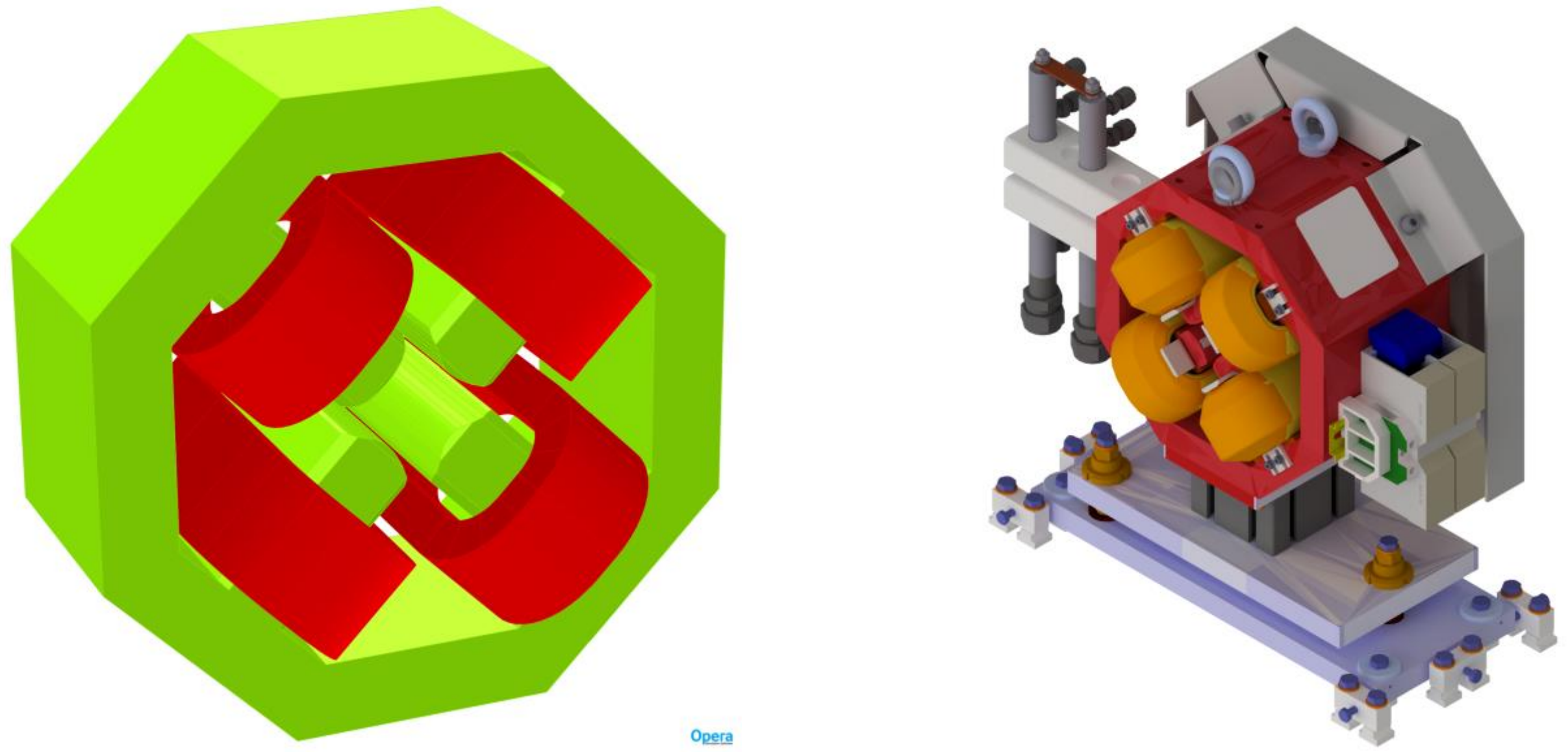

***Figure 14: Quadrupole view from an Opera 3D software simulation (left) and mechanical drawing sketch (right).***

***Table 15: STAR HEL Quadrupoles main parameters.***

| **Parameters** | |
|---|---|
| Maximum Gradient | 18 T/m |
| Overall Dim. WxHxL (including coil ends) | 246x246x177 mm |
| Magnetic Length | 100 |
| Radial Field Homogeneity | 5E-4 from n=3 to n=10 within a good field radius of 10mm |
| Aperture diameter | 30 mm |
| **COIL SPECIFICATIONS** | |
| Number of Turns per pole | 36 |
| Conductor dimensions | Ø 5 / bore 3 mm |
| **ELECTRICAL INTERFACE** | |
| Nominal Current | 47 A |
| Nominal Voltage | 4 V |
| Inductance | 7 mH |
| Resistance | 81 mΩ |
| **WATER COOLING** | |
| Water Flow Rate | 4.5 l/min |
| Temperature drop | 1 °C |
| Pressure drop | 4 bar |

Figure 15 shows a 3D view of the STAR-HEL steerers from Opera 3D software simulations and a sketch of a mechanical design while Table 16 lists all their main parameters.

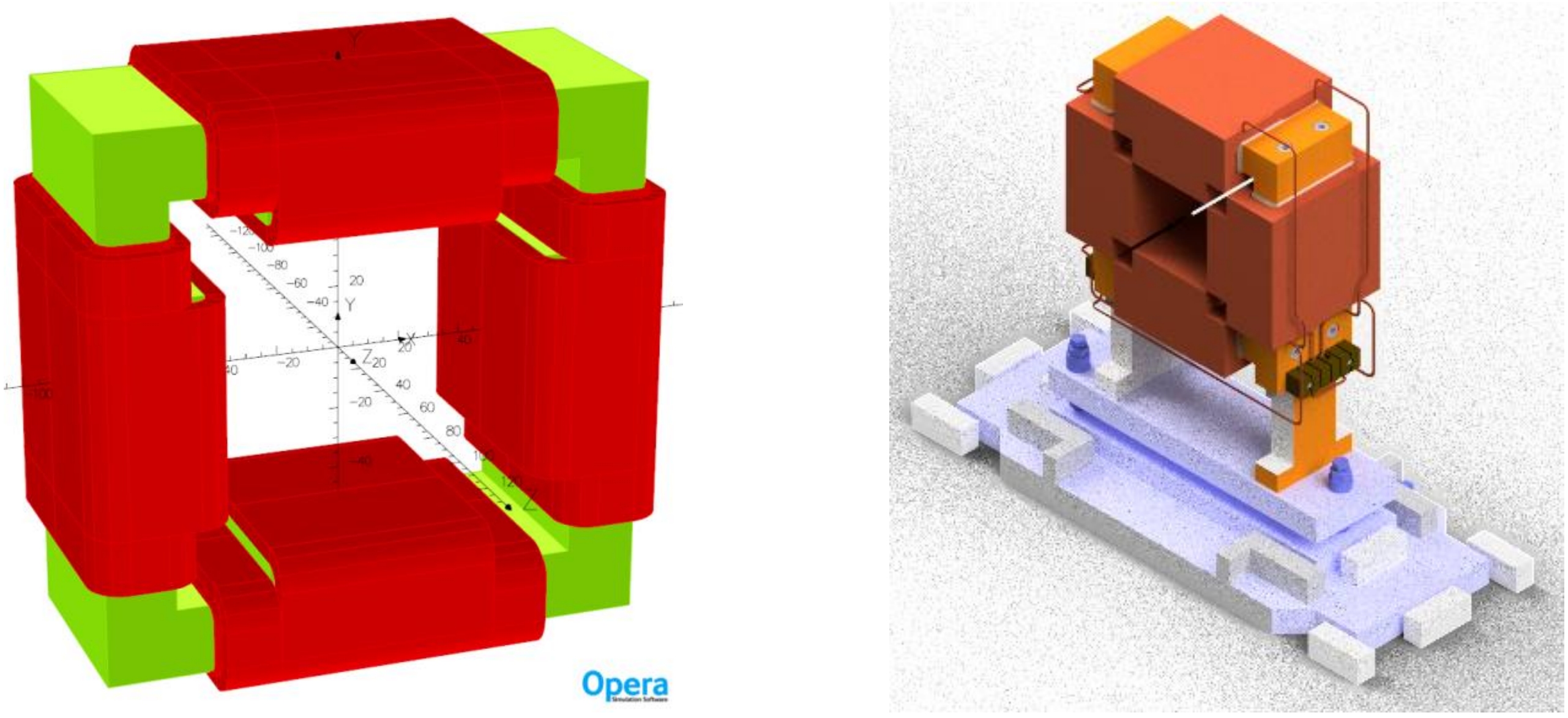

***Figure 15: Steerer view from an Opera 3D software simulation (left) and mechanical drawing sketch (right).***

*Table 16: STAR HEL Steerers main parameters.*

| **Parameters** | |
|---|---|
| Maximum Field | 70 G |
| Overall Dim. WxHxL (including coil ends) | 193x193x101 mm |
| Magnetic Length | 170 mm |
| Radial Field Homogeneity | 6e-4 over a ±10mm good field region |
| Pole gap | 70x70 mm |
| **COIL SPECIFICATIONS** | |
| Number of Turns per coil | 460 |
| Conductor dimensions | 0.9x2.12 mm |
| **ELECTRICAL INTERFACE** | |
| Nominal Current | 2.5 A |
| Nominal Voltage | 7 V |
| Inductance | 175 mH |
| Resistance | 6.8 Ω |

Regarding the solenoid AC1, several axial symmetric simulations have been performed with Poisson Superfish 2D FEA software. A sketch of the solenoid section is shown in Figure 16 while all the main parameters are listed in Table 17. It must be pointed out that the solenoid electrical parameters are fully compliant with the START 2 power supply model of Table 18.

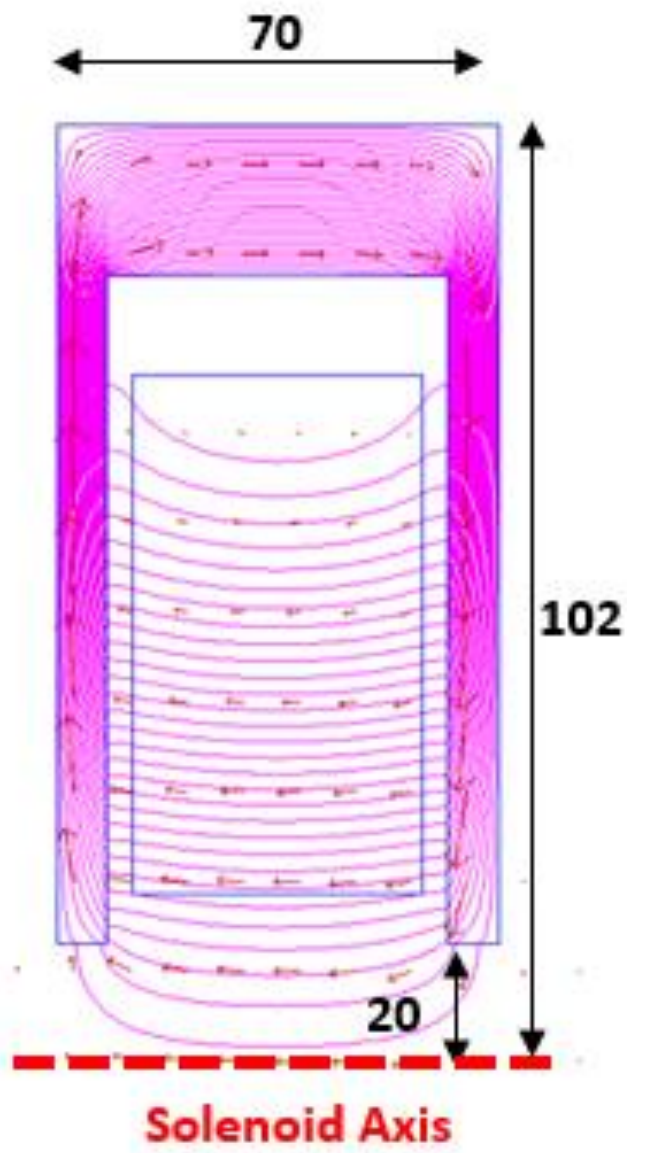

*Figure 16: Cross section of the AC1 solenoid. Dimensions are in mm.*

*Table 17: AC1 solenoid main parameters*

| | |
|---|---|
| Bmax | 2000 G |
| Yoke Material | Low Carbon Steel |
| Integrated field | 11.9 T mm |
| Good Field Radius | 10 mm |
| Integrated Field Quality | 3E-5 |
| **COIL SPECIFICATIONS** | |
| Number of Turns | 88 |
| Conductor dimension | 4x4 / bore 2 mm |
| **ELECTRICAL INTERFACE** | |
| Nominal Current | 110 A |
| Nominal Voltage | 5 V |
| Inductance | 6 mH |
| Resistance | 44 mΩ |
| **WATER COOLING** | |
| Water Flow Rate | 1.08 l/min |
| Temperature drop | 7 °C |
| Pressure drop | 1.74 bar |

All the magnets will be characterized by means of several mechanical dimension checks, electrical and hydraulic tests, such as insulation tests and hydraulic leak tests. Moreover, it will be defined the excitation curves of all magnets and a 3D magnetic field mapping will be done with a Hall probe system except for the quadrupoles. For the latter, a rotating coil measurement is most suitable to better evaluate the multipole content.

### 4.3. Power Supplies

Regarding the Power Supplies (PSs), we have defined the requirements listed in Table 18, that can guarantee the expected performances of the STAR-HEL upgrade.
Table 19 lists several PSs models that can fulfill the requirements of Table 18 and can feed all the magnets typologies presented in the previous paragraph, according to their electrical requirements.
All these PSs are air-cooled and can provide unipolar current and voltage, except for the steerers power supplies. Therefore, for all other typologies, each PS shall be equipped with a “polarity reversal” module consisting of contacts for the current polarity inversion.
Figure 17 and Figure 18 show all the power supplies typologies.

***Table 18: Power supplies main requirements***

| Parameter | Value | Unit |
|---|---|---|
| Stability, 8 hours | ± 150 | ppm |
| Accuracy | ± 500 | ppm |
| Reproducibility | ± 200 | ppm |
| Analogue programming resolution | 120 | ppm |
| Power factor | 0.92 | |
| V I ripple (resistive load). | 150 | ppm pk-pk |
| Warm-up period | 30 | minutes |

***Table 19: Power supplies for the STAR upgrade. From the 3rd column from left are listed the quantities, the maximum voltage, the maximum current, the efficiency, maximum output power, maximum power input, maximum heat air dissipated and the single crate heigth in rack units (width is the standard 19” for all the devices).***

| Power Supply Model | Magnet Powered | QTY | V [V] | I [A] | Eff. [%] | Pmax out [kW] | P in [kVA] | Pair_PS [kW] | Height [U] |
|---|---|---|---|---|---|---|---|---|---|
| SigmaPhi START11 | Dipoles | 3 | 55 | 200 | 90 | 11 | 14.38 | 1.94 | 4 |
| SigmaPhi START2 | Quadrupoles and solenoid | 7 | 15 | 135 | 90 | 2 | 2.61 | 0.35 | 3 |
| Bilt BE2811 | Steerers (1 Coil) | 14 | 18 | 5 | 85 | 0.09 | 0.12 | 0.02 | 4 |

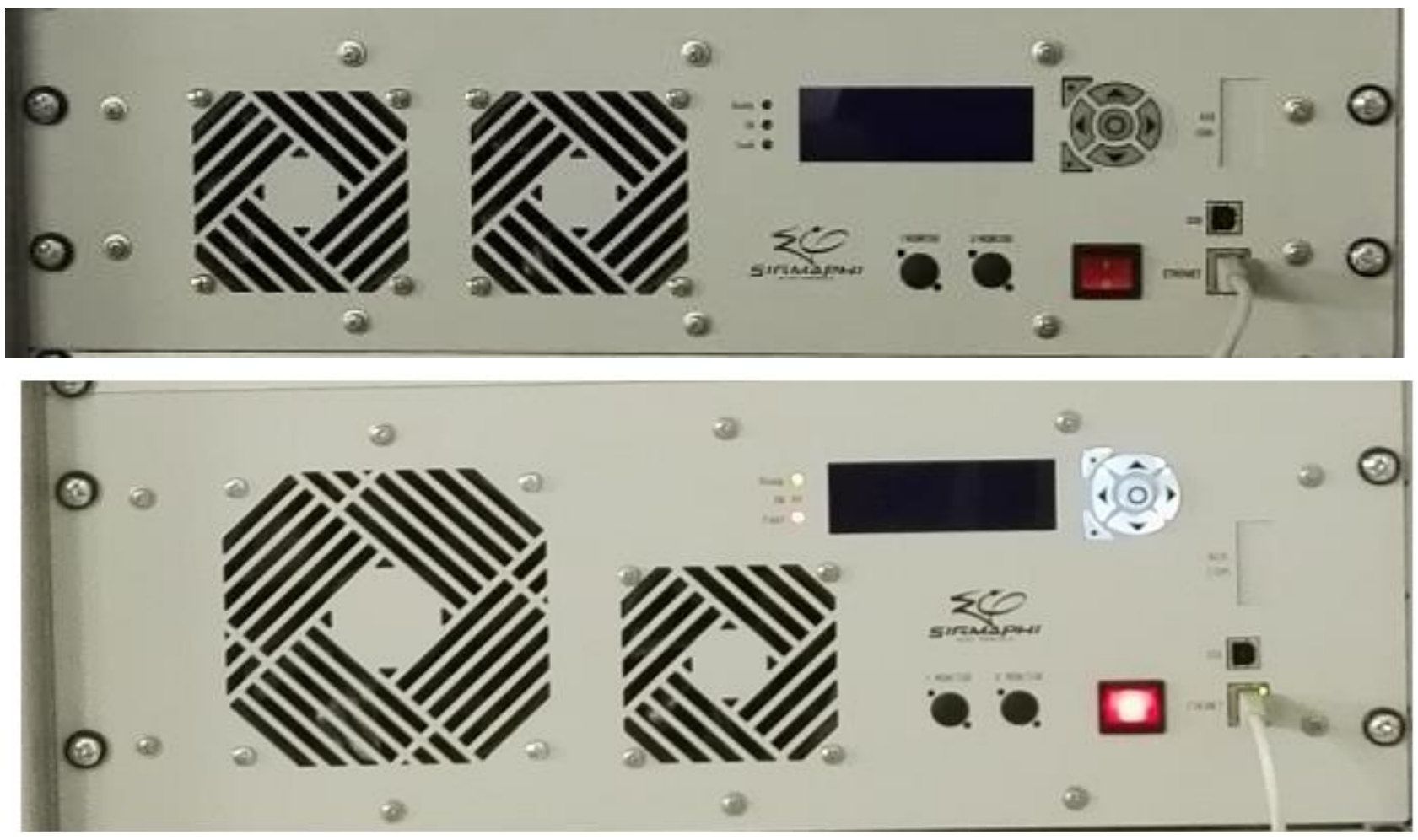

***Figure 17: START 2 power supply model (up) and START 2 power supply model (down).***

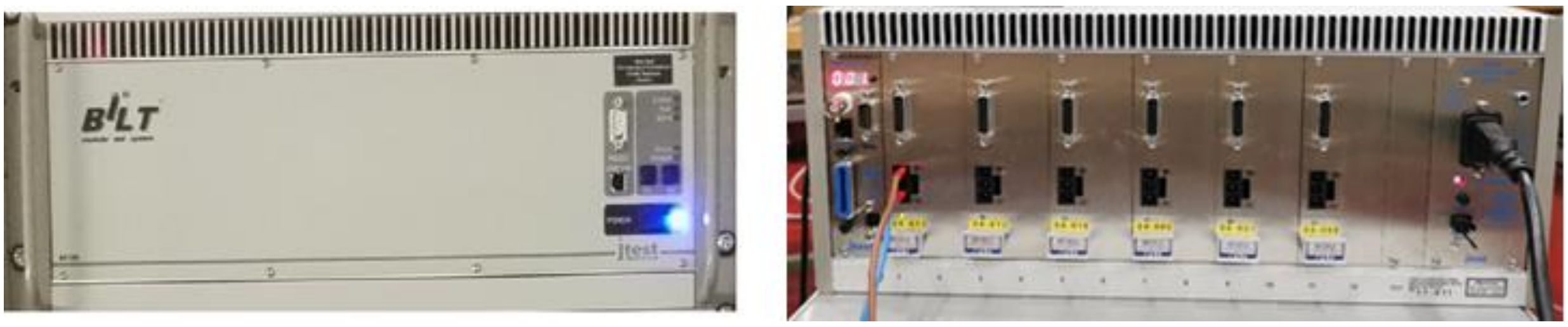

***Figure 18: Steerers power supplies main-fame front (left) and back (right) equipped with several PSs modules.***

All the power supplies will be characterized with several tests such as:

1. Long term stability
2. Current ripple
3. Resolution
4. Reproducibility
5. Interlock test
6. Efficiency

## 4.4. Magnets and Power Supply Utilities

Considering magnets and PSs main parameters reported in paragraph 4.2 and in Table 19 respectively, a preliminary estimation of magnet and power supplies utilities have been carried out in terms of electrical, water and air-cooling requirements. Table 20 reports all the utilities necessary for STAR-HEL.

*Table 20: Magnets and PSs STAR-HEL Utilities.*

| ELECTRIC POWER | |
|---|---|
| TOTAL Apparent Power [kVA] | 64.2 |
| 1-Phase Apparent Power [kVA] | 2.7 |
| 3-Phase Apparent Power [kVA] | 61.4 |
| **WATER COOLING** | |
| Magnets Water Cooling Heat Load [kW] | 16.7 |
| Total Water Flow [l/min] | 65.3 |
| **AIR COOLING** | |
| Magnets Air Cooling Heat Load [kW] | 2.6 |
| Power Supplies Air Cooling Heat Load [kW] | 8.5 |

Considering the quantities and the rack units of all the PSs typologies, we estimated a total request of four 42U racks in order to house all the power supplies listed in Table 19, including the switchboards and several units to reverse the current polarity on the magnets.

## 5. Beam Diagnostics, Electronics

Beam diagnostic devices will be installed before, between, and after the accelerating structures, as well as in the transport sections of the HE and LE Lines before the interaction points. Measurements on the electron beam include bunch charge, transverse beam position, transverse beam profile, beam energy and energy spread. Table 21 reports a list of the devices to be installed.

*Table 21: List of diagnostic devices.*

| Type | LINAC (Star1) | LINAC upgrade (Star-HEL) | LE Line (Star1) | HE line (Star-HEL) |
|---|---|---|---|---|
| Beam Position Monitor | 2 | 1 | 5 | 4 |
| Beam Current Monitor | 1 | - | 1 | 1 |
| YAG screen | 3 | 2 | 4 | 4 |
| Beam spectrometer | | | 1 | 1 |

## 5.1. Beam Position Monitor (BPM) System

A total of 12 Beam Position Monitors will be installed. The pickup selected for use is generally referred to as stripline BPM and is composed of four stainless steel electrodes of length $l$=140mm and width $w$=7.7mm, mounted with a π/2 rotational symmetry at a distance $d$=2mm from the vacuum chamber, to form a transmission line of characteristic impedance $Z_0$=50Ω with the beam pipe. Their angular width is ~26 degrees and the acceptance is Ø34mm. (see Figure 19)

Time domain reflectometry measurements will be performed to select the final strip width to get the best impedance matching.

The amplitude of the frequency response presents a sinusoidal shape with maxima at odd multiples of c/4l (~535MHz), selected to be as close as possible to the operating frequency of the detection electronics and to present nonzero response at the LINAC frequency of 2856 MHz allowing measurements of any satellite bunch, in case of multi-bunch operation.

Beam based alignment will be used to determine and adjust the centre. The estimated resolution is < 10 μm for a 0.5nC charge.

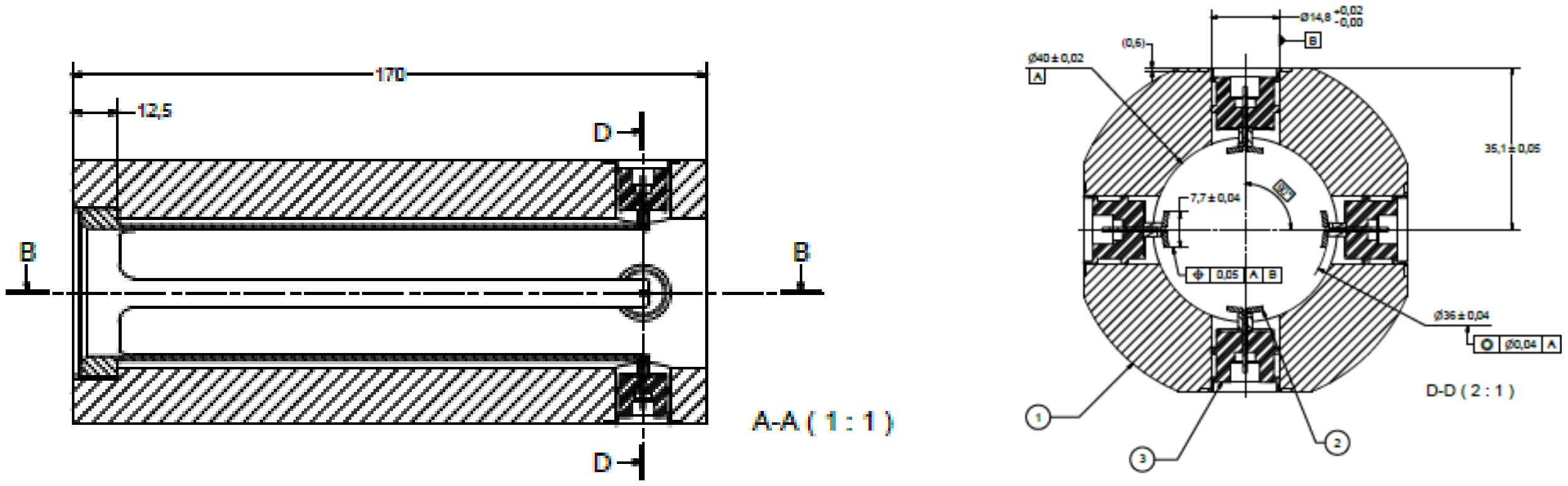

***Figure 19: CAD layout of the stripline beam position monitor.***

***Table 22: Stripline BPM parameters.***

| Parameters | Spec | Unit |
|---|---|---|
| Vacuum pipe diameter | 40 | mm |
| Stripline diameter | 35 | mm |
| Stripline angular width | 30 | degree |
| Stripline length | 140 | mm |
| Characteristic impedance | 50 | Ω |
| Strips thickness | 1 | mm |
| Max of transfer impedance | 2.08 | ohm |
| Sensitivity Δ/Sum | 1.255 | mm/dB |

The BPM detection electronics compare and process the signals induced by the beam in the four striplines to deduce beam transverse position. Each BPM will be equipped with a dedicated module (*Libera SPe*) built by Instrumentation Technologies, a Slovenian company developing systems especially for the needs of particle accelerators, and already widely used in many accelerator facilities around the world. *Libera Single Pass E* (see Figure 20) features accurate electron beam position measurements at various data rates with low crosstalk between channels, it is optimized to work with input signals from striplines and button pick-ups.
The signal from the pick-ups is processed in the signal processing chain, which is composed of analog signal processing, digitalization on fast ADCs and digital signal processing.
The analog part consists of four identical RF channels and four analog-to-digital converters. The RF channels have functions of short pulse lengthening, gain adjustment and filtering.
The processor module implements under-sampling technique. Ceramic SAW band-pass filters are implemented to suppress frequency components that are out of the chosen Nyquist zone and could cause aliasing. The output of each RF channel connects to the 16bit ADC converter suited for direct sampling of 500MHz signals.
User can access functions implemented in the unit through a control system interface. This interface is developed in order to be easily integrated into the accelerator's control system software.

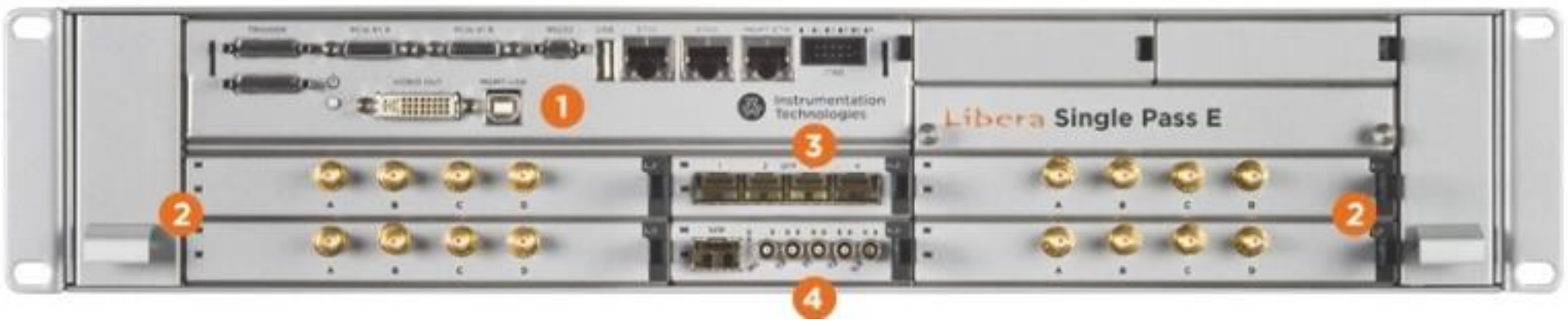

***Figure 20: Libera BPM processing module (Ref. Instrumentation Technology)***

***Table 23: diagnostic electronics installation.***

| BPM & ICT electronics | Rack BPM | Rack ICT |
|---|---|---|
| Libera SP electron | 4 units | |
| Patch Panel BPM | 13x4 input | 4 inputs |
| Ethernet Switch | 1 unit | 1 unit |
| Triggers Distribution | 2 units | 1 unit |
| Current Transformers ADC | | 4 units |

## 5.2. Charge Measurements

Monitoring of the beam charge is based on the use of integrating current transformers (ICT) and fast current transformers (FCT) manufactured by *Bergoz* company (France) and an external current monitoring electronics acquisition system. Transformers are built UHV compatible, they include a ceramic gap, shields and wall current bypass, can be mounted directly in the beam line via a stainless-steel  flange and, at least in the RF gun area, are resistant to bake-out procedure up to 150°C.
The FCT is a passive wideband AC transformer with 5ns rise time and bandwidth up to 150MHz. The ICT is a capacitively shorted transformer that stretches a beam bunch of a picosecond risetime to a calibrated output voltage pulse of around 70 nanoseconds with its area to the total charge of the beam bunch. The current transformers along the vacuum chamber are connected to the instrument room via low-loss Heliax coaxial cables.

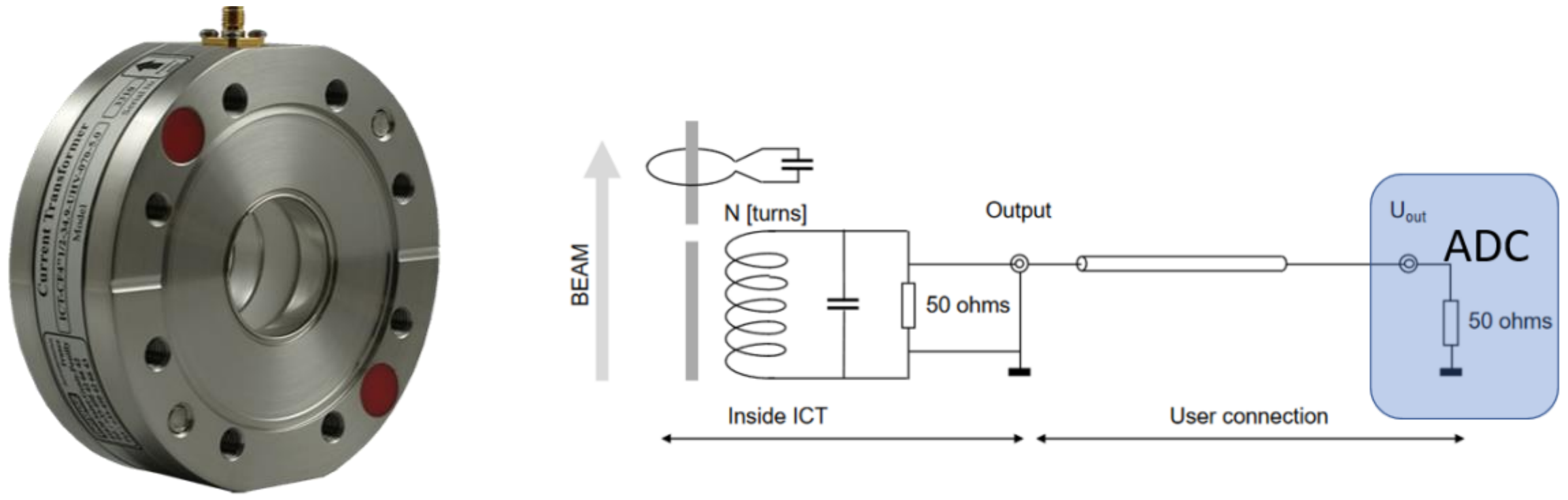

***Figure 21: integrating current transformer and equivalent circuit (Rif.  www.bergoz.com)***

The signal processing of the output waveform requires a high precision measurement of the voltage x time area of a very short pulse, which is done by commercially available multichannel ADCs. Each signal is fed into a single current monitoring electronics module which can be accessed by the control system via ethernet (see Figure 21). The total charge, including the dark current, is measured. Table 24 reports main transformer and electronics specifications

***Table 24: Bergoz ICT / FCT specifications.***

| **Parameters** | **Spec** | **Unit** |
|---|---|---|
| Sensitivity in a 50Ω load (min) | 5 | V·s/C |
| ICT Output Pulse duration | 70 | ns |
| FCT Output Pulse duration | 5 | ns |
| FCT Output signal droop | 3.59 | %/us |
| FCT low frequency cutoff | 5 | kHz |
| FCT high frequency cut off (min) | 180 | MHz |
| ICT output signal droop | 7 | %/us |
| ICT low frequency cutoff | 13 | kHz |
| ICT high frequency cut off | 2.6 | MHz |
| ADC  resolution (min) | 10 | bit |
| Sampling rate (min) | 4 | Gs/s |
| ADC Bandwidth (min) | 500 | MHz |
| ADC  input range | 0.05 - 5 (programmable) | V |

## 5.3. Beam size and profile, energy and energy spread measurements

Beam size and profile in the linac is measured by screen monitors consisting of a cerium doped yttrium aluminium garnet (Ce:YAG) crystal.  It is imaged through an 45° in-vacuum mirror onto a high-resolution camera. Digital CCD cameras (Basler SCA640), complying with GigE Vision standard, equipped with a 105 mm macro lens will be used.
Every diagnostic chamber is equipped with a mechanical actuator, with UHV feedthrough to drive inside a screen holder.
The imaging can be setup with different magnifications, accounting for the change in beam size along the accelerator. A system to read out the CCD cameras at the beam rate of 100 Hz will be available. It allows for a bunch-synchronous data acquisition and processing.
YAG screens, paired with an upstream quadrupole magnet, will be used to measure the transverse beam emittance by the quadrupole-scan method. At low bunch charges, the photocathode thermal emittance may also be measured in the same way. Due to uncertainties related to jitters and stabilities in general, we evaluated an absolute measurement error on the beam emittance of about 0.8 μm, considering a 85% content of the transverse phase space area.
The beam energy can be measured by bending the electron bunch along a known trajectory with curvature radius $r$ by a dipole magnet with magnetic field $B$. Also, by measuring the bunch width at the exit, the energy spread will be calculated knowing the dispersion function at the location of the measurement screen. For this purpose, two diagnostic chambers will be installed downstream the dipole DHPELE101 in the high and low energy line.

# 6. Radiation Safety, Beam dump

## 6.1. Shielding outlines

The new STAR layout, after upgrade with STAR-HEL, has been previously shown. Using the beam parameters listed above, the shielding has been defined in terms of size, positioning and materials to be used. The criterium adopted has been inspired by the maximum precaution and to the redundancy of the technical choices. In addition, we are also sure to guarantee all possible degrees of freedom in the event of future modifications to the machine or new needs.

High-energy electron accelerators are complex devices containing many components. All facilities contain the same basic systems:

- Accelerators structures
- RF power components
- Vacuum system
- Magnetic systems for beam steering and focusing
- Water-cooling
- Etc...

Prompt radiation and radioactivity induced by particle nuclear interaction in beam line elements and shielding structures represents the main radiation hazard of electron accelerators.
The accelerator's design parameters are of crucial importance in the determination of the nature and magnitude of radiation source. The most important parameters are:

- Particle energy

• Beam power
• Target material
•Workload
• Beam losses
as well as the physical layout

The STAR configuration, as discussed in the Introduction and Section 1, consists of a low energy line taking the electron beam to the interaction point with the laser at a maximum energy of 85 MeV, and a high energy line with a maximum electron beam energy of 150 MeV. The bunch charge is foreseen up to 500pC, with a maximum repetition rate of 100Hz. Operations with a low repetition rate, down to 1 Hz, are possible during commissioning tests, with the aim of largely reducing the beam average current taken to the beam dumps (including dark current from the accelerating sections, that scales with the repetition rate).

### 6.2. Shielding Design Criteria

The shielding design criteria have been based on the text of the Italian legislation (D.Lgs. 101/20); according to European Directives as well as the recent ICRP recommendations (ICRP 103). According to previous documents the individual limits are 20 mSv/y for radiation workers, and 1 mSv/y for the members of the public.
Moreover, the definitions of controlled and supervised areas are useful as guidelines. A controlled area is every area where 3/10 of the limits recommended for radiation worker may be exceeded. A supervised area is one area where the overcoming of 1/10 of the previous limit may occur.
Taking into account the dose levels normally found around accelerators, the thickness of the shielding was calculated maintaining the doses, within the areas outside the shield and frequented by the staff, at 1mSv/y and at 0.1 mSv/y in the other areas.
A shifting from these values could at most change the radiation classification of some areas.
In normal working condition the dose rate outside shielding should not exceed a fraction of mSv/h (about 0.02 mSv/h for 6000 h/y of operation).

### 6.3. Source Term

For shielding evaluation purposes, three components of radiation field which are produced when an electron beam, with an energy less than hundred MeV hits both a vacuum chamber wall or a thick target must be considered.

### 6.4. Bremsstrahlung

Prompt photon fields produced by Bremsstrahlung constitute the most important radiation hazard from electron machines with thin shielding. Bremsstrahlung yield is very forward peaked, and increasingly so with increasing energy.
The following equation describes this behaviour:

$$\theta_{1/2} = 100/E_0$$

where $\theta_{1/2}$ is the angle in degrees at which the intensity drops to one half of that at 0°, and $E_0$ is the energy of the initial electrons in MeV. In order to evacuate the shield thickness a "thick target", usually a target of sufficient thickness to maximize bremsstrahlung production, was considered. Photon yield from a thick target as a function of angle consists of two components: sharply varying forward component, described in the equation reported section 6.9, and a mildly varying wide-angle component. Forward (or zero-degree) bremsstrahlung contains the most energetic and penetrating photons, while bremsstrahlung at wide angles is much softer.
The source term (per unit beam power) for bremsstrahlung at 90° is independent of energy.

### 6.5. Neutrons

Photons have larger nuclear cross-sections than electrons, so neutrons and other particles resulting from inelastic nuclear reactions are produced by the bremsstrahlung radiation. Neutrons from photonuclear reactions are outnumbered by orders of magnitude by electrons and photons that form the electromagnetic shower. However, some of these neutrons constitute the most penetrating component determining factor for radiation fields behind thick shielding.

### 6.6. Giant resonance production

The giant resonance production can be seen in two steps:
1) the excitation of the nucleus by photon absorption.
2) the subsequent de-excitation by neutron emission, where memory of the original photon direction has been lost.
The cross-section displays a broad maximum around 20-23 MeV for light nuclei (mass number $A \leq 40$) and 13-18 MeV for heavier nuclei.
The angular yield of giant resonance neutrons is nearly isotropic.
The giant resonance is the dominant process of photo-neutron production at electron accelerators at any electron energy.

### 6.7. Pseudo-deuteron production

At photon energies beyond the giant resonance, the photon is more likely to interact with a neutron-proton pair rather than with all nucleons collectively. This mechanism is important in the energy interval of 30 to ~300 MeV, contributing to the high-energy end of the giant resonance spectrum. Because the cross-section is an order of magnitude lower than giant resonance, with the added weighting of bremsstrahlung spectra, this process never dominates.

### 6.8. Photopion production

The production of pions (and other particles) becomes energetically possible above the ≈140 MeV threshold. These pions decade into secondary neutrons as a by-product of their interaction with nuclei. While substantially fewer than GR neutrons, the photo-pion neutrons are very penetrating constituting the main component of the initial radiation field from a target (together with muons at very high energies) outside very thick shields.

### 6.9. Star shielding preliminary calculations

The bulk shielding for the accelerator enclosures has been calculated using the following expression:

$$\sum_i \dot{H}_i = \sum_i \frac{S_i}{r^2} e^{-d/\lambda_i} f_i$$

where

$\sum_i \dot{H}_i$ is the ambient dose equivalent rate summed over all components,

$S_i$ is the source term,
$r$ is the distance of interest,
$d$ is the bulk shielding thickness,
$\lambda_i$ is the attenuation length of the $i^{th}$ radiation component,
$f_i$ is conversion coefficients for use in radiological protection against external radiation for the corresponding radiation component ($i^{th}$ component).

The calculation has been performed at the maximum energy and current in the scenario of 1% beam loss on a thick copper target, operating therefore within the worst-case scenario and with the maximum redundancy, as previously stated.
The gas bremsstrahlung is produced by the interaction of the electron beam with residual low-pressure gas molecules in the vacuum pipe. Bremsstrahlung on residual gas is one of the main causes of beam loss in a storage ring and may represent a radiation hazard at synchrotron radiation facilities. This type of radiation has been thoroughly investigated at circular storage rings, where the beam current is much more intense. It is mainly in the straight section that a radiation problem could arise.

### 6.10. Induced Activity

Personnel exposure from radioactive components in the beam line is of concern mainly around beam lines, collimators, slots, beam stopper or beam dump, where the entire beam or a large fraction of the beam is dissipated continuously, while unplanned beam losses result from beam mis-steering due to inaccurate orbit adjustment or devices failure.
Beam losses induce activation in machine component as well as in

the beam pipes ($^{60}Co$, $^{54}Mn$, $^{51}Cr$, $^{46}Sc$, $^{22}Na$, $^{11}C$, $^{7}Be$)

the cooling water ($^{3}H$, $^{7}Be$, $^{15}O$, $^{13}N$, $^{11}C$)

the air ($^{15}O$, $^{13}N$, $^{38}Cl$, $^{41}Ar$)

the concrete walls ($^{152}Eu$, $^{154}Eu$, $^{134}Cs$, $^{60}Co$, $^{54}Mn$, $^{22}Na$)

The activation of soil as well as the groundwater by neutrons and other secondary particles can have an environmental impact.
Calculations have been also performed, following the final definition of the beam parameters and by considering the worst-case scenario beam losses, in order to quantify the possible activation mechanisms of the cooling water system, the air circulation system and the beam dump layout.

### 6.11. Machine accesses

During machine operation the STAR vault tunnel will be an excluded area.
During no operation periods the STAR vault tunnel will be a controlled area, due to the possible activation of the machine structure.
The technical areas behind the roof shield will be classified as controlled or supervised areas.
The experimental areas will be a free access area. Only areas close to the front ends or at the end of the beam line will be classified.
In order to protect workers in the experimental areas, the electron beam will be dumped below the floor after the laser interaction.

### 6.12. Beam line radiation shielding design

For each shielding situation (insertion device white beam, radiation transport, monochromator, hutches etc.) the synchrotron radiation, the gas bremsstrahlung, the high-energy bremsstrahlung, from beam halo interactions with the structures of the machine, has been calculated. In these calculations a representative geometry of the machine and of the beam line has been adopted. Also, in this case the criterium of maximum redundancy and consideration of the worst-case scenario has been adopted. Furthermore, the shielding system will be sufficiently flexible in order to be adapted to variations of the operation geometry.

### 6.13. The operational radiation safety program

The purpose of the operational safety system program is to avoid life-threatening exposure and/or to minimize inadvertent, but potentially significant, exposure to personnel. A personnel protection system can be considered as divided into two main parts: an access control system and a radiation alarm system.
The access control system is intended to prevent any unauthorized or accidental entry into radiation areas.

The access control system is composed by physical barriers (doors, shields, hutches), signs, closed circuit TV, flashing lights, audible warning devices, including associated interlock system, and a body of administrative procedures that define conditions where entry is safe. The radiation alarm system includes radiation monitors, which measure radiation field directly giving an interlock signal when the alarm level is reached.

### 6.14. Interlock design and feature

The objective of a safety interlock is to prevent injury or damage from radiation. To achieve this goal the interlock must operate with a high degree of reliability. All components should be of high grade for dependability, long life and radiation resistant. All circuits and component must be fail-safe (relay technology preferably).
To reduce the likelihood of accidental damage or deliberate tampering all cables must run in separate conduits and all logic equipment must be mounted in locked racks.
Two independent chains of interlocks must be foreseen, each interlock consisting of two micro switches in series and each micro switches consisting of two contacts.
Emergency-off buttons must be clearly visible in the darkness and readily accessible.
The reset of emergency-off buttons must be done locally.
Emergency exit mechanisms must be provided at all doors.
Warning lights must be flashing, and audible warning must be given inside radiation areas before the accelerator is turned on.
Before starting the accelerator, a radiation area search must be initiated by the activation of a "search start" button. "Search confirmation" buttons mounted along the search path must also be provided. A "Search complete" button at the exit point must also be set.
Restarting of the accelerator must be avoided if the search is not performed in the right order or if time expires.
The interlock system must prevent beams from being turned on until the audible and visual warning cycle has ended.
Any violation of the radiation areas must cause the interlocks system to render the area safe.
Restarting must be impossible before a new search. Procedures to control and keep account of access to accelerator vaults or tunnels must be implemented.

### 6.15. ELECTRON BEAM DUMP

For Star project three dumps are planned to be installed at the end of the STAR-low-energy line, STAR-High-Energy line and a dump for dark current. The layout of beam dumps as well as the size and shape of shielding materials has been defined. Integration of present radio-protection instrumentation with the access control system will be provided.
The highest goal of all beam dump systems must be a long term, faultless, safe and reliable operation, taking into account the power absorbed.
Thus, every layout decision on the dump should be taken under the hypothetical assumption that the dump cannot be exchanged and must survive the whole lifetime of the facility. Anyway, this does not at all mean to omit the exchange possibility.

The most critical design issues were to minimize the hazardous effects of radiation, both prompt and long term, and to obtain a robust thermo mechanical design.
In order to project an effective and reliable dump is necessary to take into account the following items:

- Capability to withstand the parameters of the incoming beam
- Absorption efficiency
- Lifetime
- Compactness
- Simple and Reliable Fabrication and installation Methods
- Induced Radioactivity and Radiation Shielding
- Safe dump exchange possibility
- Vacuum and cleanliness requirements
- Water cooling system
- Heating, Material Selection.

An electromagnetic shower may be characterized by

- Radiation lengths $X_0$
- Moliere radius $X_M$
- Critical Energy $E_C$

The geometry of an electron dump, usually consists of a cylindrical core in a low Z material (Aluminum or Graphite) of number of radiation lengths $X_0$ and Moliere radius $X_M$ able to absorb the electromagnetic shower produced by a very powerful beam

A dump must be surrounded by a sandwich of material as shown in Figure 22.

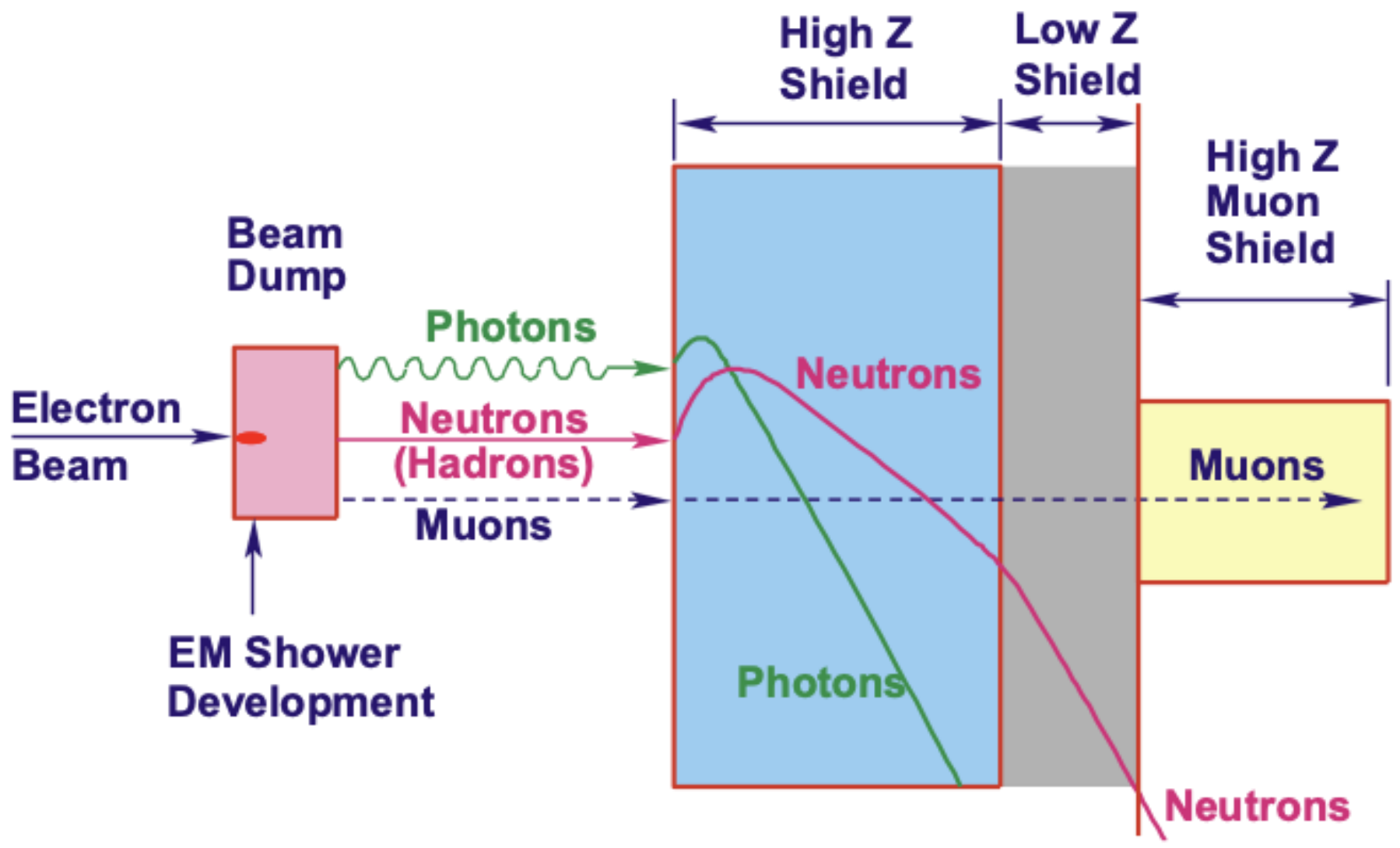

*Figure 22: Typical shield behind high energy beam dump.*

The high Z material is used to reduce the energy of high energy neutrons through the reaction (n,xn), and then absorbed the in a low z shield. An additional high Z material will be used for the absorption of both photons fron neutron capture and muons if produced.
The STAR electron beam will be dumped underground under the floor level both for high and low energy lines, while the dark current will be dumped at accelerator level.

### 6.16. OTHER RADIATION SOURCES

The RF power sources for STAR.
The klystrons will arrive already shielded from factory.
Additional shield will be installed in order eliminate and/or to reduce radiation escape. An interlocked fence around klystrons is foreseen in order to reduce as low as possible the radiation level behind the fence.

**It is hardly necessary to point out that the authorization application is the responsibility of Unical and its radiation protection Expert.**

# 7. Auxiliary Plants

## 7.1. Introduction

The STAR-1 auxiliary plants (electrical, cooling, air conditioning) must be modified to fulfill the requirements of the STAR-HEL upgrade. A preliminary review of the existing plants has been performed, based on the available documentation. The modification and enhancement of the existing plants has been dimensioned, but a complete survey and a functional check must be carried out before finalizing the implementation of the technical solutions.

## 7.2. Cooling system

The thermal plant currently existing is able to ensure the functioning of STAR-1, but in order to guarantee the functioning of STAR-HEL some of its components must upgraded. To make up for this deficit, it is considered appropriate to adopt a scheme in which the various users currently served by the heat engine are decoupled by installing for some of them dedicated thermal machines. In detail, we intend to install a separate chiller for the accelerating sections. This solution allows to guarantee a better control of flow stability, and therefore the achievement of a better thermal stability, a particularly critical parameter as regards the accelerating sections, whose relevance is illustrated in Chapter 1.5.

The estimate of the power to be dissipated for the upgrade is reported in the table.

***Table 25: STAR-HEL cooling power demand***

| Cooling power | (kW) |
|---|---|
| RF STATIONS (2) | 90 |
| MAGNETS | 17 |
| TOTAL | 107 |

In the next table a preliminary analysis of the available capacity of the existing plant is reported.

***Table 26: cooling capacity STAR-1***

| Heat exchanger capacity STAR-1 | |
|---|---|
| Cooling power (kW) | 233 |
| Water flow (l/min) | 308 |

The heat exchanger capacity is not adequate to the upgrade demand, and the water flow seems already barely fulfilling the STAR-1 requirements.
Note that the refrigerating power of the installed Trane RTWD120 chiller (428 kW. 1225 l/min) is quite larger than the heat exchanger one, possibly due to the additional requests of the site's air handling units.
An updated power and flow budget must be assessed to evaluate the best upgrade solution for the cooling system as well for the air conditioning.

## 7.3. Electric plant

The site electric station has already a suitable spare. In the table the assessment of the upgrade is reported.

*Table 27: HEL upgrade Electric load estimation*

| Electric load | Active power [kW] | Effective current Ib [A] | Note |
|---|---|---|---|
| RF Stations | 108 | **173.41** | Feeder available in QGBT CABINA TR1 (spare 250 A) |
| Magnets power supplies | 63 | **101.16** | Feeder available in QGBT CABINA TR1 (spare 200 A) |
| Vacuum pumps | 9 | **43.48** | Supply available in distribution cabinets |
| Cooling Upgrade | - | **-** | To be assessed, depending on the cooling system upgrade |
| **TOTAL** | | **318.04** | |

A preliminary assessment of the activities for the electric plant upgrade is summarized in the next table.

*Table 28: Electric plant upgrade activities*

| **Description** | **Quantity** |
|---|---|
| **line 250 A 3,5x 95 mm2 in exist. conduit** | **200 m** |
| **Busway arrival cabinet** | **2** |
| **Busway 250 A 3m** | **6** |
| **Busway feeder** | **2** |
| **Busway derivation** | **15** |
| **Magnets power supplies AC cabling** | **1** |
| **Magnets power supplies DC cabling** | **2000 m** |
| **Auxiliaries cabling** | **1** |
| **(HW/SW) PLC and SCADA update and integration** | **1** |
| **Equipotenzial network (CBN) for EMC** | **1** |
| **Active filtering for RF Stations** | **2** |

# 8. Control System

The control system architecture for the STAR-HEL upgrade will be compatible with the EPICS (Experimental Physics and Industrial Control System) framework chosen as control system platform for the STAR-1 project.

EPICS is a set of Open-Source software tools, libraries, and applications developed collaboratively and with an international user-base. The EPICS framework is widely used as the control system infrastructure to create distributed soft real-time control systems for large scientific installations. The strength of EPICS is in its ability to allow communications over a large computer network to provide control interface and receive feedbacks from distributed parts and components of a single installation.

Through the EPICS framework the STAR command and control system was designed using a three layers architecture which consists of:

1) the device interface layer, that include all the interaction with equipment and devices;
2) the central supervision, monitoring and data handling layer that operates all of the services needed to run independently from user's interactions, the alarm and interlocks systems and the slow feedbacks loop;
3) the presentation layer in which all the user interfaces and high-level software are located.

In STAR-HEL the devices to be integrated in the existing control system are very similar to the STAR-1 ones: the radio frequency I/O controller manages the high and low power RF signals reading back all the modulator signals, the interlocks and RF diagnostics devices. Some drivers will have to be installed to control the C-Band LLRF and the modulators functionalities.
For the magnets power supplies the drivers will have to be adapted to the firms' protocol but no special issues are expected to connect to the existing STAR control system. The same considerations are valid for the Vacuum system.
Diagnostics systems are also similar to those already installed. The Beam Position Monitor (BPM) are strip-line monitors that will use the same drivers for both the position measurement and the reconstruction of the trajectory through the acquisition electronics. The transverse dimension monitors are composed by removable view screens; the CCD cameras will be similar to the installed ones in order to replicate the driver and the emittance measurements tools. The current monitors have their own electronics and will also be compatible with the existing acquisition drivers.

Therefore, the STAR-HEL control system is an implementation of the existing STAR-1 one.
An integration of the new devices will be performed in the interface level and in the high-level software including the presentation level.

### 8.1. Synchronization and Timing

MTG (Master Time Generator) IOC is in charge to generate event signals to synchronize data control/monitoring and to fulfill strictly the timing sequences.
The timing system performs two relevant tasks: fast timing and event signal. It provides a time stamp mechanism for the events that require a shot-by-shot synchronization and monitoring.
The aim of the fast timing is to trigger devices with fine time resolution and low jitter. These devices are, for example, the photo-injector, the beamline transport components, the LLRF, beam diagnostic (beam position monitors and current transformers), Laser Systems. All of these systems must be, as a matter of fact, synchronized with the bunch arrival time at IP better than <1 ps. The nominal maximum rep. rate at which the STAR source shall operate is 100 Hz.

State-of-the-art and commercially available Master Timing Generator IOC will be used. The TTL signals are transmitted to the other equipment through a TTL-Fiber converter. All delays, duty cycles, etc. are totally controlled by the software.

# 9. Vacuum

The vacuum system is one of the key components in the STAR-HEL upgrade. The performances of the linac depend critically on the vacuum pressure. Extreme care must be then adopted during each step of design, construction and assembling of all the vacuum system. An accurate ultra-high vacuum technology practice must be adopted during each step of the design of each part of the vacuum chamber; only all metal components and devices are permitted together with oil free vacuum pumping systems. Special care must be adopted for the design of the RF Gun vacuum system, because of the very high pollution sensitivity of the photo cathode.
The system has been designed using several ion pumps of different nominal vacuum speed, in order to reach a final vacuum level of about $10^{-8}$ mbar, that is compatible with these kinds of accelerators.
STAR is going to be a user facility: therefore, reliability and easy maintenance during operation is of paramount importance. For this reason, we will install several vacuum valves in order to have the possibility to work in a specific section of the linac keeping the other sections under vacuum.
A fast valve has been foreseen on the line to preserve vacuum level of the electron Gun, that is the most critical part of the accelerator, in case of failure or incorrect operation sequences.
Another critical point is the beam–laser interaction chamber. Being its engineering design in progress, a preliminary setup using one or two turbo molecular pumps has been adopted, but specific simulations must be performed in order to better understand the vacuum profile inside the chamber.
In Figure 23 we show the distribution of the on-axis vacuum pressure along the full STAR machine, including STAR-HEL.

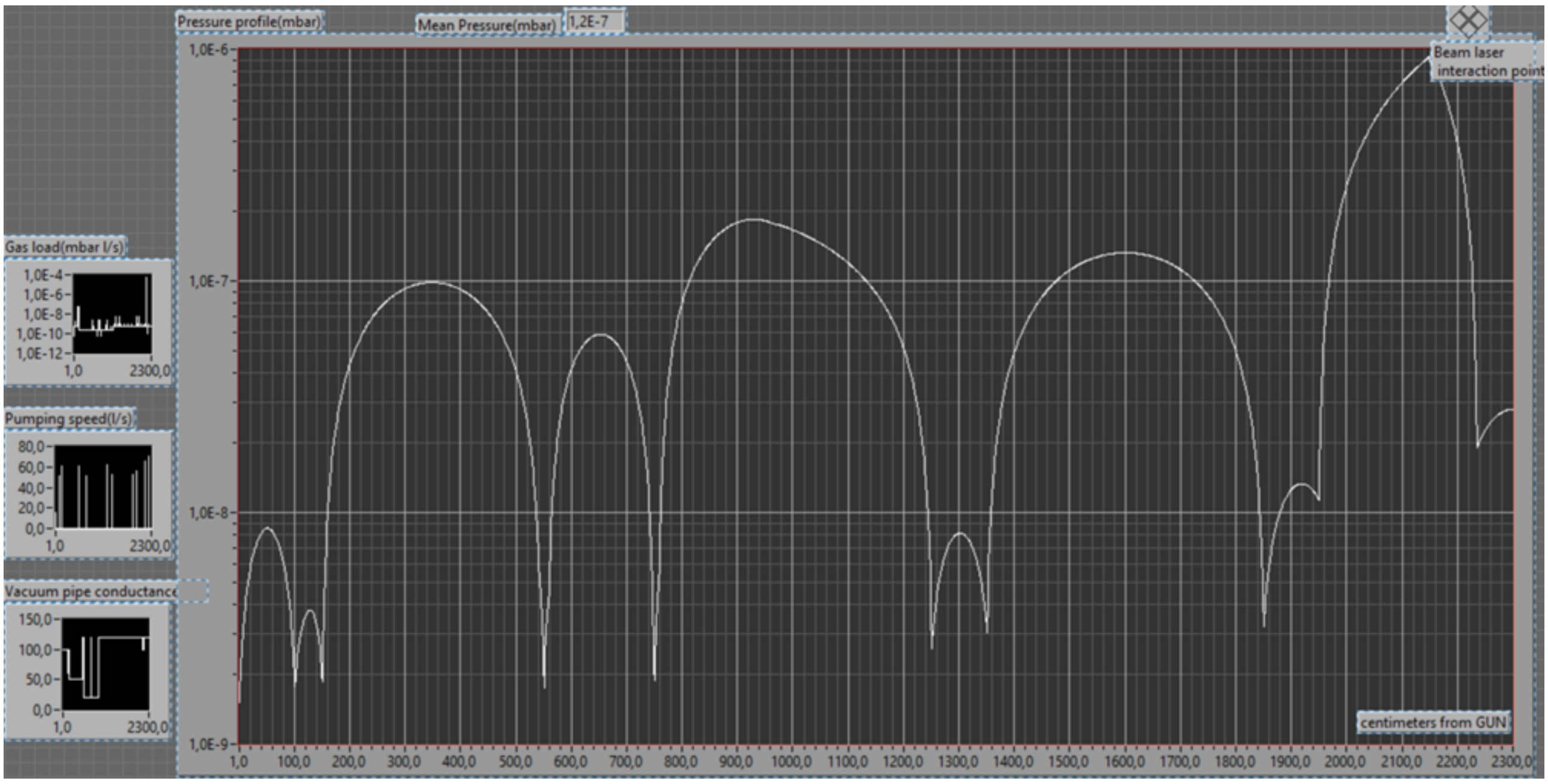

***Figure 23: the distribution of the on-axis vacuum pressure along the full STAR machine, including STAR-HEL.***

These simulations have been computed using a conservative out-gassing rate for copper, stainless steel and aluminum. These values are conservative for materials without bake-out treatment. With a baking out procedure, they should improve by a factor up to 100 or more.

***Table 29: List of vacuum pumps (in black the new ones for STAR-HEL upgrade). Ion pumps for waveguides are not included in the list.***

| | |
|---|---|
| PS-GUN00 | VacIon 2 L/s Pump Agilent Technology |
| PS-IOPGUN01 | VacIon Plus 300 Pump Agilent Technology |
| PS-IOPGUN02 | VacIon Plus 150 Pump Agilent Technology |
| PS-IOPLIN01 | VacIon Plus 300 Pump Agilent Technology |
| PS-IOPLIN02 | VacIon Plus 300 Pump Agilent Technology |
| PS-IOPDGL01 | VacIon Plus 300 Pump AgilentTtechnology |
| PS-IOPDGL02 | VacIon Plus 300 Pump Agilent Technology |
| PS-IOPIPL01 | VacIon Plus 300 Pump Agilent Technology |
| PS-IOPIPL02 | VacIon Plus 300 Pump Agilent Technology |
| PS-IOPDUM01 | VacIon Plus 300 Pump Agilent Technology |
| PS-IOPHEL01 | VacIon Plus 300 Pump Agilent Technology |
| PS-IOPHEL02 | VacIon Plus 300 Pump Agilent Technology |
| PS-IOPHEL03 | VacIon Plus 300 Pump Agilent Technology |
| PS-IOPHEL04 | VacIon Plus 300 Pump Agilent Technology |
| PS-IOPHEL04 | VacIon Plus 300 Pump Agilent Technology |

# 10. Engineering

## 10.1. Organization Breakdown Structure

A short version of the organization structure is represented in the Figure 24: Organization schematic.

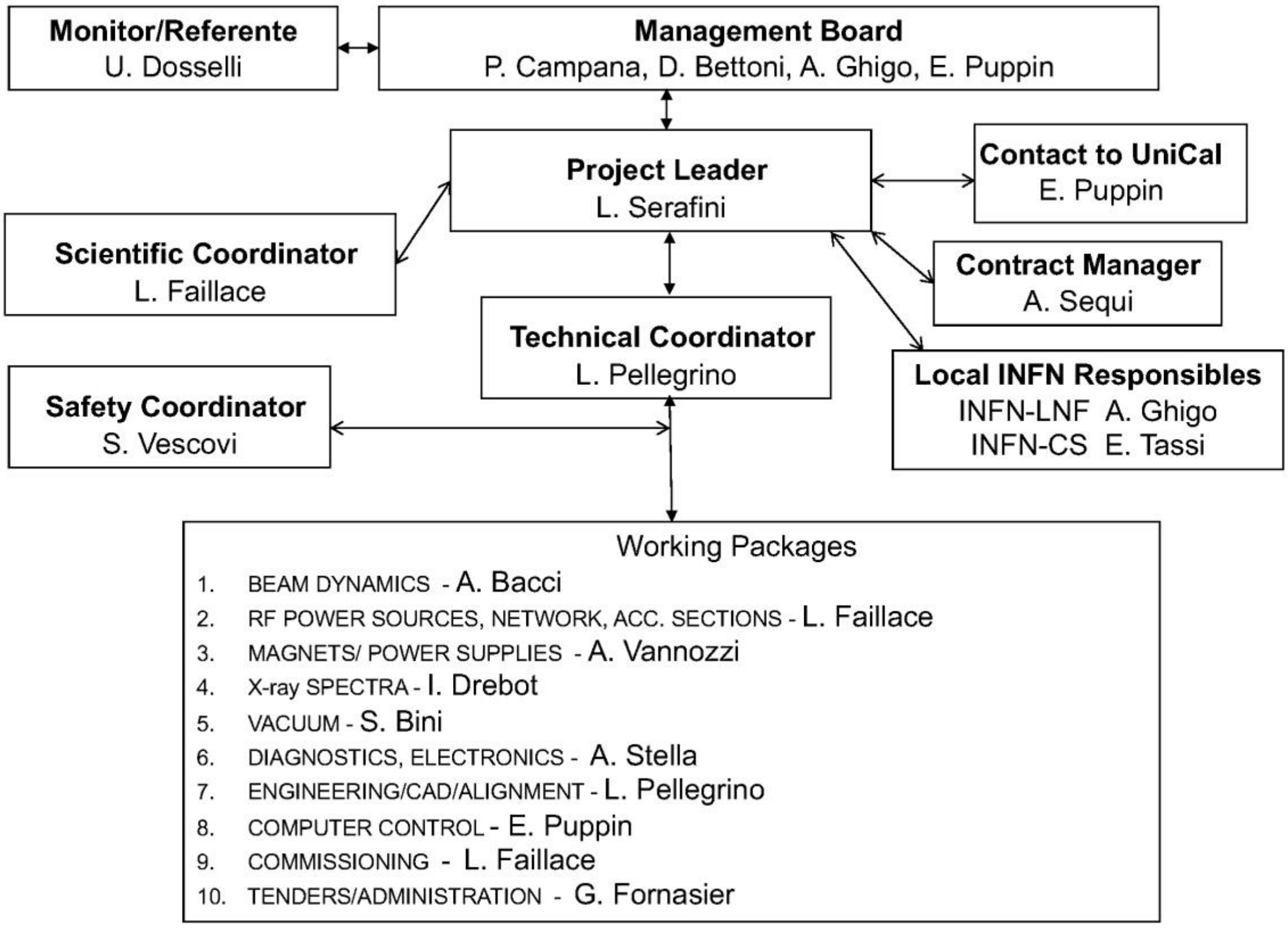

*Figure 24: Organization schematic*

## 10.2. Accelerator Breakdown Structure

The list of the components of the accelerator is reported in the Table 30. The blue lines refer to the STAR-1 installation and are therefore not included in the present contract. Nevertheless, they are presented for sake of clarity, to specify the function and location of the components of STAR-HEL upgrade.

A matrix form has been preferred in the context of the CDDR to have a functional view of the complex. The naming is defined in the next paragraph. We list a summary of the component's quantity at the end of the Table 30.

*Table 30: Accelerator Breakdown structure – matrix schematic*

| LINE COMPONENT | | HARDWARE | | | | |
|---|---|---|---|---|---|---|
| DESCRIPTION | NAMING | MAGNET POWER SUPPLY | POLARITY REVERSAL | VACUUM PUMP POWER SUPPLY | BPM DAQ / YAG SCREEN hardware / BCM DIGITIZER | LLRF SYSTEM |
| **LINAC STAR-1** | | | | | | |
| GUN ADDITIONAL VACUUM PUMP | IOPGUN00 | | | PS-GUNadd | | |
| GUN | GUNGUN01 | | | | | |
| SOLENOID | SOLGUN01 | PS-SOLGUN01 | | | | |
| GATE VALVE | GVLGUN01 | | | | | |
| INJECTION CHAMBER | INJGUN01 | | | | YAG-INJGUN01 | |
| ION PUMP | IOPGUN01 | | | PS-IOPGUN01 | | |
| STEERING MAGNET | CORGUN01 | PS-CORGUN01 | | | | |
| BEAM CURRENT MONITOR | BCMGUN01 | | | | DIG-BCMGUN01 | |
| BEAM POSITION MONITOR | BPMGUN01 | | | | DAQ-BPMGUN01 | |
| FAST VALVE | FVLGUN01 | | | | | |
| STEERING MAGNET | CORGUN02 | PS-CORGUN02 | | | | |
| DIAG CHAMBER | FLGGUN01 | | | | YAG-FLGGUN01 | |
| ION PUMP | IOPGUN02 | | | PS-IOPGUN02 | | |
| SOLENOID (before AC1LIN01) | SOLLIN01 | PS-SOLLIN01 | | | | |
| ACCELERATOR SECTION | AC1LIN01 | | | | | |
| STEERING MAGNET | CORLIN01 | PS-CORLIN01 | | | | |
| STEERING MAGNET | CORLIN02 | PS-CORLIN02 | | | | |
| STEERING MAGNET | CORLIN03 | PS-CORLIN03 | | | | |
| STEERING MAGNET | CORLIN04 | PS-CORLIN04 | | | | |
| | | | | | | |
| **UPGRADE LINAC** | | | | | YAG-FLGLIN01 | |
| DIAG CHAMBER | FLGLIN01 | | | | | |
| ION PUMP | IOPLIN01 | | | PS-IOPLIN01 | | |
| STEERING MAGNET | CORLIN05 | PS-CORLIN05 | | | | |
| ACCELERATOR SECTION | ACCLIN02 | | | | YAG-FLGLIN02 | |
| DIAG CHAMBER | FLGLIN02 | | | | | |
| ION PUMP | IOPLIN02 | | | PS-IOPLIN02 | | |
| BEAM POSITION MONITOR | BPMLIN01 | | | | DAQ-BPMLIN01 | |
| GATE VALVE | GVLDRF01 | | | | | |
| STEERING MAGNET | CORLIN06 | PS-CORLIN06 | | | | |
| ACCELERATOR SECTION | ACCLIN03 | | | | | |
| STEERING MAGNET | CORLIN07 | PS-CORLIN07 | | | | |
| QUADRUPOLE | QUATRL01 | PS-QUATRL01 | | | | |
| QUADRUPOLE | QUATRL02 | PS-QUATRL02 | | | | |
| QUADRUPOLE | QUATRL03 | PS-QUATRL03 | | | | |
| GATE VALVE | GVLTRL01 | | | | | |
| STEERING MAGNET | CORLIN08 | PS-CORLIN08 | | | | |
| BEAM POSITION MONITOR | BPMLIN02 | | | | DAQ-BPMLIN02 | |
| DIAG CHAMBER | FLGTRL01 | | | | | |
| ION PUMP | IOPTRL01 | | | | | |

| LINE COMPONENT | | HARDWARE | | | | |
|---|---|---|---|---|---|---|
| DESCRIPTION | NAMING | MAGNET POWER SUPPLY | POLARITY REVERSAL | VACUUM PUMP POWER SUPPLY | BPM DAQ / YAG SCREEN hardware / BCM DIGITIZER | LLRF SYSTEM |
| DIPOLE | DHPHEL01 | PS-DHPHEL01 | | | | |
| | | | | | | |
| LE LINE STAR-1 (MOVED LEFT) | | | | | | |
| DIAG CHAMBER | FLGDGL01 | | | | YAG-FLGDGL01 | |
| ION PUMP | IOPDGL01 | | | PS-IOPDGL01 | | |
| QUADRUPOLE | QUADGL01 | PS-QUADGL01 | | | | |
| BEAM POSITION MONITOR | BPMDGL01 | | | | DAQ-BPMDGL01 | |
| STEERING MAGNET | CORDGL01 | PS-CORDGL01 | | | | |
| QUADRUPOLE | QUADGL02 | PS-QUADGL02 | | | | |
| BEAM POSITION MONITOR | BPMDGL02 | | | | DAQ-BPMDGL02 | |
| STEERING MAGNET | CORDGL02 | PS-CORDGL02 | | | | |
| QUADRUPOLE | QUADGL03 | PS-QUADGL03 | | | | |
| BEAM POSITION MONITOR | BPMDGL03 | | | | DAQ-BPMDGL03 | |
| DIAG CHAMBER | FLGDGL02 | | | | YAG-FLGDGL02 | |
| ION PUMP | IOPDGL02 | | | PS-IOPDGL02 | | |
| DIPOLE | DHPDGL02 | PS-DHPDGL02 | | | | |
| GATE VALVE | GVLIPL01 | | | | | |
| BEAM CURRENT MONITOR | BCMIPL01 | | | | DIG-BCMIPL01 | |
| STEERING MAGNET | CORIPL01 | PS-CORIPL01 | | | | |
| DIAG CHAMBER | FLGIPL01 | | | | YAG-FLGIPL01 | |
| ION PUMP | IOPIPL01 | | | PS-IOPIPL01 | | |
| GATE VALVE | GVLIPL02 | | | | | |
| QUADRUPOLE | QUAIPL01 | PS-QUAIPL01 | | | | |
| BEAM POSITION MONITOR | BPMIPL01 | | | | DAQ-BPMIPL01 | |
| QUADRUPOLE | QUAIPL02 | PS-QUAIPL02 | | | | |
| QUADRUPOLE | QUAIPL03 | PS-QUAIPL03 | | | | |
| BEAM POSITION MONITOR | BPMIPL02 | | | | DAQ-BPMIPL02 | |
| STEERING MAGNET | CORIPL02 | PS-CORIPL02 | | | | |
| INTERACTION CHAMBER | INTIPL01 | | | | | |
| ION PUMP | IOPIPL02 | | | PS-IOPIPL02 | | |
| GATE VALVE | GVLIPL03 | | | | | |
| VERTICAL DIPOLE | DPVDUM01 | PS-DPVDUM01 | | | | |
| QUADRUPOLE | QUADUM01 | PS-QUADUM01 | PR-QUADUM01 | | | |
| DIAG CHAMBER | FLGDUM01 | | | | YAG-FLGDUM01 | |
| ION PUMP | IOPDUM01 | | | PS-IOPDUM01 | | |
| | | | | | | |
| HIGH ENERGY LINE | | | | | | |
| DIAG CHAMBER | FLGHEL01 | | | | YAG-FLGHEL01 | |
| ION PUMP | IOPHEL01 | | | PS-IOPHEL01 | | |
| QUADRUPOLE | QUAHEL01 | PS-QUAHEL01 | PR-QUAHEL01 | | | |
| BEAM POSITION MONITOR | BPMHEL01 | | | | DAQ-BPMHEL01 | |
| STEERING MAGNET | CORHEL01 | PS-CORHEL01 | | | | |
| QUADRUPOLE | QUAHEL02 | PS-QUAHEL02 | PR-QUAHEL02 | | | |

| LINE COMPONENT | | HARDWARE | | | | |
|---|---|---|---|---|---|---|
| **DESCRIPTION** | **NAMING** | **MAGNET POWER SUPPLY** | **POLARITY REVERSAL** | **VACUUM PUMP POWER SUPPLY** | **BPM DAQ / YAG SCREEN hardware / BCM DIGITIZER** | **LLRF SYSTEM** |
| BEAM POSITION MONITOR | BPMHEL02 | | | | DAQ-BPMHEL02 | |
| STEERING MAGNET | CORHEL02 | PS-CORHEL02 | | | | |
| QUADRUPOLE | QUAHEL03 | PS-QUAHEL03 | PR-QUAHEL03 | | | |
| DIAG CHAMBER | FLGHEL02 | | | | YAG-FLGHEL02 | |
| ION PUMP | IOPHEL02 | | | PS-IOPHEL02 | | |
| GATE VALVE | GVLHEL01 | | | | | |
| DIPOLE | DHPHEL02 | PS-DHPHEL02 | PR-DHPHEL02 | | | |
| DIAG CHAMBER | FLGHEL03 | | | | YAG-FLGHEL03 | |
| ION PUMP | IOPHEL03 | | | PS-IOPHEL03 | | |
| STEERING MAGNET | CORHEL03 | PS-CORHEL03 | | | | |
| GATE VALVE | GVLHEL03 | | | | | |
| STEERING MAGNET | CORHEL04 | PS-CORHEL04 | | | | |
| BEAM POSITION MONITOR | BPMHEL03 | | | | DAQ-BPMHEL03 | |
| QUADRUPOLE | QUAHEL04 | PS-QUAHEL04 | PR-QUAHEL04 | | | |
| BEAM CURRENT MONITOR | BCMHEL01 | | | | DIG-BCMHEL01 | |
| QUADRUPOLE | QUAHEL05 | PS-QUAHEL05 | PR-QUAHEL05 | | | |
| STEERING MAGNET | CORHEL05 | PS-CORHEL05 | | | | |
| QUADRUPOLE | QUAHEL06 | PS-QUAHEL06 | PR-QUAHEL06 | | | |
| BEAM POSITION MONITOR | BPMHEL04 | | | | DAQ-BPMHEL04 | |
| INTERACTION CHAMBER | INTHEL01 | | | | | |
| ION PUMP | IOPHEL04 | | | PS-IOPHEL04 | | |
| GATE VALVE | GVLHEL04 | | | | | |
| STEERING MAGNET | CORHEL06 | | | | | |
| DIAG CHAMBER | FLGHEL04 | | | | YAG-FLGHEL04 | |
| ION PUMP | IOPHEL04 | | | PS-IOPHEL04 | | |
| VERTICAL DIPOLE | DPVHEL01 | PS-DPVHEL01 | PR-DPVHEL01 | | | |
| DUMP | DUMHEL01 | | | | | |
| | | | | | | |
| **RADIOFREQUENCY SYSTEM** | | | | | | |
| MODULATOR | MODLIN01 | | | | | LLRF 1 |
| C-Band Klystron | KLYLIN01 | | | | | |
| Pumping Unit | WPULIN01 | | | PS-WPULIN01 | | |
| Pumping Unit | WPULIN02 | | | PS-WPULIN02 | | |
| Pumping Unit | WPULIN03 | | | PS-WPULIN03 | | |
| Pumping Unit | WPULIN04 | | | | | |
| Pumping Unit | WPULIN05 | | | | | |
| Pumping Unit | WPULIN06 | | | | | |
| Pumping Unit | WPULIN07 | | | | | |
| 4-P Hybrid Couplers -7.4 dB | COULIN01 | | | | | |
| 4-P Hybrid Couplers: -3 dB | COULIN02 | | | | | |
| Variable Attenuator | ATTLIN01 | | | | | |
| High Power Phase Shifter | PHSLIN01 | | | | | |
| S-BAND CIRCULATOR | CIRLIN01 | | | | | |

| LINE COMPONENT | | HARDWARE | | | | |
|---|---|---|---|---|---|---|
| **DESCRIPTION** | **NAMING** | **MAGNET POWER SUPPLY** | **POLARITY REVERSAL** | **VACUUM PUMP POWER SUPPLY** | **BPM DAQ / YAG SCREEN hardware / BCM DIGITIZER** | **LLRF SYSTEM** |
| Directional Coupler (60dB) | WDCLIN02 | | | | | |
| RF WAVEGUIDES LINE | WGLLIN01 | | | | | |
| RF window | RFWLIN01 | | | | | |
| MODULATOR | MODHEL02 | | | | | Libera LLRF 2 |
| C-Band Klystron | KLYHEL02 | | | | | |
| waveguide Directional Coupler (60dB) | WDCHEL03 | | | | | |
| Pumping Unit | WPUHEL04 | | | PS-WPUHEL04 | | |
| Pumping Unit | WPUHEL08 | | | PS-WPUHEL08 | | |
| Pumping Unit | WPUHEL09 | | | PS-WPUHEL09 | | |
| waveguide Directional Coupler (60dB) | WPUHEL10 | | | | | |
| RF WAVEGUIDES LINE | WGLHEL02 | | | | | |
| RF window | RFWHEL02 | | | | | |
| MODULATOR | MODHEL03 | | | | | Libera LLRF 3 |
| C-Band Klystron | KLYHEL03 | | | | | |
| waveguide Directional Coupler (60dB) | WDCHEL05 | | | | | |
| Pumping Unit | WPUHEL11 | | | PS-WPUHEL11 | | |
| Pumping Unit | WPUHEL12 | | | PS-WPUHEL12 | | |
| Pumping Unit | WPUHEL13 | | | PS-WPUHEL13 | | |
| waveguide Directional Coupler (60dB) | WDCHEL06 | | | | | |
| RF WAVEGUIDES LINE | WGLHEL03 | | | | | |
| RF window | RFWHEL03 | | | | | |
| | | | | | | |
| **SYNCHRONIZATION AND TIMING SYSTEM** | | | | | | |
| | | | | | | |
| **TOTALE COMPONENTI** | 130 | 42 | 9 | 23 | 27 | 3 |
| **TOTALE COMPONENTI STAR-1** | 70 | 25 | 0 | 11 | 15 | 1 |
| **TOTALE COMPONENTI STAR-HEL** | 60 | 17 | 9 | 12 | 12 | 2 |

## 10.3. Naming

The name of the components in the accelerator lines are composed by:

- Family name+zone+2digits

The names of the related electronic hardware are composed by:

- a prefix + the component name.

The following tables give the key for reading the names.

*Table 31: Naming families*

| SUBSYSTEM | FAMILY | DESCRIPTION |
|---|---|---|
| MAGNETS | QUA | Quadrupole |
| | DPH | Horizontal Dipole |
| | DPV | Vertical Dipole |
| | COR | Steering Magnet |
| | | |
| DIAGNOSTIC | BCM | Beam Current Monitor |
| | FLG | Screen viewer |
| | BPM | Beam Position Monitor |
| | | |
| DUMP | DUM | Dump |
| | | |
| RADIOFREQUENCY | WGL | Waveguide lines |
| | KLY | Klystron |
| | ACC | Accelerating section |
| | WPU | Waveguide pumping unit |
| | MOD | Modulator |
| | RFW | Window |
| | ATT | Attenuator |
| | PHS | Phase shifter |
| | CIR | Circulator |
| | COU | Coupler |
| VACUUM | | |
| | IOP | Ion Pump |
| | GVL | Gate Valve |
| | FVL | Fast Valve |
| | CHA | Vacuum Chamber |
| | GAU | Vacuum Gauge |

*Table 32: Naming Zones*

| ZONE | DESCRIPTION |
|---|---|
| **STAR-1** | |
| GUN | Gun |
| LIN | Linac |
| TRL | Transfer Line |
| STR | Straight Line |
| DGL | Dogleg |
| IPL | Interaction Point Line |
| DUM | Dump Line |
| | |
| **STAR-HEL** | |
| HEL | High Energy Line |

*Table 33: Naming prefixes*

| PREFIX | DESCRIPTION |
|---|---|
| DAQ | BPM Data Acquisition Hardware |
| YAG | Yag screen Hardware |
| DIG | Digitalizer |
| PS | Power Supply |
| PR | Power reversal |

## 10.4. Work Breakdown Structure

A preliminary list of activities for the WBS and for the Time Schedule has been prepared. The final version of the WBS will follow.

*Table 34: WBS preliminary draft*

| |
|---|
| **DESIGN AND COSNTRUCTION STAR-HEL** |
| **CDDR** |
| **Milestone: CDDR APPROVED** |
| **PREPARATORY ACTIVITIES** |
| **SAFETY** |
| personnel fitness for work, training, safety equipment |
| Safety plan PSCG (UNICAL), notifica cantiere, CSP e CSE |
| safety plan POS (INFN) |
| Green light for INFN personnel |
| Sub-contractor safety plan POS |
| Green light for Sub contractor |
| **Radioprotection** |
| Dump LE, STL e HE definition |
| **LOGISTICS** |
| Preliminary inspections |
| Green light to main components procurement |
| Handling and delivery contract assignment |
| Provisionary Storage at LNF organization |
| Material delivery and handling at LNF |
| Preparation, pre-assembly and test |
| Material delivery to UNICAL |
| **Milestone: LE LINE MAIN COMPONENTS DELIVERED** |
| **LINAC AND LOW ENERGY LINE MODIFICATION** |
| **DESIGN** |
| Updating accelerator layout |
| Support design revision |
| Interaction vacuum chamber design |
| New RF lines design (AC2 e 3) and revision (AC1) |
| **CONSTRUCTION** |
| Alignment network check |
| Floor lining (laser tracker) |
| Floor anchor holes drilling, dump pits excavation |
| Cooling distribution upgrade |
| Vacuum breaking |
| Components dismounting and installation in the new position |
| Accelerating sections 1 and 2 installation |
| New interaction vacuum chamber installation |
| Alignment |
| Vacuum Leak test vacuum pumping |
| Waveguide line AC1 modification |
| Modulators and Klystrons installation |
| Waveguide lines AC2 and AC3 installation |
| DC cabling |
| Cooling hoses mounting |
| Automation and diagnostic cabling |
| Final alignment check |
| **NEW HIGH ENERGY LINE** |
| **DETAILED DESIGN** |
| HE CAD Layout definition |
| Definition of the Utility Matrix |
| Interconnection system routing design |
| Vacuum system detailed design |
| Power supplies detailed design |
| Lines and dipoles vacuum chamber detailed design |
| Interaction chamber detailed design |
| Dump detailed design |
| Supports detailed design |
| Diagnostics detailed design |
| Auxiliary plants detailed design |
| **Milestone: "TO BUILD" DESIGNED FINISHED** |
| **PROCUREMENTS AND PROCEDURES** |
| Professional appointment for auxiliary plant detailed design |
| Professional appointment for construction site direction |
| Long term procedure approval |
| RF stations procurement and SAT |
| HP Load S-band (water cooled) procurement |
| Supports procurement |
| vacuum chamber procurement |
| electric plant upgrade bid |
| cooling plant upgrade bid |
| Handling and mechanical mounting contract assignment |
| Civil work contract assignment |
| Control system upgrade contract assignment |
| Other minor procedures |
| **Milestone: RF STATIONS DELIVERED** |
| **CONSTRUCTION** |
| Civil work |
| HE line installation |
| RF station installation |
| Racks, power supply, electronics installation |
| Cooling system upgrade and hoses installations |
| electric plant upgrade and cabling |
| Other infrastructural work |
| Control system upgrade |
| Milestone: INSTALLATION COMPLETE |
| **START-UP & SITE ACCEPTANCE TEST (SAT)** |
| Plants |
| Electrical plant upgrade start-up |
| Cooling system upgrade start-up |
| Power supplies Start-up e SAT |
| Accelerator |
| Control system start-up |
| Diagnostic system Start-up |
| C-band power units SAT |
| Accelerator Module SAT |
| **END OF SAT** |
| **"as built" documentation** |
| **Milestone: "AS BUILT" DOCUMENTATION DELIVERED** |

## 10.5. 3D CAD lay-out

A complete CAD lay-out has been designed based on the beam dynamics simulation and including all the subsystems.
The position of the STAR-1 installation has been modified to accommodate the STAR-HEL upgrade.
In the next pages the table of the position of the accelerator components and a set of drawings and renderings are reported.
The reference frame is centered on the gun cathode surface, X is the axis along the beam line, Z is the vertical axis and the Y the horizontal axis, consequently. Note that beam dynamics results illustrated in section 1 use, as traditional in the particle accelerator community, the Z coordinate for the longitudinal propagation axis of the electron beam.

*Table 35: Accelerator lay-out coordinates*

| **LINAC** | | X | Y | Z |
|---|---|---|---|---|
| GUNGUN01 | | 0.00 | 0.00 | 0.00 |
| SOLGUN01 | | 200.00 | 0.00 | 0.00 |
| GVLGUN01 | | 400.00 | 0.00 | 0.00 |
| INJGUN01 | IOPGUN01 | 605.00 | 0.00 | 0.00 |
| CORGUN01 | | 896.00 | 0.00 | 0.00 |
| BCMGUN01 | | 976.00 | 0.00 | 0.00 |
| BPMGUN01 | | 1149.40 | 0.00 | 0.00 |
| FVLGUN01 | | 1330.00 | 0.00 | 0.00 |
| CORGUN02 | | 1410.00 | 0.00 | 0.00 |
| FLGGUN01 | IOPGUN02 | 1600.00 | 0.00 | 0.00 |
| SOLLIN01 | | 1780.00 | 0.00 | 0.00 |
| AC1LIN01 | | 1843.86 | 0.00 | 0.00 |
| CORLIN01 | | 2125.06 | 0.00 | 0.00 |
| CORLIN02 | | 2375.06 | 0.00 | 0.00 |
| CORLIN03 | | 4425.06 | 0.00 | 0.00 |
| CORLIN04 | | 4675.06 | 0.00 | 0.00 |
| FLGLIN01 | IOPLIN01 | 5050.30 | 0.00 | 0.00 |
| CORLIN05 | BPMLIN01 | 5252.80 | 0.00 | 0.00 |
| ACCLIN02 | | 5372.30 | 0.00 | 0.00 |
| FLGLIN02 | IOPLIN01 | 7400.00 | 0.00 | 0.00 |
| GVLDRF01 | | 7550.00 | 0.00 | 0.00 |
| CORLIN06 | | 7620.00 | 0.00 | 0.00 |
| ACCLIN03 | | 7660.10 | 0.00 | 0.00 |
| CORLIN07 | | 9715.65 | 0.00 | 0.00 |
| QUATRL01 | | 9900.00 | 0.00 | 0.00 |
| QUATRL02 | | 10210.00 | 0.00 | 0.00 |
| QUATRL03 | | 10520.00 | 0.00 | 0.00 |
| GVLTRL01 | | 10800.00 | 0.00 | 0.00 |
| CORLIN08 | BPMLIN02 | 11000.00 | 0.00 | |
| FLGTRL01 | IOPTRL01 | 11300.00 | 0.00 | 0.00 |
| DHPHEL01 | | 11670.00 | 0.00 | 0.00 |
| FLGSTR01 | IOPSTR01 | | | |
| | | | | |
| **LE LINE** | | linea LE | | |
| FLGDGL01 | IOPDGL01 | 12350.00 | 246.99 | 0.00 |
| QUADGL01 | BPMDGL01 | 12844.55 | 427.14 | 0.00 |
| CORDGL01 | | 13430.00 | 640.41 | 0.00 |
| QUADGL02 | BPMDGL02 | 14019.09 | 855.00 | 0.00 |
| CORDGL02 | | 14650.00 | 1084.83 | 0.00 |
| QUADGL03 | BPMDGL03 | 15193.64 | 1283.24 | 0.00 |
| FLGDGL02- | IOPDGL02 | 15700.00 | 1467.32 | 0.00 |
| DHPDGL02 | | 16368.19 | 1710.00 | 0.00 |
| GVLIPL01 | | 16762.66 | 1710.00 | 0.00 |
| BCMIPL01 | | 16883.29 | 1710.00 | 0.00 |
| CORIPL01 | | 16960.00 | 1710.00 | 0.00 |
| FLGIPL01 | IOPIPL01 | 17145.64 | 1710.00 | 0.00 |
| GVLIPL02 | | 17365.41 | 1710.00 | 0.00 |
| QUAIPL01 | BPMIPL01 | 17518.19 | 1710.00 | 0.00 |
| QUAIPL02 | | 17828.19 | 1710.00 | 0.00 |
| QUAIPL03 | BPMIPL02 | 18138.19 | 1710.00 | 0.00 |
| CORIPL02 | | 18260.00 | 1710.00 | 0.00 |
| INTIPL01 | IOPIPL02 | 18900.00 | 1710.00 | 0.00 |
| GVLIPL03 | | 20444.03 | 1710.00 | 0.00 |
| DPVDUM01 | | 21238.42 | 1710.00 | 0.00 |
| QUADUM01 | | 22500.00 | 1710.00 | -631.251 |
| FLGDUM01 | IOPDUM01 | 22900.00 | 1710.00 | -776.839 |
| | | | | |
| **HE LINE** | | linea HE | | |
| FLGHEL01 | IOPHEL01 | 12500.00 | -302.10 | 0.00 |
| QUAHEL01 | | 12844.55 | -427.14 | 0.00 |
| CORHEL01 | BPMHEL01 | 13450.00 | -647.87 | 0.00 |
| QUAHEL02 | | 14019.09 | -855.00 | 0.00 |
| CORHEL02 | BPMHEL02 | 14620.00 | -1073.71 | 0.00 |
| QUAHEL03 | | 15193.64 | -1283.24 | 0.00 |
| FLGHEL02 | IOPHEL02 | 15780.00 | -1495.92 | 0.00 |
| GVLHEL01 | | 16000.00 | -1575.99 | 0.00 |
| DHPHEL02 | | 16368.19 | -1710.00 | 0.00 |
| FLGHEL03 | IOPHEL03 | 16690.00 | -1710.00 | 0.00 |
| CORHEL03 | | 16906.00 | -1710.00 | 0.00 |
| GVLHEL03 | | 17064.54 | -1710.00 | 0.00 |
| CORHEL04 | BPMHEL03 | 17296.00 | -1710.00 | 0.00 |
| QUAHEL04 | | 17518.19 | -1710.00 | 0.00 |
| BCMHEL01 | | 17685.00 | -1710.00 | 0.00 |
| QUAHEL05 | | 17828.19 | -1710.00 | 0.00 |
| CORHEL05 | | 17990.00 | -1710.00 | 0.00 |
| QUAHEL06 | | 18138.19 | -1710.00 | 0.00 |
| BPMHEL05 | | 18400.00 | -1710.00 | 0.00 |
| INTHEL01 | IOPHEL04 | 18900.00 | -1710.00 | 0.00 |
| CORHEL06 | | 19900.00 | -1710.00 | 0.00 |
| GVLHEL04 | | 20100.00 | -1710.00 | 0.00 |
| FLGHEL04 | IOPHEL04 | 20350.00 | -1710.00 | 0.00 |
| DPVHEL01 | | 21298.21 | -1710.00 | -247.79 |
| DUMHEL01 | | 21546.00 | -1710.00 | -1200.00 |

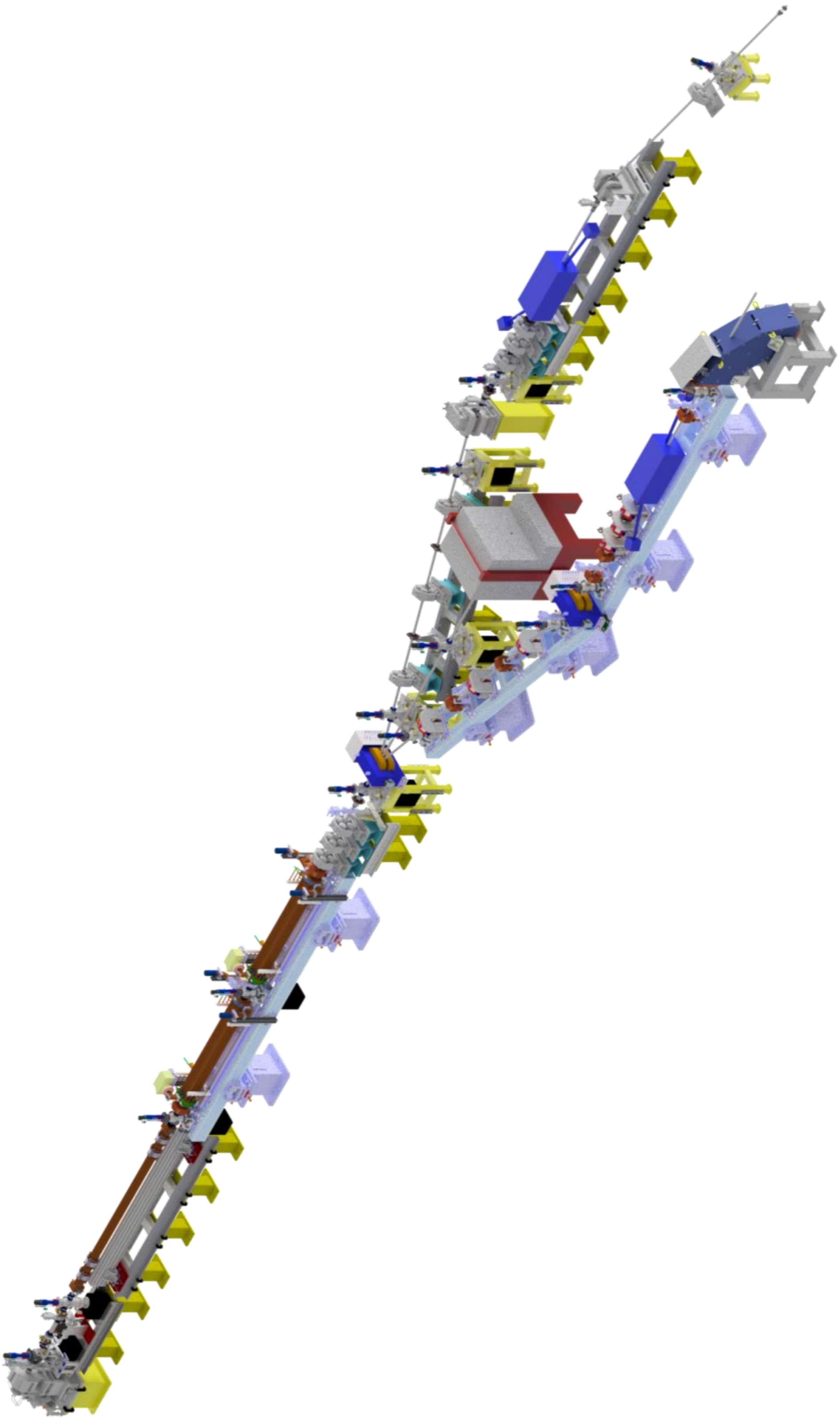

*Figure 25: Accelerator CAD rendering*

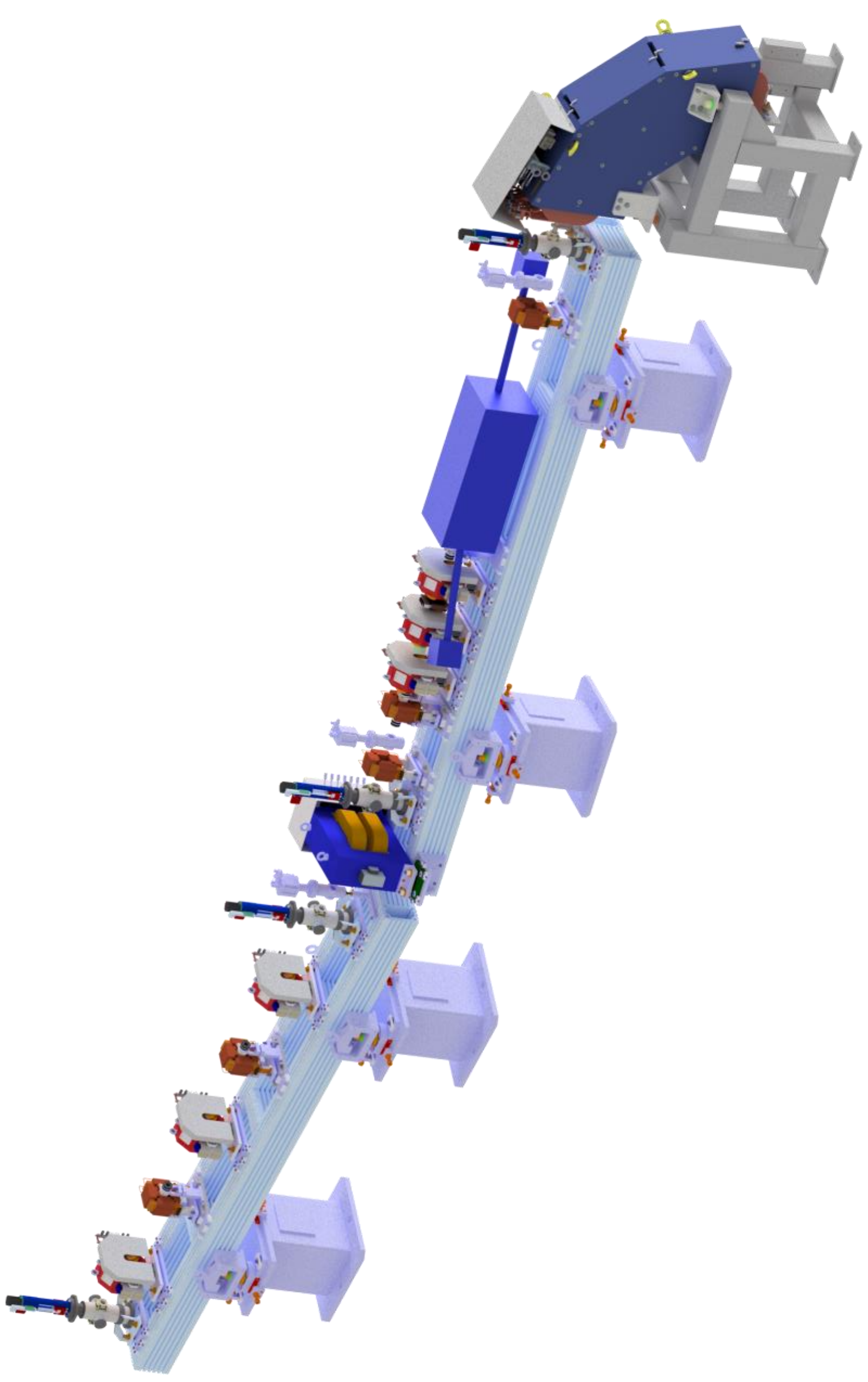

***Figure 26: HE-line CAD rendering***

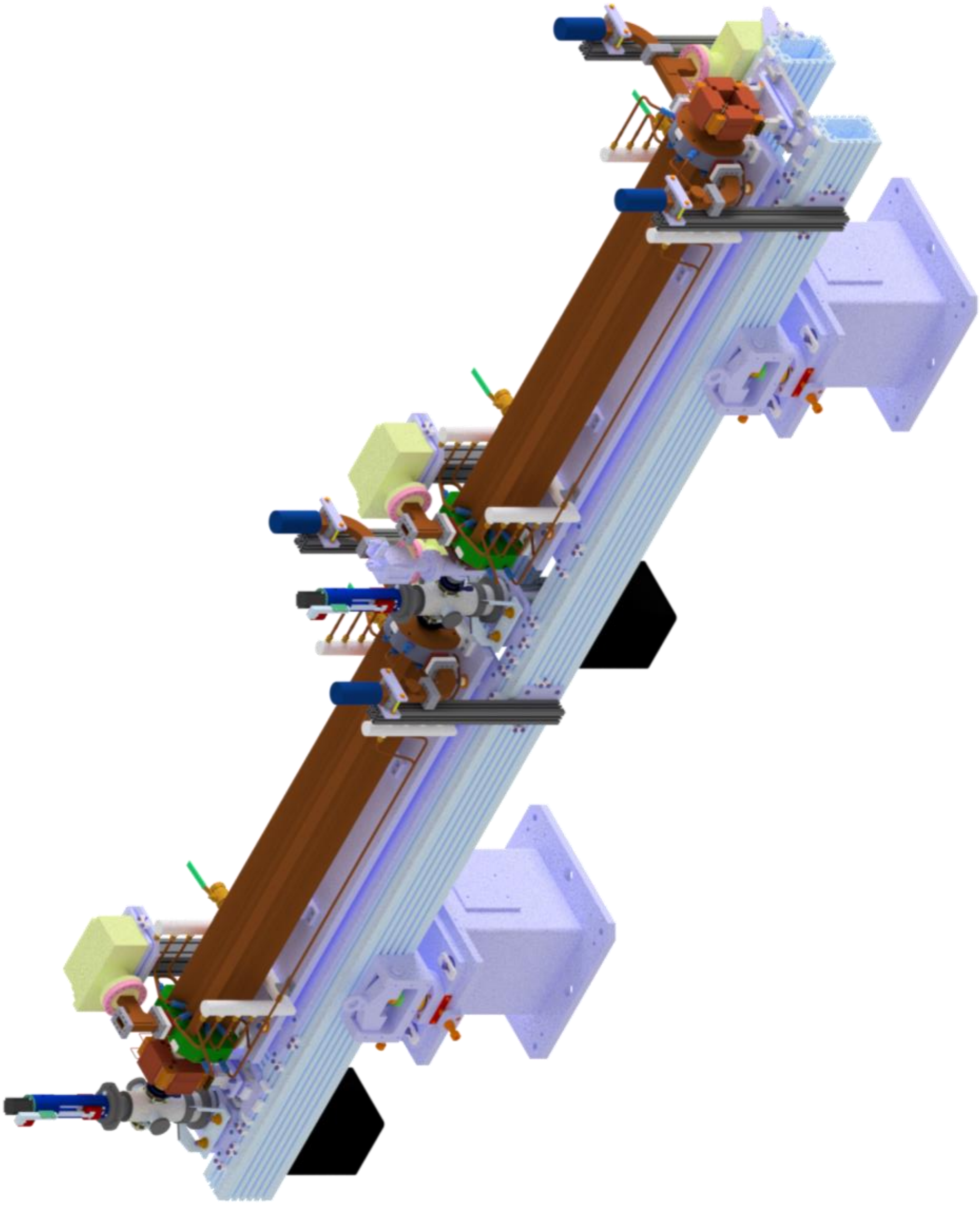

***Figure 27: RF module CAD rendering***

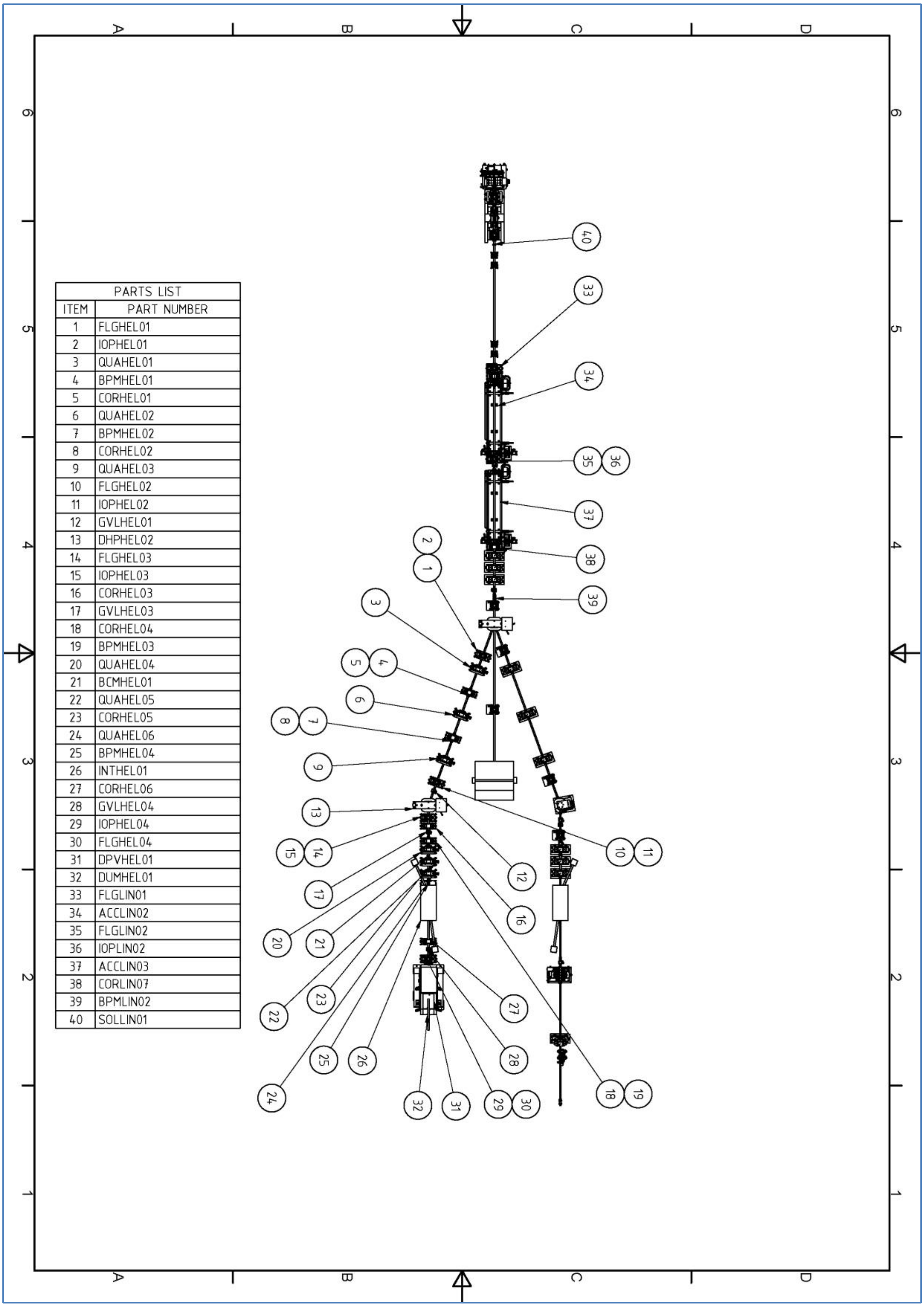

| PARTS LIST | |
|---|---|
| ITEM | PART NUMBER |
| 1 | FLGHEL01 |
| 2 | IOPHEL01 |
| 3 | QUAHEL01 |
| 4 | BPMHEL01 |
| 5 | CORHEL01 |
| 6 | QUAHEL02 |
| 7 | BPMHEL02 |
| 8 | CORHEL02 |
| 9 | QUAHEL03 |
| 10 | FLGHEL02 |
| 11 | IOPHEL02 |
| 12 | GVLHEL01 |
| 13 | DHPHEL02 |
| 14 | FLGHEL03 |
| 15 | IOPHEL03 |
| 16 | CORHEL03 |
| 17 | GVLHEL03 |
| 18 | CORHEL04 |
| 19 | BPMHEL03 |
| 20 | QUAHEL04 |
| 21 | BCMHEL01 |
| 22 | QUAHEL05 |
| 23 | CORHEL05 |
| 24 | QUAHEL06 |
| 25 | BPMHEL04 |
| 26 | INTHEL01 |
| 27 | CORHEL06 |
| 28 | GVLHEL04 |
| 29 | IOPHEL04 |
| 30 | FLGHEL04 |
| 31 | DPVHEL01 |
| 32 | DUMHEL01 |
| 33 | FLGLIN01 |
| 34 | ACCLIN02 |
| 35 | FLGLIN02 |
| 36 | IOPLIN02 |
| 37 | ACCLIN03 |
| 38 | CORLIN07 |
| 39 | BPMLIN02 |
| 40 | SOLLIN01 |

***Figure 28: accelerator naming***

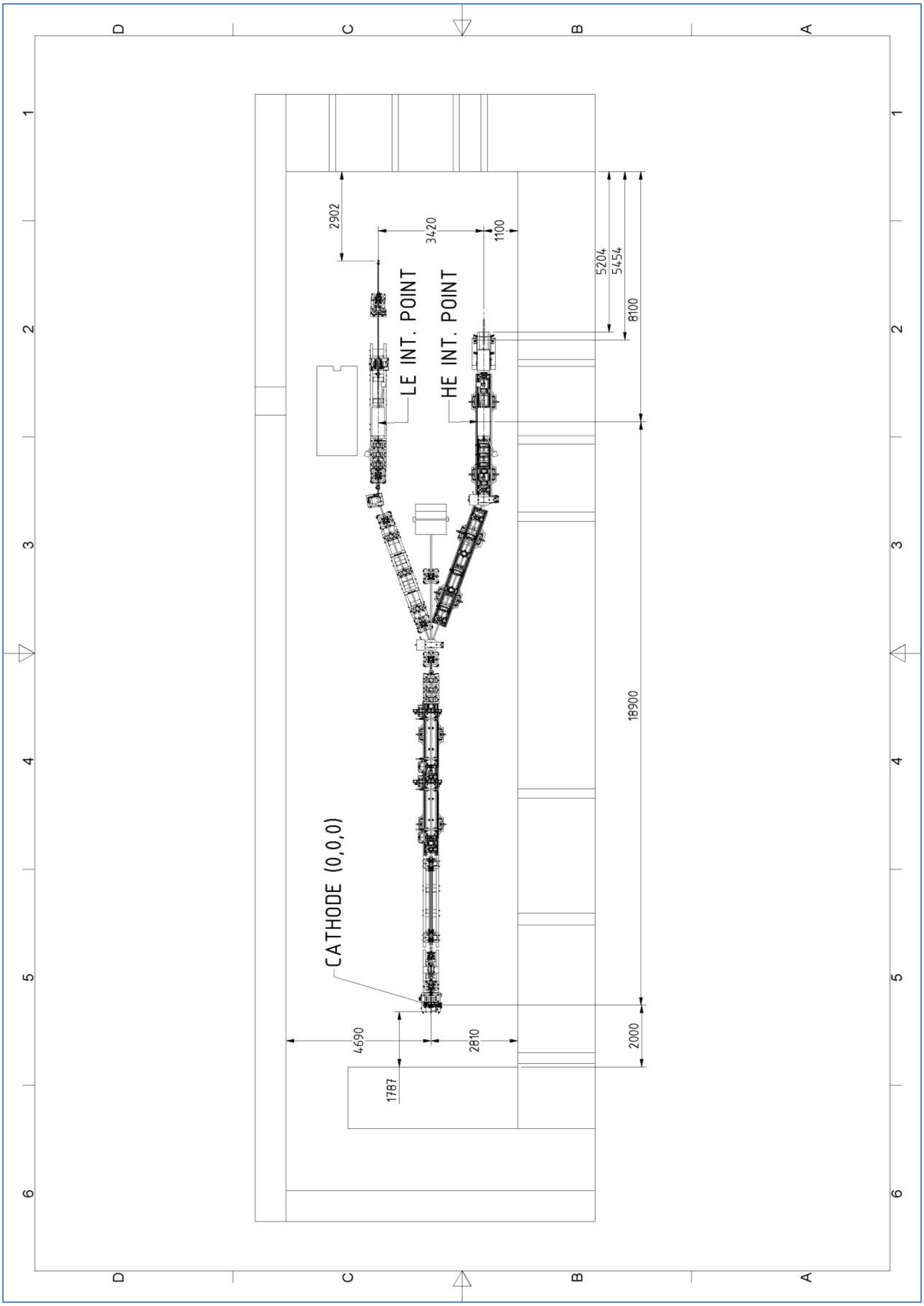

***Figure 29: accelerator placing in the Bunker***

## 10.6. Civil work

The necessary civil works include:
- The vertical dump (HEL) hole drilling;
- The LE dump pit excavation in the new position;
- The existing LE pit closing;
- The LE hole drilling in the end wall in the new position.

The design of each of the above items has been defined in detail. The foreseen civil works will take place inside the bunker; therefore, no interference is foreseen with other parties.

## 10.7. Alignment

In 2016 the LNF alignment group realized the alignment network, by laser tracker measurements inside the bunker. Few reference points have been taken also in the clean room and in the experimental hall.
The 1.5" targets for the laser tracker have been mounted on permanent wall-mounted sockets, floor sockets and cantilevered sockets (at the inlet of the holes in the walls).
With the network framework reference the STAR-1 LINAC and the LE-line has been positioned. The reference network is represented in Figure 30, The points effectively measured are summarized in the Table 36 and represented in the Figure 31. A typical accuracy of component positioning in the bunker obtained with this technique is within 0.1 mm.
The first operation to be carried out for the STAR-HEL upgrade, just before any activity, must be the network check. Small movements, given by concrete settlement within years, are expected, and should be taken in to account in order to perform the alignment of the new installation as well as the check of the old one.

***Table 36: Points of the alignment network***

| Nome Punto | X | Y | Z | Nome Punto | X | Y | Z |
|---|---|---|---|---|---|---|---|
| 1 | -2033,94 | -292,96 | 892,22 | 32 | 24909,74 | -2737,2 | 188,9 |
| 2 | -2035,9 | -294,06 | 192,07 | 33 | 21911,95 | -2742,4 | 898,85 |
| 3 | -2039,41 | -299,67 | -707,57 | 34 | 21908,95 | -2742,27 | 193,72 |
| 4 | -2031,76 | 2206,49 | 897,88 | 35 | 18913,54 | -2735,96 | 892,05 |
| 5 | -2035,95 | 2205,39 | 191,61 | 36 | 18907,57 | -2738,06 | 193,42 |
| 6 | -2039,87 | 2213,17 | -707,81 | 37 | 12914,05 | -2741,35 | 896,52 |
| 7 | 933,22 | 4621,2 | 894,8 | 38 | 12913,08 | -2742,16 | 196,35 |
| 8 | 931,06 | 4633,99 | 198,31 | 39 | 9917,85 | -2742,91 | 895,15 |
| 9 | 936,46 | 4642,08 | -713,12 | 40 | 9912,14 | -2742,65 | 191,18 |
| 10 | 4180,51 | 4629,85 | 898,06 | 41 | 6907,92 | -2744,84 | 899,67 |
| 11 | 4181,09 | 4637,68 | 200,49 | 42 | 6909,94 | -2748,29 | 203,41 |
| 12 | 4186,06 | 4643,28 | -707,17 | 43 | 3908,7 | -2738,48 | 898,56 |
| 13 | 6936,25 | 4630,84 | 905,77 | 44 | 3907,97 | -2740 | 203,21 |
| 14 | 6936,06 | 4641,42 | 199,82 | 45 | 912,32 | -2736,6 | 893,73 |
| 15 | 6933,71 | 4647,3 | -705,62 | 46 | 909,15 | -2738,02 | 190,71 |
| 16 | 9938,24 | 4624,71 | 893,53 | 47 | 710,51 | -2739,15 | 195,08 |
| 17 | 9938,48 | 4635,7 | 190,06 | 48 | 935,19 | 1704,89 | -1207,93 |
| 18 | 9939,1 | 4646,74 | -710,68 | 49 | 3932,02 | 1724,18 | -1202,26 |
| 19 | 12937,13 | 4632,62 | 893,52 | 50 | 6930,28 | 1789,26 | -1200,06 |
| 20 | 12933,34 | 4636,45 | 190,89 | 51 | 9933,85 | 1817,17 | -1205,99 |
| 21 | 12935,67 | 4643,36 | -715,3 | 52 | 12934,82 | 1845,2 | -1207,51 |
| 22 | 15935,73 | 4637,53 | 895,13 | 53 | 15933,87 | 1904,2 | -1207,73 |
| 23 | 15930,32 | 4644,03 | 185,86 | 54 | 18933,92 | 1978,49 | -1206,08 |
| 24 | 15929,88 | 4652,54 | -716,85 | 55 | 21934,69 | 2000,57 | -1204,17 |
| 25 | 21931,84 | 4641,81 | 890,3 | 56 | 24932,48 | 2065,45 | -1205,67 |
| 26 | 21931,02 | 4645,89 | 187,41 | 57 | 26959,04 | 3205,17 | 444,24 |
| 27 | 21926,77 | 4651,03 | -717,19 | 58 | 26975,66 | 1205,38 | 465,39 |
| 28 | 25373,02 | 4645,67 | 891,98 | 59 | 26977,13 | -804,33 | 472,85 |
| 29 | 25368,58 | 4650,75 | 188,62 | 60 | 20699,62 | -2767,65 | 429,84 |
| 30 | 25353,99 | 4656,22 | -702,69 | 61 | 18199,91 | -2761,86 | 424,34 |
| 31 | 24908,97 | -2736,13 | 887,95 | 62 | 15699,98 | -2767,04 | 414,84 |
| Cp_1 | 20777,91 | -13434,5 | 454,95 | Ext_1 | 29518,11 | 3218,64 | 463,95 |
| Cp_2 | 18120,91 | -13440,7 | 451,68 | Ext_2 | 29526,44 | 1191,12 | 462,87 |
| Cp_3 | 15705,21 | -13441,6 | 466,61 | Ext_3 | 29539,88 | -800,88 | 462,76 |
| Foro1 | 26978,63 | -1708,24 | -38,68 | Foro3 | 2714,13 | -2769,23 | 1721,34 |
| Foro2 | 6709,63 | -2765,33 | 1692,51 | Foro4 | -178381 | -2757,15 | 1705,14 |
| | | | | | | | |

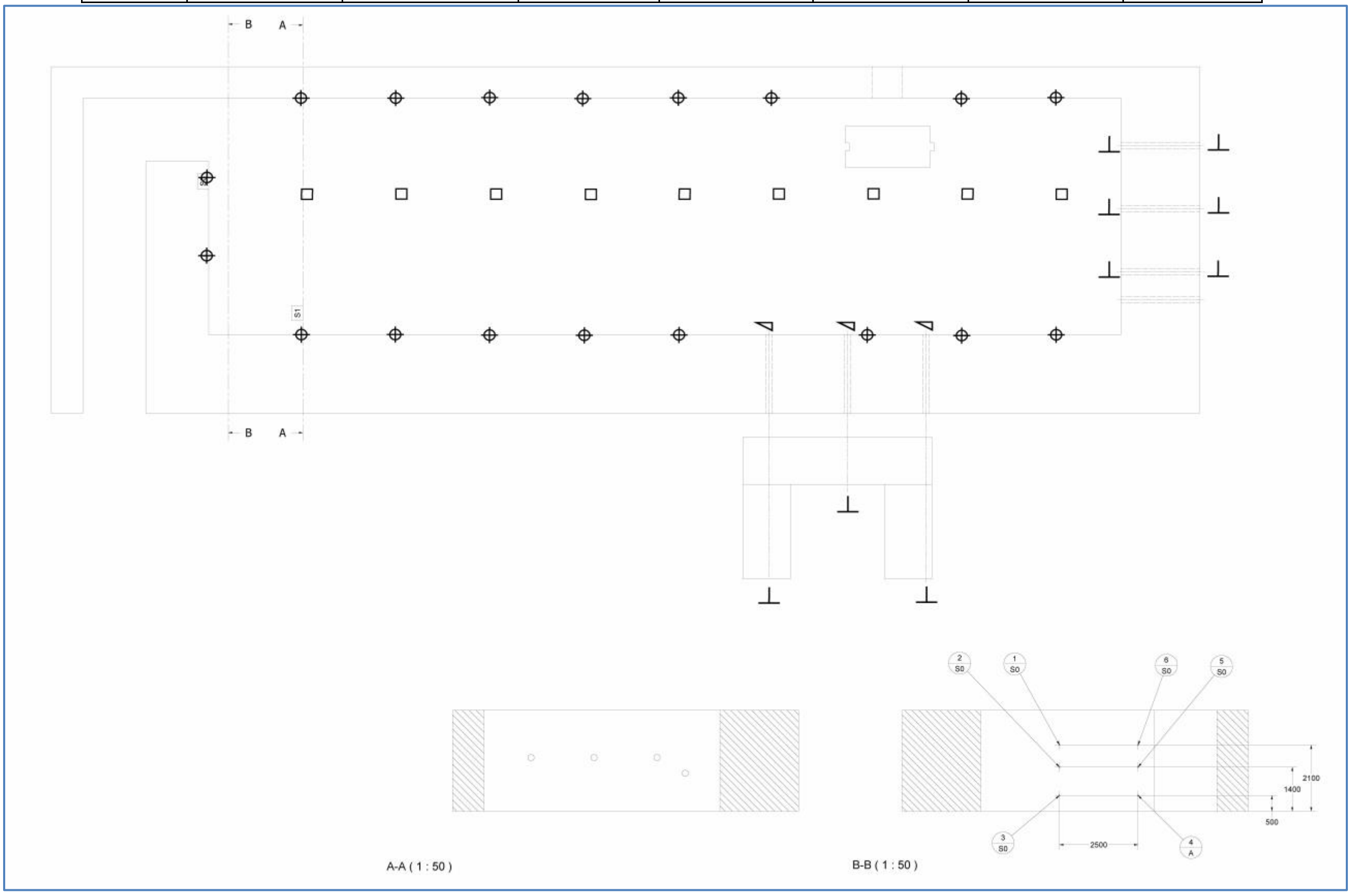

***Figure 30: the design of the STAR-1 reference network***

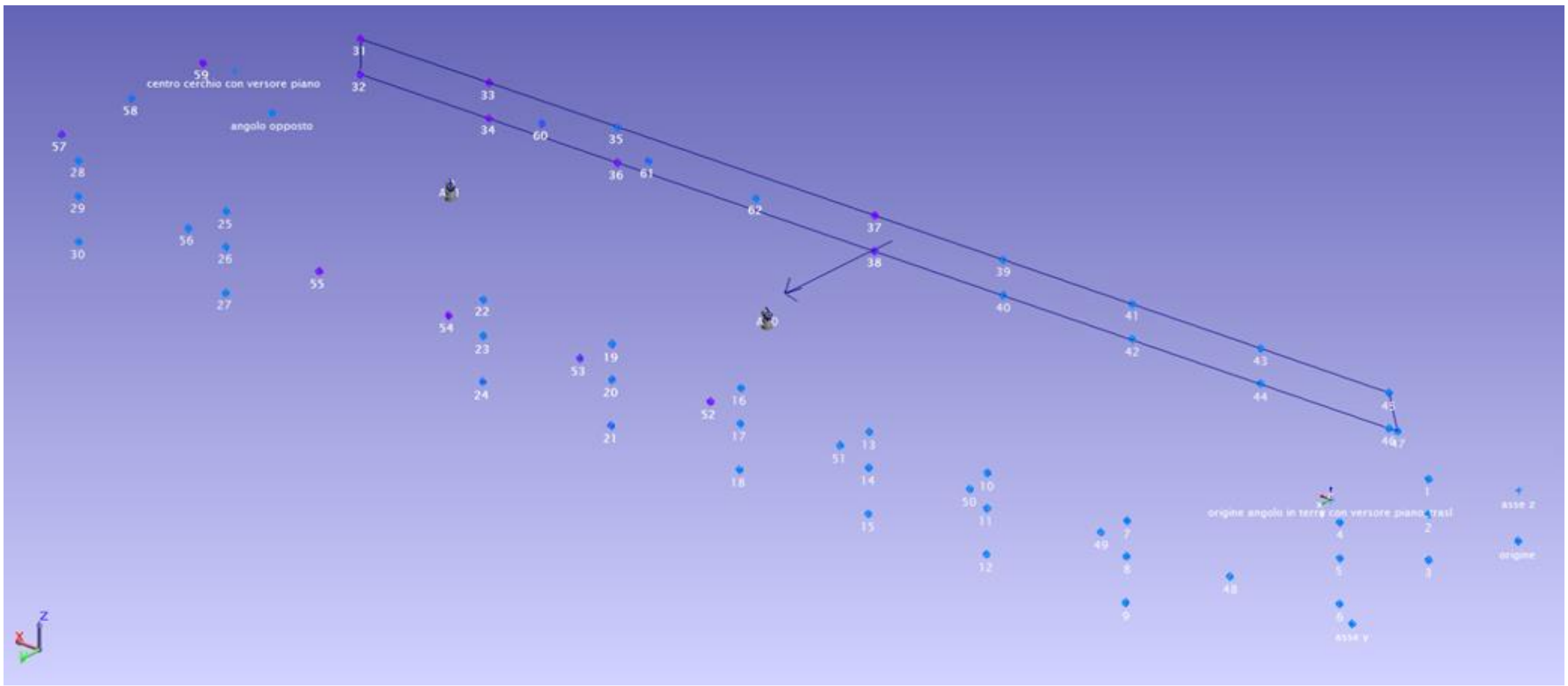

***Figure 31: The measured points inside the STAR bunker***

### 10.8. Time Schedule

A detailed analysis of all the phases of the STAR-HEL project has been performed. The final version of the time schedule will be issued after the CDDR approval and the kick-off meeting, as well as after all the necessary on-site surveys and the functional check of the existing infrastructure.
The present draft of the Master Plan is reported in Figure 32.

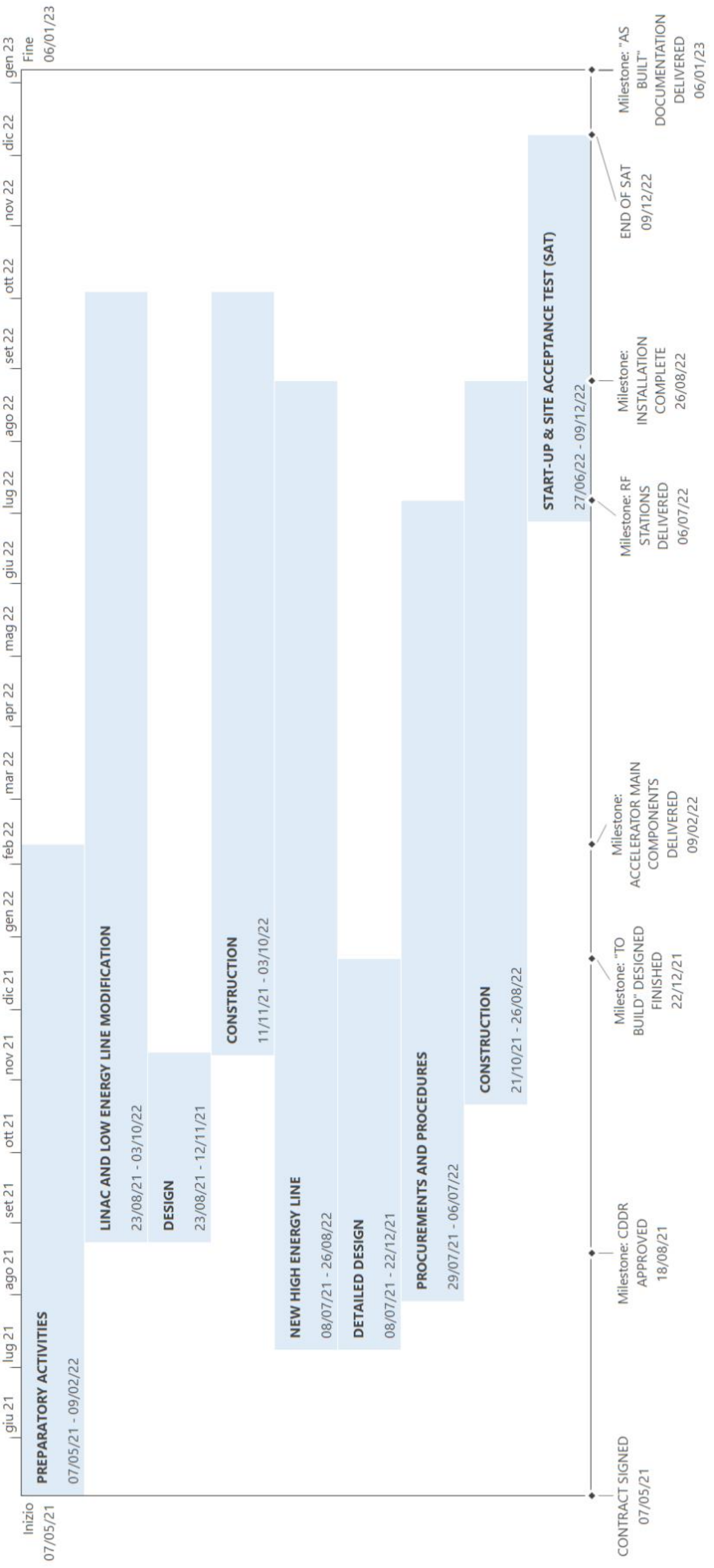

***Figure 32: Master Plan***

# 11. Site Acceptance Tests

## 11.1. C-band power units

The RF power unit will undergo a first series of tests at the manufacturer's premises, including the RF tests, prior to delivery, and a second series of tests, constituting the acceptance tests, at the STAR infrastructure. Representatives of the RF power unit manufacturer and INFN will participate in both tests.
The tests at the STAR infrastructure, which will be carried out within 20 weeks of delivery of the RF power unit, are followed by a non-stop 100-hour run at rated output power into a high-power load. After the successful completion of the 100-hour run the unit is considered as accepted.

## 11.2. Accelerator Module RF Conditioning

The RF conditioning of the two C-Band accelerating sections will be to achieve high-gradient operation at the klystron nominal power of 42 MW at a repetition rate of 100 Hz with the nominal pulse width of 1μs. Specifically, the forward and reflected RF power signals from the directional couplers located at the input and output linac couplers will be measured by using diodes and oscilloscopes. A LabView GUI interface will be used to read these signals and output the corresponding power levels during the high-power testing process.
The klystron power will be gradually increased together with the RF pulse repetition rate and length. During all these steps, the current signals from the ion pumps and the RF signal from the pickups will be monitored. The conditioning procedure will be possible in a semi-automatic way. Indeed, the interlock of the HV modulator will be enabled in one of the three following cases:

- manually by the operators in the control room;
- automatically, if the signal of the current absorption from the ion pumps exceeds a certain level corresponding to a pressure of $1x10^{-7}$ mbar;
- automatically, if the reflected RF power back to the klystron exceeds about 10%.

The time estimation for the full conditioning of both C-Band accelerating sections is about 200 hours.

## 11.3. Magnets and Power Supplies

The magnets S.A.T. foresees a magnetic field measurement by means of a portable hall probe, aiming to check the polarity of the magnetic field and to have a very rough estimation of its order of magnitude.
Regarding the power supplies, they will undergo to the following tests:

- Current Ripple
- Long term stability
- External interlock test

Moreover, the remote-control capability shall be tested verifying a matching between the remote current set and the measured current output.